\documentclass[12pt, letterpaper]{article}
\usepackage[utf8]{inputenc}
\usepackage{graphicx, tikz}
\usepackage{amsfonts, amsmath, amssymb}
\usepackage{array, setspace, mathrsfs, yfonts, dsfont, bbm, colonequals, amscd, mathtools}
\usepackage{relsize, suffix, cancel, bbm, capt-of, upgreek, verbatim, xcolor}
\usepackage{dsfont}
\usepackage{jheppub}

\numberwithin{equation}{section}
\allowdisplaybreaks

\newcommand{\cA}{\mathcal{A}}

\newcommand{\cC}{\mathcal{C}}

\newcommand{\cF}{\mathcal{F}}
\newcommand{\cG}{\mathcal{G}}
\newcommand{\cH}{\mathcal{H}}

\newcommand{\cK}{\mathcal{K}}
\newcommand{\cL}{\mathcal{L}}
\newcommand{\cM}{\mathcal{M}}
\newcommand{\cN}{\mathcal{N}}

\newcommand{\cW}{\mathcal{W}}

\newcommand{\Z}{\mathbb{Z}}

\newcommand{\R}{\mathbb{R}}

\newcommand{\m}{\mathfrak{m}}
\newcommand{\n}{\mathfrak{n}}

\def\bea{\begin{align}}
\def\eea{\end{align}}
\def\be{\begin{equation}}
\def\ee{\end{equation}}

\begin{document}

\title{\Large 3D-3D Correspondence from Seifert Fibering Operators}

\author{Yale Fan}

\affiliation{Department of Physics, Princeton University, Princeton, NJ 08544, USA}

\abstract{Using recently developed Seifert fibering operators for 3D $\mathcal{N} = 2$ gauge theories, we formulate the necessary ingredients for a state-integral model of the topological quantum field theory dual to a given Seifert manifold under the 3D-3D correspondence, focusing on the case of Seifert homology spheres with positive orbifold Euler characteristic.  We further exhibit a set of difference operators that annihilate the wavefunctions of this TQFT on hyperbolic three-manifolds, generalizing similar constructions for lens space partition functions and holomorphic blocks.  These properties offer intriguing clues as to the structure of the underlying TQFT.}

\maketitle
\setcounter{page}{1}

\section{Introduction}

A broad goal of the supersymmetric localization program is to exploit the locality of quantum field theory to find fundamental building blocks of supersymmetric partition functions and observables.  In this regard, a powerful point of view is that line operators in the field theory can be used to modify the background geometry on which it resides.  Such an idea traces back at least to work of Blau and Thompson \cite{Blau:1993tv, Blau:2006gh} on Chern-Simons theory, but has recently been shown to generalize to arbitrary three-dimensional quantum field theories with at least $\mathcal{N} = 2$ supersymmetry \cite{Closset:2018ghr}.  These ``fibering operators'' allow one to interpret observables in a 3D $\mathcal{N} = 2$ theory as observables in an auxiliary two-dimensional topological quantum field theory, and are related to special limits of holomorphic blocks \cite{Beem:2012mb}.  They form the crux of recently developed methods to compute partition functions of 3D $\mathcal{N} = 2$ theories on Seifert manifolds, as reviewed in \cite{Closset:2019hyt}.

A complementary organizing principle to the idea of decomposing observables into simpler pieces --- one that comes from the top down rather than the bottom up --- is that of deriving lower-dimensional field theories and their dualities by compactifying the 6D $(2, 0)$ superconformal field theory on various backgrounds.  One realization of this approach is the 3D-3D correspondence \cite{Dimofte:2011ju, Dimofte:2011py}, which posits a duality between a 3D $\mathcal{N} = 2$ SCFT $T[M]$ associated to a three-manifold $M$ when placed on a different three-manifold $\mathcal{M}_3$ and a (non-supersymmetric) TQFT on $M$, associated to $\mathcal{M}_3$.  Roughly, it goes as follows.  Compactifying the 6D $(2, 0)$ theory on $M$ with a suitable topological twist leads to an effective 3D $\mathcal{N} = 2$ theory $T[M]$.  Under favorable conditions, this theory flows to an SCFT that depends only on the topology of $M$.  We may then use the $U(1)$ $R$-symmetry to couple this 3D $\mathcal{N} = 2$ theory to another three-manifold $\mathcal{M}_3$ and compute its partition function (among other observables), which should give rise to a topological invariant of $M$.  In other words, we expect the theory $T[M]$ on $\mathcal{M}_3$ to be dual to a TQFT on $M$, associated to $\mathcal{M}_3$.

As we review below, the structure of this correspondence between geometry and field theory, and that of the theories $T[M]$, has been probed for an enormous variety of three-manifolds $M$ --- indeed, for ``most'' three-manifolds.  By comparison, the correspondence has thus far been understood for a sparse list of three-manifolds $\mathcal{M}_3$, comprised of lens spaces and special cases thereof.  In light of this asymmetry, the techniques of \cite{Closset:2018ghr} vastly expand the arena of computability of the $T[M]$ partition functions, and thereby hold promise for the construction of previously unexplored topological field theories in three dimensions.  A natural question then arises:
\begin{quote}
What is the 3D TQFT dual to $\mathcal{M}_3$ under the 3D-3D correspondence, in the case that $\mathcal{M}_3$ is a Seifert manifold?
\end{quote}
This is the question that we aim to address in this paper.

We should warn the reader from the start that we will not be able to answer this question directly.  Instead, we set ourselves the more modest goal of leveraging the tools of \cite{Closset:2018ghr} to understand some concrete properties of the putative TQFTs dual to Seifert manifolds under the 3D-3D correspondence.  For simplicity, we restrict to $M$ hyperbolic throughout this paper and focus primarily on the theories $T_2[M]$ descending from the $(2, 0)$ theory of type $A_1$.  We address both analytic and algebraic aspects of the correspondence.

On the analytic side, we construct a state-integral model for (a subsector of) this TQFT at the level of the $\mathcal{M}_3$ partition function of $T_2[M]$, along the lines of \cite{Dimofte:2014zga} for lens spaces.  An important element of our construction is that for $\mathcal{M}_3$ to have a TQFT interpretation, gluing of $T_2[M]$ partition functions should be implemented by a theory-independent integration contour.  We derive the linear integral identities for the elementary mirror symmetries needed to define a state-integral model for an arbitrary Seifert manifold, and show that the existence of this contour imposes constraints on which Seifert geometries admit a straightforward TQFT dual.  While such mirror symmetry identities have been studied on Seifert manifolds from various perspectives, the novelty here is to formulate these identities in terms of linear integration contours that lend themselves to the construction of a state-integral model, as opposed to discrete sums over Bethe vacua.  For technical reasons, our conclusions are most well-established (and most easily formulated) for Seifert homology spheres whose orbifold bases have positive Euler characteristic, although we expect many of them to hold more generally.

On the algebraic side, we show that the $T_2[M]$ partition functions (or equivalently, the ``$\mathcal{M}_3$-TQFT'' wavefunctions on $M$) can be characterized as solutions to a finite set of difference equations.  A similar characterization has previously been established for holomorphic blocks \cite{Beem:2012mb}.  In particular, for the simplest such theory $T_2[\Delta]$ associated to a hyperbolic tetrahedron $\Delta$, we arrive at two main points:
\begin{enumerate}
\item There are as many difference equations as exceptional fibers in the Seifert geometry $\mathcal{M}_3$.
\item The Hilbert space of the $\mathcal{M}_3$-TQFT on $\partial\Delta$ results from quantizing a classical phase space with two noncompact directions and a number of compact directions determined by the first homology group of $\mathcal{M}_3$.
\end{enumerate}
These statements generalize similar ones pertaining to lens spaces and to noncompact Chern-Simons theory with rank-one gauge group.  For example, the phase space of the latter on $\partial\Delta$ is $\mathbb{R}^2$ for $SL(2, \mathbb{R})$ and $(\mathbb{C}^\ast)^2$ for $SL(2, \mathbb{C})$.

We begin in Sections \ref{3d3dbackground} and \ref{seifertbackground} by giving some necessary background.  In Section \ref{stateint}, we derive the integral identities that form the basis of a state-integral model and analyze the conditions for their convergence.  In Section \ref{diffeqs}, we present a detailed study of the difference equations for $\mathcal{M}_3$ partition functions, generalizing the observations of \cite{Dimofte:2011ju, Dimofte:2011py, Beem:2012mb} for lens spaces.  This understanding yields some hints as to the canonical structure of the associated TQFTs, which we comment on in Section \ref{quantization}, where we also interpret the difference equations as algebraic line operator identities.  It remains to understand how the Hilbert spaces of these putative TQFTs might arise from a Lagrangian formulation, as well as their potential interpretation in terms of analytically continued Chern-Simons theory.  We conclude in Section \ref{outlook} by sketching some of the many open problems that remain.

\section{3D-3D Correspondence} \label{3d3dbackground}

\subsection{Overview}

Our work takes place in the context of many known results and open questions about the 3D-3D correspondence (many related developments are summarized in the reviews \cite{Dimofte:2014ija, Dimofte:2016pua}).  Here, we survey the most relevant ones.

The 3D-3D correspondence, as usually formulated, goes as follows.  The 6D $(2, 0)$ theory is labeled by an $ADE$ Lie algebra $\mathfrak{g} = \operatorname{Lie}(G)$ and has a $USp(4)$ $R$-symmetry with an $SU(2)\times U(1)$ subgroup.  We may use the $SU(2)$ subgroup to couple the theory to $M\times \mathbb{R}^3$ for an arbitrary smooth three-manifold $M$ via the Rozansky-Witten topological twist along $M$ \cite{Gukov:2016gkn}.  This preserves a residual 3D $\mathcal{N} = 2$ supersymmetry in $\mathbb{R}^3$, with the $U(1)$ commutant of $SU(2)$ being the $R$-symmetry, leading to a 3D $\mathcal{N} = 2$ theory $T_{\mathfrak{g}}[M]$ (equivalently, for $\mathfrak{g} = A_{n-1}$, this is the theory $T_n[M]$ obtained by wrapping $n$ M5-branes on $M$).  More precisely, one can label the theory as $T[M; G]$.\footnote{Even more precisely, the fact that the $(2, 0)$ theories are relative QFTs means that their compactifications depend on data associated to the topology of $M$ that go beyond specifying the global form of $G$ \cite{Eckhard:2019jgg}.  We will ignore such global issues.}  We will often write $T[M]$ for simplicity.  We may then use the $U(1)$ $R$-symmetry to couple this 3D $\mathcal{N} = 2$ theory to another three-manifold $\mathcal{M}_3$.  In the case of a Seifert manifold, we perform a partial topological twist along the Riemann surface base of $\mathcal{M}_3$.  In other words, we effectively place the 6D theory on $M\times \mathcal{M}_3$ via a twist by $SU(2)\times U(1)\subset USp(4)$.

The main claim to fame of the theory $T[M]$ is that its dependence on the metric chosen for coupling to $M$ is often irrelevant, in the RG sense.  That is, one generally expects $T[M]$ to flow in the IR to an SCFT that is independent of the metric on $M$.  In this way, the SCFT defined by $T[M]$ is a topological invariant of $M$, which we expect to be dual to a TQFT.  This expectation is borne out very explicitly when $M$ is closed and hyperbolic (admits a metric of constant negative curvature).\footnote{While most three-manifolds are hyperbolic, Seifert manifolds are notable exceptions.}  In this case, the Mostow rigidity theorem states that the hyperbolic metric on $M$ is unique.  The IR SCFT then manifestly depends only on the topology of $M$.  This SCFT may have many different UV descriptions.  Henceforth, we use $T[M]$ to refer to either the IR SCFT or any of its UV descriptions.

In this paper, we work within the conceptual framework introduced by Dimofte, Gaiotto, and Gukov (DGG) \cite{Dimofte:2011ju, Dimofte:2011py}, which associates an SCFT $T[M]$ to a hyperbolic three-manifold $M$ via an \emph{ideal triangulation} of $M$.  Many preliminary results were obtained for the theories $T_2[M]$, linking various lens space partition functions of $T_2[M]$ to $SL(2, \mathbb{C})$ Chern-Simons theory.  A unified view of these results follows from decomposing these lens space partition functions into \emph{holomorphic blocks} \cite{Beem:2012mb}, which compute the path integral of analytically continued $SU(2)$ Chern-Simons theory over integration cycles labeled by irreducible flat $SL(2, \mathbb{C})$ connections $\mathcal{A}^\alpha$ on $M$ \cite{Witten:2010cx}.\footnote{Flat connections are critical points of the Chern-Simons functional and label middle-dimensional cycles in the space of connections over which the analytically continued path integral converges.  The holonomies of a flat $G$-connection on $M$ specify a representation $\rho : \pi_1(M)\to G$ whose reducibility is characterized by the commutant of the image of $\rho$ inside $G$.  A flat connection is referred to as irreducible if this commutant is the center of $G$.  In particular, a flat $SL(2, \mathbb{C})$ connection is completely reducible (or abelian) if its holonomies mutually commute, and irreducible otherwise.}  The DGG algorithm for $T_2[M]$ was extended to $T_{n > 2}[M]$ in \cite{Dimofte:2013iv}.

There are two complementary questions that one could ask regarding the role of Seifert manifolds in the 3D-3D correspondence:
\begin{enumerate}
\item What is the theory $T[M]$ for $M$ a Seifert manifold?\footnote{This paragraph and the next are the only ones in this paper, minus the introduction, where $M$ possibly denotes a non-hyperbolic three-manifold.} \label{Q1}
\item What is the partition function of $T[M]$ (for $M$ hyperbolic) on a Seifert manifold $\mathcal{M}_3$, and what is the 3D TQFT (associated to $\mathcal{M}_3$) whose partition function on $M$ this is equal to? \label{Q2}
\end{enumerate}
The first question has been the subject of many recent studies --- for instance, \cite{Gukov:2015sna, Pei:2015jsa, Gukov:2017kmk, Alday:2017yxk, Eckhard:2019jgg}.  In this paper, we focus on the second question.

In fact, the answer to Question \ref{Q1}, in its most basic form, is more or less completely known.  In, e.g., Section 5 of \cite{Chung:2014qpa} (see also \cite{Gadde:2013sca} and Chapter V of \cite{Pei:2016rmn}), a procedure is given for constructing the answer.  Specific results in this and follow-up work include the identification of $T[L(k, 1); G]$ as a 3D $\mathcal{N} = 2$ $G$-Chern-Simons theory coupled to matter and the identification of $T[L(k, p); G]$ for $G = U(1)$ (and presumably, in general) as a quiver Chern-Simons theory, roughly interpreted as a Chern-Simons-mat\-ter theory at fractional level.  In the process, \cite{Chung:2014qpa} addresses a shortcoming of the ideal triangulation algorithm of DGG.  Namely, the theories constructed by DGG and in most subsequent work are only subsectors of the theories $T_n[M]$ whose vacua correspond to \emph{irreducible} flat $SL(n, \mathbb{C})$ connections on $M$.  The DGG theories are thus dual to consistent truncations of analytically continued Chern-Simons theory that are sensitive only to the corresponding integration cycles \cite{Dimofte:2014zga}.  On the other hand, one would expect the moduli space of vacua of the full theory $T[M; G]$ on $\mathbb{R}^2\times S^1$ to coincide with the space of \emph{all} flat $G_{\mathbb{C}}$-connections on $M$:
\begin{equation}
\mathcal{M}_\text{SUSY}(T[M; G]) = \mathcal{M}_\text{flat}(M; G_{\mathbb{C}}).
\end{equation}
This relation is essential to the cutting and gluing construction of the theories $T[M; G]$.  Using this fact, \cite{Chung:2014qpa} explains how to construct the \emph{full} theories $T_2[M]$ for Seifert manifolds and knot complements $M = S^3\setminus K$, from which the DGG theories can be obtained by Higgsing a non-generic $U(1)$ symmetry that is specific to the $T[M]$ theories for non-hyperbolic $M$.\footnote{By contrast, it has been argued \cite{Gang:2018wek} (see also \cite{Gang:2018hjd, Gang:2019uay, Benini:2019dyp}) that for hyperbolic $M$, the insensitivity of the DGG theories to reducible flat connections is a non-issue.  This is related to the fact that reducible flat connections contribute trivially to the partition function of complex, as opposed to compact, Chern-Simons theory \cite{Chung:2014qpa}.  The reasoning is that the proper definition of the 3D SCFT $T[M]$ via twisted compactification of the 6D theory on hyperbolic $M$ (which has contributions from all flat connections) involves choosing a particular point on the moduli space of the compactified theory on $\mathbb{R}^3$.  For the theory on $\mathbb{R}^2\times S^1$, this point becomes a discrete set of Bethe vacua corresponding only to irreducible flat $SL(n, \mathbb{C})$ connections on $M$, leading to precisely the DGG theory.  The number of Bethe vacua is given by the Witten index of the 3D SCFT.}

Our interest lies in Question \ref{Q2}, for which partial answers are known when $\mathcal{M}_3$ is a (squashed) lens space $L(k, p)_b$:
\begin{align*}
(k, p) &= (1, 1): & Z_{S_b^3}(T_n[M]) &= Z_\text{CS}^{SL(n, \mathbb{C})}(M), & k &= 1, & \sigma &= \frac{1 - b^2}{1 + b^2}, \\
(k, p) &= (0, 1): & Z_{S^2\times_\sigma S^1}(T_n[M]) &= Z_\text{CS}^{SL(n, \mathbb{C})}(M), & k &= 0, & \sigma &= \sigma, \\
(k, p) &= (k, 1): & Z_{L(k, 1)_b}(T_n[M]) &= Z_\text{CS}^{SL(n, \mathbb{C})}(M), & k &= k, & \sigma &= k\left(\frac{1 - b^2}{1 + b^2}\right).
\end{align*}
The parameters $k, \sigma$ on the right denote the levels of the corresponding Chern-Simons theory, one quantized and one continuous.  The first line was proposed in \cite{Dimofte:2011ju} and derived in \cite{Cordova:2013cea} (many other aspects have also been explored; see, e.g., \cite{Gang:2014ema, Gang:2015wya}).  The second line was proposed in \cite{Dimofte:2011py} and derived in \cite{Yagi:2013fda, Lee:2013ida}.  The last line is discussed systematically in \cite{Dimofte:2014zga, Mikhaylov:2017ngi}, related earlier results having been obtained in \cite{Terashima:2011qi, Benini:2011nc}.

In particular, Dimofte \cite{Dimofte:2014zga} shows that $T_n[M]$ on $L(k, 1)_b$ (which is a Seifert manifold for $b^2\in \mathbb{Q}$) is equivalent to $SL(n, \mathbb{C})$ Chern-Simons theory at quantized level $k$ on $M$, while also proposing a state-integral model for $L(k, p)_b$ invariants of $M$.  The TQFT for the latter can morally be interpreted as a twisted version of complex Chern-Simons theory, analogous to the result of setting the parameter $q$ in the compact case to a primitive root of unity $e^{2\pi ip/k}$ rather than the standard $e^{2\pi i/k}$.  We follow the basic approach of \cite{Dimofte:2014zga} to determine the TQFT dual of a given Seifert manifold $\mathcal{M}_3$, which involves first evaluating the $\mathcal{M}_3$ partition function of the basic tetrahedron theory (a free chiral multiplet) and then identifying copies of this object in a way appropriate to an arbitrary three-manifold $M$.  The final answer is subject to the same limitations as in \cite{Dimofte:2014zga}: namely, it captures only a subsector of the full TQFT.  In the general case, our TQFT wavefunctions do not exhibit a simple factorization into holomorphic and antiholomorphic blocks as in \cite{Dimofte:2014zga}.  One might expect that at least for certain $\mathcal{M}_3$, this TQFT can be thought of as a (non-supersymmetric) quiver Chern-Simons theory with noncompact gauge nodes.

\subsection{DGG Construction} \label{DGGreview}

The foundation of our approach was laid by DGG.  To orient the reader, we describe their construction qualitatively, glossing over many caveats (precise details can be found in the original papers \cite{Dimofte:2011ju, Dimofte:2011py}).

We start with geometry.  A hyperbolic metric on $M$ is essentially equivalent to a flat $SL(2, \mathbb{C})$ connection, $(P)SL(2, \mathbb{C})$ being the isometry group of $\mathbb{H}^3$.  $M$ can have two types of boundaries: geodesic boundaries (possibly with punctures) where the induced metric is hyperbolic, and cusp boundaries where the induced metric is Euclidean.  An ideal triangulation is a decomposition of $M$ into tetrahedra whose vertices lie at the punctures and whose faces are glued pairwise.  We write it as $M = \bigcup_{i=1}^N \Delta_i$.  In practice, one truncates the vertices of an ideal tetrahedron to Euclidean triangles; then the geodesic boundaries are triangulated by faces and the cusp boundaries are triangulated by vertices.  Let $\mathcal{P}_{\partial M}$ be the moduli space of flat $SL(2, \mathbb{C})$ connections on $\partial M$, and let $\mathcal{L}_M$ be the Lagrangian submanifold of $\mathcal{P}_{\partial M}$ comprised of those flat connections that extend to all of $M$.  We denote by $\Pi$ a polarization of the boundary phase space $\mathcal{P}_{\partial M}$.

\begin{figure}[!htb]
\begin{center}
\includegraphics[width=0.8\textwidth]{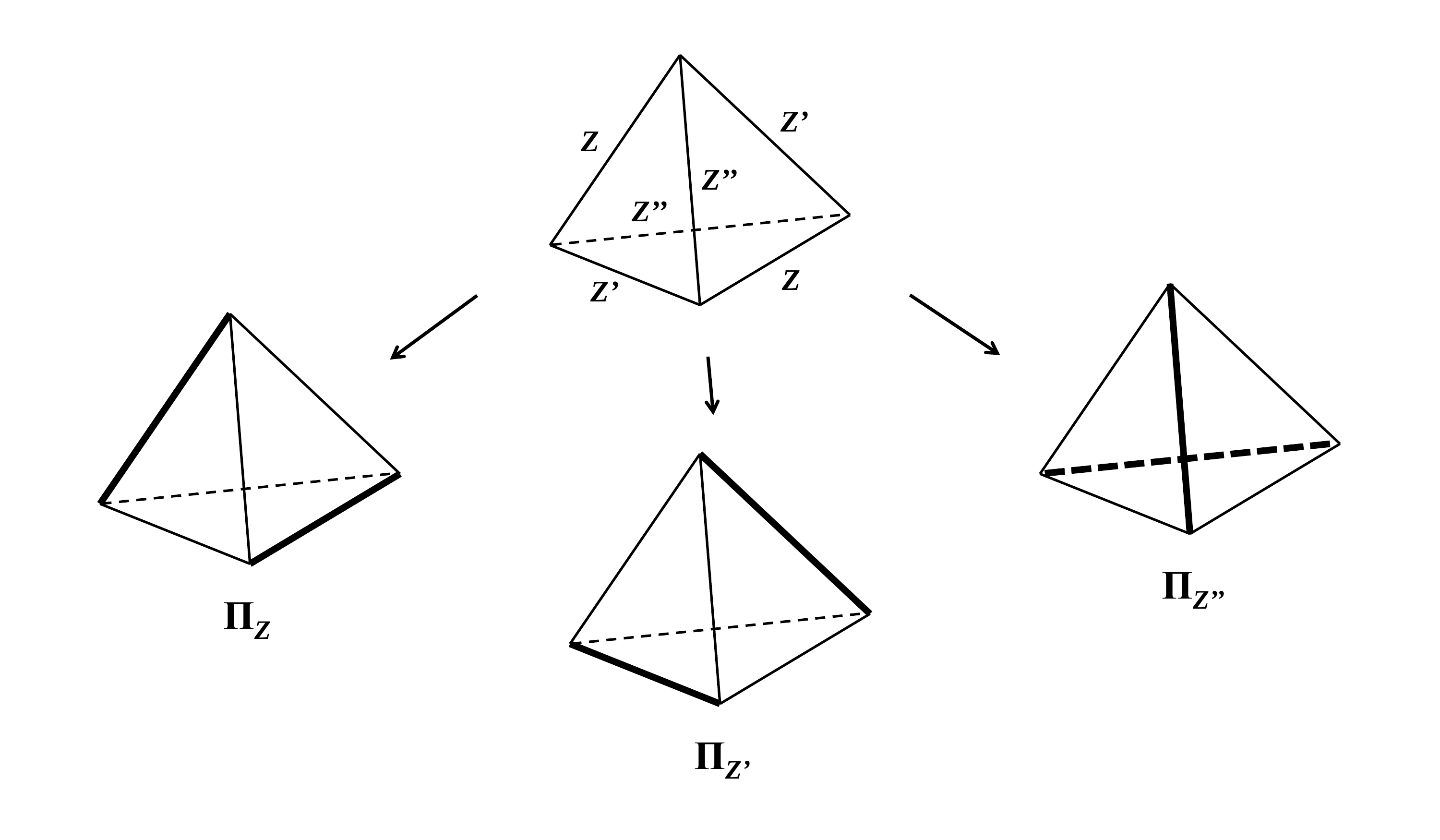}
\caption{An ideal tetrahedron labeled by three (logarithmic) edge parameters, along with its three canonical polarizations (position coordinates are indicated by bold lines).}
\label{figuretetrahedron}
\end{center}
\end{figure}

The most basic such $M$ is a tetrahedron $\Delta$ itself.  An ideal tetrahedron is specified by three complex edge parameters $z, z', z''$ associated to pairs of opposite edges (for the tetrahedron to be nondegenerate, we require these parameters not to be $0, 1, \infty$).  Then the boundary phase space $\mathcal{P}_{\partial\Delta}$ is the locus 
\begin{equation}
zz'z'' = -1
\end{equation}
inside $(\mathbb{C}^\ast\setminus \{1\})^3$, while the Lagrangian submanifold $\mathcal{L}_\Delta$ is the following locus inside $\mathcal{P}_{\partial\Delta}$:
\begin{equation}
z + z'^{-1} - 1 = 0 \Longleftrightarrow z' + z''^{-1} - 1 = 0 \Longleftrightarrow z'' + z^{-1} - 1 = 0.
\end{equation}
Write $z = e^Z, z' = e^{Z'}, z'' = e^{Z''}$.  In light of the boundary symplectic form
\begin{equation}
\omega_{\partial\Delta} = dZ\wedge dZ' = dZ'\wedge dZ'' = dZ''\wedge dZ,
\end{equation}
there are three natural polarizations for the boundary phase space, which we denote by $\Pi_Z, \Pi_{Z'}, \Pi_{Z''}$.  These have position coordinates $Z, Z', Z''$ and conjugate momenta $Z'', Z, Z'$, respectively (see Figure \ref{figuretetrahedron}).  The space of polarizations $(X, P)$ admits an action of $ISp(2, \mathbb{Z})\cong SL(2, \mathbb{Z})\ltimes \mathbb{Z}^2$, given by taking invertible integer linear combinations and shifts by multiples of $i\pi$ (since $Z + Z' + Z'' = i\pi$).

Similarly, a collection of $N$ tetrahedra admits an action of the affine symplectic group $ISp(2N, \mathbb{Z})\cong Sp(2N, \mathbb{Z})\ltimes \mathbb{Z}^{2N}$ on the space of polarizations.  $\mathcal{P}_{\partial M}$ and $\mathcal{L}_M$ can be constructed as certain symplectic quotients of the products $\prod_i \mathcal{P}_{\partial\Delta_i}$ and $\prod_i \mathcal{L}_{\Delta_i}$, with the moment map constraints ensuring that the gluing is smooth.

In field theory, there exists a corresponding $Sp(2N, \mathbb{Z})$ action on the set of 3D CFTs with $U(1)^N$ flavor symmetry \cite{Witten:2003ya}, which lifts to the action of electric-magnetic duality on a 4D abelian gauge theory coupled to the 3D CFT at a boundary.  We describe it for the case of a $U(1)$ gauge theory coupled to a background gauge field $A$:
\begin{itemize}
\item The $T$ generator adds a Chern-Simons term for $A$ at unit level.
\item The $S$ generator makes $A$ dynamical (in the process, generating a dual topological flavor symmetry to which we couple a background gauge field $A'$) and adds a mixed $AA'$ Chern-Simons term at unit level.
\end{itemize}
We need the $\mathcal{N} = 2$ supersymmetric version of this action, which is defined \emph{mutatis mutandis}.\footnote{To define the $\mathcal{N} = 4$ version of this symplectic action, one should further define an operation $F$, satisfying $F^2 = 1$, on 3D $\mathcal{N} = 2$ theories with a superpotential coupling to a background chiral multiplet that simply makes this background chiral multiplet dynamical. \label{Foperation}}  In the $\mathcal{N} = 2$ language, the shifts that extend $Sp(2N, \mathbb{Z})$ to $ISp(2N, \mathbb{Z})$ act on the $U(1)_R$ charges of operators in the theory.

We are now ready to define the theory $T_{M, \Pi}$ associated to $M$ and a given polarization $\Pi$ of $\mathcal{P}_{\partial M}$.  The central claim is that $T_{M, \Pi}$ is a topological invariant of $(M, \Pi)$.  $T_{M, \Pi}$ couples naturally to the IR degrees of freedom on the Coulomb branch of a 4D $\mathcal{N} = 2$ gauge theory, with gauge group $U(1)^N$.  The $ISp(2N, \mathbb{Z})$ action on $\Pi$ is then identified with the electric-magnetic duality group of the 4D theory, affinely extended in the presence of a boundary.  The choice of duality frame is a choice of polarization, and affine symplectic transformations on the polarization are associated with corresponding actions on the 3D theory at the boundary of the 4D theory.

We first define the tetrahedron theory associated to $\Delta$ with boundary polarization $\Pi_Z$, and denoted by $T_{\Delta, \Pi_Z}$: it is a free chiral multiplet coupled to a background $U(1)$ gauge field in the $U(1)_{1/2}$ quantization.\footnote{By this, we mean (borrowing the terminology of \cite{Closset:2018ghr}) that there exists a UV Chern-Simons contact term at level $1/2$ for this symmetry.}  Now fix a triangulation of $M$, and consider $N$ decoupled copies of the tetrahedron theory $\bigotimes_{i=1}^N T_{\Delta_i, \Pi_i}$.  We first perform an $Sp(2N, \mathbb{Z})$ transformation on the set of polarizations $\{\Pi_i\}$ to obtain a polarization compatible with $\Pi$.  This amounts to gauging flavor symmetries and adding Chern-Simons terms.  We then add a term to the superpotential for each internal edge of the triangulation, thus breaking the $U(1)$ flavor symmetries under which they are charged.  This implements the gluing (moment map) constraints.  The result is that the theory $T_{M, \Pi}$ has $\frac{1}{2}\dim\mathcal{P}_{\partial M}$ unbroken $U(1)$ flavor symmetries carried by chiral operators associated to the position coordinates in $\Pi$.  From now on, we leave the $\Pi$-dependence implicit and denote $T_{M, \Pi}$ by $T_2[M]$, as well as the tetrahedron theory by $T_\Delta = T_2[\Delta]$.

While the UV Lagrangian description of $T_2[M]$ as an abelian Chern-Simons-matter theory (i.e., as a theory of ``class $\mathcal{R}$'' \cite{Dimofte:2011py}) depends on the triangulation of $M$, the IR SCFT does not, by virtue of 3D mirror symmetry.  Two elementary mirror symmetries play a distinguished role.

\begin{figure}[!htb]
\begin{center}
\includegraphics[width=0.8\textwidth]{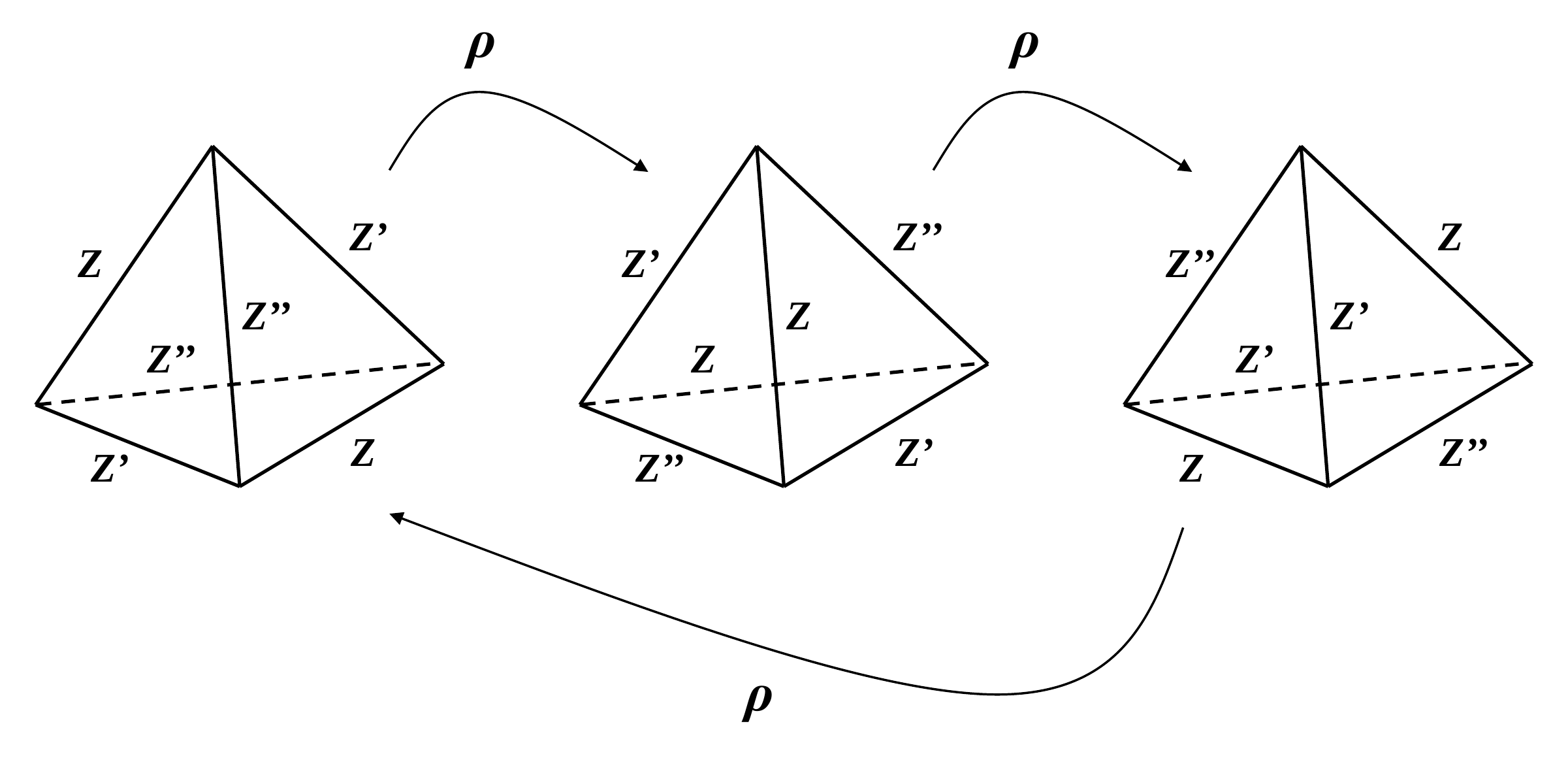}
\caption{Geometrical meaning of the $\rho$-transformation on a tetrahedron.}
\label{figurerho}
\end{center}
\end{figure}

First, a certain affine $ST$-transformation $\rho$, defined more precisely in Appendix \ref{DGGcomparison}, generates a $\mathbb{Z}_3$ subgroup of $ISp(2, \mathbb{Z})$.  Physically, the $\rho$-invariance of the tetrahedron theory expresses the mirror symmetry between a free and a gauged chiral multiplet.  Geometrically, $\rho$ is a triality symmetry of the tetrahedron that cyclically permutes the three canonical polarizations of $T_\Delta$ as well as the quantum operators $\hat{z}, \hat{z}'', \hat{z}'$, which can be identified with Wilson, 't Hooft, and dyonic line operators of charges $(e, m) = (1, 0)$, $(0, 1)$, and $(-1, 1)$ in the 4D $\mathcal{N} = 2$ abelian gauge theory, respectively \cite{Dimofte:2011ju}.  Hence $\rho$-invariance of $T_\Delta$ implies that $T_2[M]$ is independent of relabelings of the edge parameters of the tetrahedra that preserve the desired orientation.  See Figure \ref{figurerho}.

\begin{figure}[!htb]
\begin{center}
\includegraphics[width=0.8\textwidth]{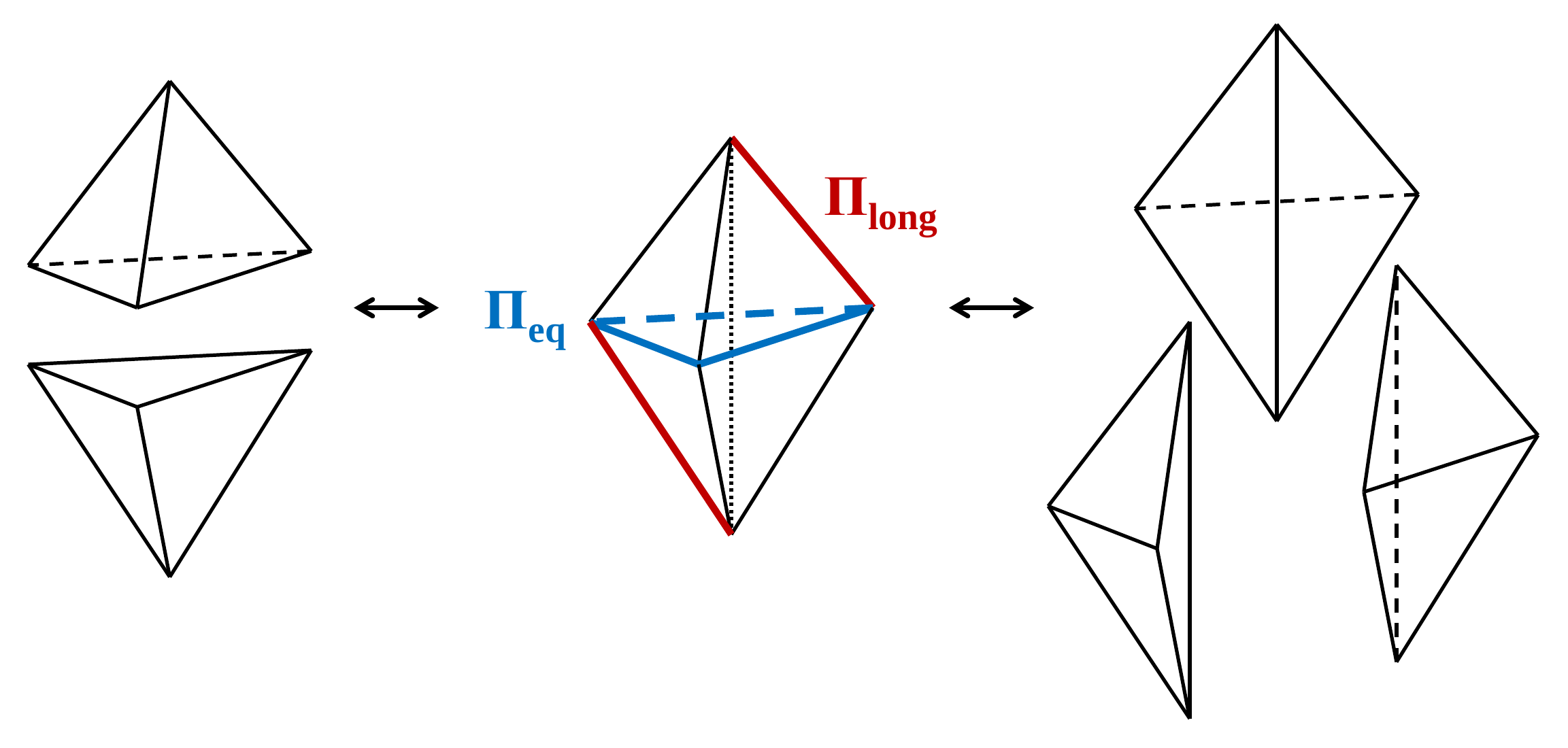}
\caption{Geometrical meaning of the 2-3 move on a bipyramid, with equatorial and longitudinal polarizations indicated in blue and red, respectively.}
\label{figure23}
\end{center}
\end{figure}

Second, for a fixed triangulation of the geodesic boundaries, any two triangulations of $M$ are related by a sequence of 2-3 Pachner moves.\footnote{Changing the triangulation of a geodesic boundary entails the gluing of a tetrahedron onto the boundary to create a new internal edge.  This corresponds to the $F$ operation of Footnote \ref{Foperation}.}  Hence triangulation-invariance of $T_2[M]$ comes down to a statement of mirror symmetry between two descriptions of the ``bipyramid'' theory.  The decomposition into two tetrahedra gives $\mathcal{N} = 2$ SQED with $N_f = 1$, while the decomposition into three tetrahedra gives the XYZ model.\footnote{More precisely, for $T[\text{bipyramid}]$, we can consider the equatorial polarization $\Pi_\text{eq}$ or the longitudinal polarization $\Pi_\text{long}$.  These correspond to maximal sets of independent edges that share no common faces (including internal faces).  The 2-3 move can be thought of as corresponding either to the $\mathcal{N} = 2$ SQED$_1$/XYZ mirror symmetry ($\Pi_\text{eq}$) or to the $\mathcal{N} = 4$ SQED$_1$/free hyper mirror symmetry ($\Pi_\text{long}$).}  See Figure \ref{figure23}.

\section{Seifert Fibering Operators} \label{seifertbackground}

\subsection{Formalism}

Having described the conceptual basis for our work, we turn to the technical ma\-chi\-nery, which is provided by the formalism of \cite{Closset:2018ghr} for computing partition functions of 3D $\mathcal{N} = 2$ gauge theories on Seifert manifolds.  We briefly review this formalism and state our conventions along the way.  The basic idea is to interpret observables in a half-BPS 3D $\mathcal{N} = 2$ geometry as observables in an auxiliary 2D TQFT, the ``3D A-model.''

A Seifert manifold $\cM_3$ (assumed closed and oriented) comes with an $S^1$ fibration $\pi : \cM_3\to \hat{\Sigma}_g$ over an orbifold Riemann surface $\hat{\Sigma}_g$, and is characterized by a finite set of integers (Seifert symbols):
\begin{equation}
\mathcal{M}_3\cong [\mathbf{d}; g; (q_1, p_1), \ldots, (q_n, p_n)].
\end{equation}
Here, $\mathbf{d}$ is the degree (first Chern number) of the Seifert fibration, $g$ is the genus of the base $\hat{\Sigma}_g$ (which has $n$ marked points), and $(q_i, p_i)$ are the Seifert invariants of the $n$ exceptional fibers (we will always assume that $q_i > 0$).  The neighborhood of the special fiber above the $i^\textrm{th}$ orbifold point is modeled on the space
\begin{equation}
(D^2\times S^1)/\Z_{q_i}, \quad (z, \theta)\sim (z, \theta + 2\pi)\sim \left(e^{2\pi i/q_i}z, \theta + \frac{2\pi p_i}{q_i}\right).
\end{equation}
$\cM_3$ is smooth if and only if $\operatorname{gcd}(q_i, p_i) = 1$, in which case if a generic $S^1$ fiber has radius $\beta$, then the fiber over the $i^\textrm{th}$ marked point has radius $\beta/q_i$.

The Seifert manifold $\cM_3$ can equivalently be defined as the circle bundle associated to a choice of line bundle $\cL_0$ over $\hat{\Sigma}_g$, called the ``defining line bundle.''  Holomorphic line bundles over $\hat{\Sigma}_g$ are classified by the Picard group $\operatorname{Pic}(\hat{\Sigma}_g)$, which is in general finitely but not freely generated.\footnote{Our nomenclature follows that of \cite{Closset:2018ghr}, but what we call the Picard group is really what mathematicians would call the N\'eron-Severi group $\operatorname{Pic}(\hat{\Sigma}_g)/\operatorname{Pic}^0(\hat{\Sigma}_g)$.}  Any such line bundle can be written in terms of elementary line bundles and ``ordinary'' or ``fractional'' fluxes of its connection (an abelian gauge field) localized at smooth or orbifold points, respectively.  Concretely, an element of $\operatorname{Pic}(\hat{\Sigma}_g)$ may be represented by a choice of integer fluxes 
\begin{equation}
L\cong [\n_0; \n_1, \ldots, \n_n]
\label{linebundle}
\end{equation}
up to the identifications
\begin{equation}
[\n_0; \ldots, \n_i, \ldots]\sim [\n_0 - 1; \ldots, \n_i + q_i, \ldots].
\end{equation}
The defining line bundle $\cL_0$ is represented by
\begin{equation}
\cL_0\cong [\mathbf{d}; p_1, \ldots, p_n].
\end{equation}
Thus we see that the Seifert presentation is invariant under the identifications
\begin{equation}
[\mathbf{d}; g; \ldots, (q_i, p_i), \ldots]\sim [\mathbf{d} - 1; g; \ldots, (q_i, p_i + q_i), \ldots].
\label{seifident}
\end{equation}
In addition, we can always add or remove a ``trivial'' special fiber of type $(1, 0)$:
\begin{equation}
[\mathbf{d}; g; \ldots, (1, 0), \ldots]\sim [\mathbf{d}; g; \ldots, \ldots].
\label{seifident2}
\end{equation}
Using these relations, we can fix $0\leq p_i < q_i$ and eliminate all trivial special fibers, in which case we call the Seifert presentation ``normalized.''  In fact, it will often be more convenient to set $\mathbf{d} = 0$ by adding a $(1, \mathbf{d})$ special fiber.

The partition function of a 3D $\mathcal{N} = 2$ theory on $\mathcal{M}_3$ can be written in terms of the following basic defect line operators, which are functions of Coulomb branch (gauge symmetry) parameters $u$ and flavor (global symmetry) parameters $\nu$:
\begin{itemize}
\item the handle-gluing operator $\mathcal{H}(u, \nu)$ shifts the genus of $\hat{\Sigma}_g$,
\item the ordinary fibering operator $\mathcal{F}(u, \nu)$ shifts the degree $\mathbf{d}$,
\item and the $(q, p)$-fibering operator $\mathcal{G}_{q, p}(u, \nu)_{\mathfrak{m}}$ introduces an exceptional fiber of type $(q, p)$.
\end{itemize}
Here, $\mathfrak{m}$ labels fractional global symmetry fluxes.  More precisely, we should consider the ``elementary'' $(q, p)$-fibering operator $\mathcal{G}_{q, p}(u, \nu)_{\mathfrak{n}, \mathfrak{m}}$ where the integers $\mathfrak{n}, \mathfrak{m}$ label fractional fluxes of the gauge and flavor symmetry gauge fields localized at the exceptional fiber.  The ``full'' $(q, p)$-fibering operator is then a sum over fractional gauge fluxes of these elementary operators, including contributions from Chern-Simons terms, chiral multiplets, and vector multiplets: schematically, 
\begin{equation}
\mathcal{G}_{q, p}(u, \nu)_{\mathfrak{m}} = \sum_{\mathfrak{n}} \mathcal{G}_{q, p}^{\text{CS}}(u, \nu)_{\mathfrak{n}, \mathfrak{m}}\mathcal{G}_{q, p}^{\text{matter}}(u, \nu)_{\mathfrak{n}, \mathfrak{m}}\mathcal{G}_{q, p}^{\text{vector}}(u)_{\mathfrak{n}}.
\end{equation}
Finally, gauge and flavor fluxes are incorporated via powers of the gauge and flavor flux operators, denoted by $\Pi_a(u, \nu)$ and $\Pi_\alpha(u, \nu)$ ($a = 1, \ldots, \operatorname{rank}(G)$).  With these ingredients in place, the partition function takes the form
\begin{equation}
Z_{\mathcal{M}_3}(\nu)_{\mathfrak{m}} = \left\langle\mathcal{H}(u, \nu)^g\prod_{i=0}^n \mathcal{G}_{q_i, p_i}(u, \nu)_{\mathfrak{m}_i}\right\rangle_{S^2\times S^1},
\label{partition}
\end{equation}
where the $S^2\times S^1$ background corresponds to the topologically twisted index.  Here, we adopt the convention that $(q_0, p_0) = (1, \mathbf{d})$ (while $q_i > 1$ for $i > 0$), so that we always include a fiber over a smooth marked point.  This is equivalent to the effect of a nonzero Chern degree of the $S^1$ bundle, $\mathbf{d} = p_0$, as mentioned after \eqref{seifident2}, and also allows us to insert ordinary gauge and flavor flux via the relation \cite{Closset:2017zgf}
\begin{equation}
\cG_{1, \mathbf{d}}(u, \nu)_{\m_0, \n_0} = \cF(u, \nu)^{\mathbf{d}}\prod_a \Pi_a(u, \nu)^{\m_{0, a}}\prod_\alpha \Pi_\alpha(u, \nu)^{\n_{0, \alpha}}.
\end{equation}
Hence it is understood that $\m = (\m_i)$ in \eqref{partition} includes both the ordinary fluxes $\m_0$ and the fractional fluxes $\m_{i>0}$.

In more detail, the Coulomb branch effective action of the effective 2D $\mathcal{N} = (2, 2)$ gauge theory is determined (up to $Q$-exact terms) by two functions, which follow from the UV action: the twisted superpotential $\mathcal{W}(u, \nu)$ and the effective dilaton $\Omega(u, \nu)$.  In terms of them, we have the half-BPS line (or in 2D, twisted chiral) operators \cite{Nekrasov:2014xaa, Closset:2017zgf}
\begin{align}
\mathcal{H}(u, \nu) &= e^{2\pi i\Omega(u, \nu)}\det_{a, b}\frac{\partial^2\mathcal{W}(u, \nu)}{\partial u_a\partial u_b}, \\
\mathcal{F}(u, \nu) &= \exp\left[2\pi i\left(\mathcal{W}(u, \nu) - u_a\frac{\partial\mathcal{W}(u, \nu)}{\partial u_a} - \nu_\alpha\frac{\partial\mathcal{W}(u, \nu)}{\partial\nu_\alpha}\right)\right],
\end{align}
as well as the gauge and flavor flux operators
\begin{equation}
\Pi_a(u, \nu) = \exp\left(2\pi i\frac{\partial\mathcal{W}(u, \nu)}{\partial u_a}\right), \quad \Pi_\alpha(u, \nu) = \exp\left(2\pi i\frac{\partial\mathcal{W}(u, \nu)}{\partial\nu_\alpha}\right).
\end{equation}
While the general $\mathcal{G}_{q, p}(u, \nu)_{\mathfrak{n}, \mathfrak{m}}$ does not have a simple expression in terms of $\mathcal{W}(u, \nu)$ and $\Omega(u, \nu)$, it can still be constructed from the data of the UV Lagrangian.

The partition function \eqref{partition} on $\cM_3$ was shown in \cite{Closset:2018ghr} to be given by several equivalent formulas.  First, it can be derived in terms of the 3D A-model.  From this point of view, we identify the Bethe vacua \cite{Nekrasov:2009uh} of the 2D TQFT, which correspond to solutions of the Bethe equations $\Pi_a = 1$ for all $a$, excluding solutions not acted on freely by the Weyl group, and modulo the action of the Weyl group.  Then observables in the 3D A-model (correlation functions of half-BPS line operators $\mathscr{L}(u, \nu)$ in the A-twisted\footnote{In previous work \cite{Closset:2017zgf, Closset:2018ghr}, the 3D $\mathcal{N} = 2$ supersymmetric background preserving two supercharges is the pullback of the 2D $(2, 2)$ topological A-twist \cite{Witten:1988xj} on $\hat{\Sigma}_g$.  Further, for a 3D $\mathcal{N} = 4$ theory, the twists by $SU(2)_H$ and $SU(2)_C$ that preserve four scalar supercharges on any three-manifold are referred to as the A- and B-twists in \cite{Closset:2016arn}; the latter is the Rozansky-Witten twist.

In \cite{Closset:2016arn}, the A-twist is always meant in the 2D $(2, 2)$ sense, which can then specialize to either the A- or the B-twist in the 3D $\mathcal{N} = 4$ sense.  Note that twisting by $U(1)_\text{vector}$ in 2D uplifts to twisting by $U(1)_R$ in 3D $\mathcal{N} = 2$, but $U(1)_\text{axial}$ in 2D has no lift.} $\hat{\Sigma}_g\times S^1$ geometry) are sums (traces) over the finite set of 2D Bethe vacua $S_\text{vac}$:
\begin{equation}
\langle\mathscr{L}_i\mathscr{L}_j\cdots\rangle_{\hat{\Sigma}_g\times S^1} = \sum_{\hat{u}\in S_\text{vac}} \mathscr{L}_i(\hat{u}, \nu)\mathscr{L}_j(\hat{u}, \nu)\cdots \mathcal{H}(\hat{u}, \nu)^{g-1}.
\end{equation}
The power of $g - 1$ comes from tracing on $T^2$.  In particular, we have
\begin{equation}
Z_{\mathcal{M}_3}(\nu)_{\mathfrak{m}} = \sum_{\hat{u}\in S_\text{vac}} \mathcal{H}(\hat{u}, \nu)^{g-1}\prod_{i=0}^n\sum_{\n_i\in (\Lambda_W^\vee)_{q_i}} \mathcal{G}_{q_i, p_i}(\hat{u}, \nu)_{\n_i, \mathfrak{m}_i}
\label{bethesum}
\end{equation}
where $(\Lambda_W^\vee)_q\equiv \Lambda_W^\vee/q\Lambda_W^\vee\cong (\Z_q)^{\operatorname{rank}(G)}$, with $\Lambda_W^\vee$ being the coweight lattice of $G$.  This ``Bethe-sum formula'' has been shown to hold for gauge groups that are products of unitary and compact, simply connected, simple Lie groups.\footnote{Given \eqref{bethesum}, a refined statement of the matching of $Z_{\mathcal{M}_3}(\nu)_{\mathfrak{m}}$ between IR-dual theories $\mathcal{T}, \mathcal{T}^D$ is that the $(q, p)$-fibering operators agree on dual Bethe vacua: $\mathcal{G}_{q, p}^{\mathcal{T}}(\hat{u}, \nu)_{\mathfrak{m}} = \smash{\mathcal{G}_{q, p}^{\mathcal{T}^D}(\hat{u}^D, \nu)_{\mathfrak{m}}}$.}  Two other equivalent expressions for the partition function on $\cM_3$ were derived in \cite{Closset:2018ghr} from the supersymmetric localization point of view.\footnote{In fact, the precise forms of these expressions are only well-understood for general $\cM_3$ in the case of abelian gauge groups, and for nonabelian gauge groups only when $\cM_3$ is a lens space.  Below, we will focus on abelian gauge theories.}  The first is a contour integral formula given by
\begin{equation}
Z_{\cM_3}(\nu)_\m = \frac{(-1)^{\operatorname{rank}(G)}}{|\cW|}\sum_{\substack{(\n_0, \n_i) \\ \in \Lambda_W^\vee\otimes \operatorname{Pic}(\hat{\Sigma}_g)}}\oint_\text{JK} d^{\operatorname{rank}(G)}u\, \cH(u, \nu)^{g - 1}H(u, \nu)\prod_{i=0}^n \cG_{q_i, p_i}(u, \nu)_{\n_i, \m_i},
\end{equation}
where $H(u, \nu)$ is the Hessian determinant of the twisted superpotential:
\begin{equation}
H(u, \nu) = \det_{a, b}\frac{1}{2\pi i}\frac{\partial\log\Pi_a}{\partial u_b}.
\label{hessdef}
\end{equation}
The sum is over fractional fluxes taking values in the Picard group of the base space $\hat{\Sigma}_g$, and the integral is over the Jeffrey-Kirwan integration contour in the domain $u\in \mathfrak{h}_{\mathbb{C}}/\Lambda_W^\vee$ (see \cite{Closset:2018ghr} for details).  The factor of $(-1)^{\operatorname{rank}(G)}$ entails a choice of orientation for the contour.  Finally, in cases where the defining line bundle $\cL_0$ of $\cM_3$ has nonzero orbifold Chern degree\footnote{For a general holomorphic line bundle \eqref{linebundle} over $\hat{\Sigma}_g$, with $\n_0\in \mathbb{Z}$ and $\n_i\in \mathbb{Z}_{q_i}$, we have $c_1(L) = \n_0 + \sum_{i=1}^n \frac{\n_i}{q_i}$.}
\begin{equation}
c_1(\mathcal{L}_0) = \mathbf{d} + \sum_{i=1}^n \frac{p_i}{q_i},
\end{equation}
one finds a related integral expression involving a noncompact contour in the domain $u\in \mathfrak{h}_{\mathbb{C}}$ called the ``$\sigma$-contour'' $\cC_\sigma$, which connects the regions $\operatorname{Im}(u)\to \pm \infty$:
\begin{equation}
Z_{\cM_3}(\nu)_\m = \frac{(-1)^{\operatorname{rank}(G)}}{|\cW|}\sum_{\substack{(\n_0, \n_i) \\ \in \Lambda_W^\vee\otimes \widetilde{\operatorname{Pic}}(\cM_3)}}\oint_{\cC_\sigma} d^{\operatorname{rank}(G)}u\, \cH(u, \nu)^{g - 1}H(u, \nu)\prod_{i=0}^n \cG_{q_i, p_i}(u, \nu)_{\n_i, \m_i}.
\label{sigmacontour}
\end{equation}
Now the sum is over the Picard group of $\cM_3$, which can be identified as the quotient
\begin{equation}
\widetilde{\operatorname{Pic}}(\mathcal{M}_3)\cong \operatorname{Pic}(\hat{\Sigma}_g)/\langle[\mathcal{L}_0]\rangle.
\label{3dpicard}
\end{equation}
When $g = 0$, we have $\widetilde{\operatorname{Pic}}(\mathcal{M}_3)\cong H_1(\mathcal{M}_3, \mathbb{Z})$, so the fluxes take values in this homology group.  More generally, there would be an additional contribution to the homology from the cycles of $\hat{\Sigma}_g$, but these do not contribute to the partition function.

There is one more ingredient that we must describe before properly defining the various operators mentioned above, which is the $R$-symmetry line bundle $\mathbf{L}_R$ on which the $U(1)_R$ gauge field lives.  We may specify this bundle by its fractional fluxes:
\begin{equation}
\mathbf{L}_R\cong [\n_0^R; \n_1^R, \ldots, \n_n^R].
\end{equation}
To preserve supersymmetry, it must satisfy (as a relation in $\operatorname{Pic}(\hat{\Sigma}_g)$)
\begin{equation}
\mathbf{L}_R^{\otimes 2}\cong \cK\otimes \cL_0^{\otimes 2\nu_R}
\label{LRrelation}
\end{equation}
where $\cK$ is the canonical bundle of $\hat{\Sigma}_g$, given by
\begin{equation}
\cK\cong [2(g - 1); q_1 - 1, \ldots, q_n - 1]
\end{equation}
(a spin structure on $\hat{\Sigma}_g$ is a line bundle $\sqrt{\mathcal{K}}$, whereas a spin$^c$ structure corresponds to introducing an $L$ such that $\mathcal{K}\otimes L$ has a square root).  The relation \eqref{LRrelation} may allow several possible choices of $\n_i^R\in \Z$ and $\nu_R\in \frac{1}{2}\Z$.  These choices are correlated with that of a spin structure on $\cM_3$, as discussed in \cite{Closset:2018ghr}.  It is convenient to parametrize the possible $R$-symmetry line bundles by
\begin{equation}
\mathfrak{n}_0^R = g - 1 + \frac{\ell_0^R}{2} + \nu_R\mathbf{d}, \quad \mathfrak{n}_i^R = \frac{q_i - 1}{2} + \frac{\ell_i^R q_i}{2} + \nu_R p_i, \quad \sum_{i=0}^n \ell_i^R = 0,
\label{rsymbackground}
\end{equation}
where the integers $\ell_0^R$ and $\ell_i^R$ are chosen so that the fluxes $\mathfrak{n}_0^R$ and $\mathfrak{n}_i^R$ are integers.  In fact, only the choice of each $\ell_i^R$ (mod $2$) is meaningful.

Finally, explicit formulas for the fibering operators can be found in Appendix \ref{explicit}.  The $(q_i, p_i)$-fibering operators may not be individually well-defined for every choice of $\mathbf{L}_R$ in the half-BPS background $(\mathcal{M}_3, \mathbf{L}_R)$, although their product must be.  The $R$-symmetry line bundle also determines the allowed $R$-charges of matter fields, $r$, via the requirement that $\mathbf{L}_R^{\otimes r}$ be well-defined for all $r$.

Note that all half-BPS lens space backgrounds allow for one-parameter deformations by $b\in \mathbb{C}$.  When $b^2\in \mathbb{Q}$, $L(p, q)_b$ admits a presentation as a Seifert fibration over $S^2$ with two exceptional fibers.  For such rationally squashed lens spaces, the Seifert fibering formalism reproduces known results while clarifying the dependence on spin structure.  In particular, in the limit of rational squashing, holomorphic blocks reduce to $(q, p)$-fibering operators and the gluing formula for the former reduces to the Bethe-sum formula for the latter.

\subsection{A 2D TQFT for \texorpdfstring{$T[M]$}{T[M]}?}

At this point, we pause to make a few motivating remarks.  One approach to de\-ter\-mi\-ning the TQFT dual to a Seifert manifold $\mathcal{M}_3$ is to understand the map from general (hyperbolic) $M$ to the structure of the 3D A-model for $T[M]$, namely the set of Bethe vacua, operator algebra, and so on.  Since the A-model controls the partition function and loop operator expectation values on $\mathcal{M}_3$, such a map would allow one to compute these observables directly from $M$.  One of the earliest attempts to describe the A-model data of $T_n[M]$ in terms of topological data of $M$ can be found in \cite{Gukov:2016gkn}, and the most basic dictionary entry for the correspondence between these sets of data is the identification of Bethe vacua of the DGG theory $T_n[M]$ with irreducible flat $SL(n, \mathbb{C})$ connections on $M$.

As observed in \cite{Closset:2018ghr}, essentially all of the data in the $\mathcal{M}_3$ partition functions are implicitly contained in the holomorphic blocks for $T[M]$.  The latter are known to be dual to the contributions to the complex Chern-Simons partition function from given saddles/flat connections on $M$ \cite{Beem:2012mb}.  However, the blocks are more complicated objects than the A-model observables: the latter are determined by the twisted superpotential, with simple polynomial vacuum equations.  So it may be technically simpler to work in the limit where the A-model is relevant.  In particular, there should be a prescription for gluing tetrahedra by solving corresponding systems of polynomial equations, which determine the appropriate A-model data and correspond to some interesting topological invariants of $M$ that might be explicitly computable.

One hint in this direction is that interesting simplifications of state integrals relevant to complex Chern-Simons theory (i.e., $S_b^3$ partition functions) were noticed in \cite{Garoufalidis:2014ifa} in the case of rational squashing (i.e., when the parameter $q$ is a root of unity).\footnote{The cyclic quantum dilogarithm and the $b = 1$ quantum dilogarithm in \cite{Garoufalidis:2014ifa} combine to give the $(q, p)$-fibering operator for a free chiral.  Their $\Phi_b$ is also called the noncompact quantum dilogarithm $e_b$, which is precisely the partition function of the tetrahedron theory as defined in \cite{Dimofte:2011ju} and is related to the partition function of a free chiral (the double sine function $s_b$) by an exponential factor (Chern-Simons contact term).  The basic state-integral formula that they consider is the $S_b^3$ partition function of a particular theory of class $\mathcal{R}$ (abelian Chern-Simons-matter theory); see Section 7.2 of \cite{Closset:2018ghr}.  The results of \cite{Closset:2018ghr} suggest that the basic statements of \cite{Garoufalidis:2014ifa} generalize to class-$\mathcal{R}$ theories of arbitrary rank.}  Namely, the quantum invariant from a flat connection is in general given by a complicated integral (holomorphic block), while in this case it simplifies to a finite sum.  In other words, complex Chern-Simons invariants (which map to $S_b^3$ partition functions under the 3D-3D correspondence) specialize to ordinary Chern-Simons (Wit\-ten-Re\-she\-tik\-hin-Tu\-ra\-ev) invariants in this limit.  This is precisely the limit relevant for Seifert manifolds, suggesting that the Seifert fibering operators give new basic building blocks of observables.  There is then the question of why and whether the particular ways these are combined to form closed Seifert manifold partition functions are interesting on the Chern-Simons side.  Recent work \cite{Cheng:2018vpl, Chun:2019mal} may help to shed light on this question.

We should note that there already exists a partial dictionary between perturbative $SL(n, \mathbb{C})$ Chern-Simons invariants of $M$ and the A-model data for $T_n[M]$ that has found applications in several contexts \cite{Gang:2018hjd, Gang:2019uay, Benini:2019dyp}.\footnote{The operators appearing in this dictionary have also cropped up in completely different constructions of 3D TQFTs than the one we consider here \cite{Dedushenko:2018bpp}.}  To wit, the 3D-3D dictionary entries for the handle-gluing and ordinary fibering operators are
\begin{equation}
\mathcal{H}^\alpha = \exp(-2S_1^\alpha), \quad \mathcal{F}^\alpha = \exp\left(\frac{iS_0^\alpha}{2\pi}\right).
\label{dictionary}
\end{equation}
The objects on the left are used to assemble the partition function of $T_n[M]$ on $\mathcal{M}_3$, and the objects on the right are terms in the perturbative expansion of the $SL(n, \mathbb{C})$ Chern-Simons partition function on $M$ around the flat connection $\mathcal{A}^\alpha$ corresponding to the Bethe vacuum $\alpha$:
\begin{equation}
Z_\text{CS}^\alpha[M] = \exp\left(\sum_{j=0}^\infty \hbar^{j-1} S_j^\alpha\right).
\end{equation}
These relations are deduced by combining two facts from the literature: the asymptotic relation between $Z_\text{CS}^\alpha[M]$ and a holomorphic block $B^\alpha$ of $T_n[M]$ \cite{Beem:2012mb}, and the asymptotic relation between holomorphic blocks and $\mathcal{H}, \mathcal{F}$ for any given theory \cite{Closset:2018ghr}.

Let us examine the derivation of the dictionary \eqref{dictionary}, and in the process, see how the na\"ive extension to $(q, p)$-fibering operators fails.  We know from \cite{Beem:2012mb} that as $\hbar\to 0$,
\begin{equation}
Z_\text{CS}^\alpha(\hbar)\sim B^\alpha(\mathfrak{q}\equiv e^\hbar)
\label{CStoblock}
\end{equation}
where $B^\alpha$ is a holomorphic block.  Above, we have left the dependence on $M$ implicit and suppressed the $\nu$-dependence of $Z_\text{CS}^\alpha(\hbar)$, which should really be written as $Z_\text{CS}^\alpha(y, \mathfrak{q})$ where $y\equiv e^{2\pi i\nu}$ and $\mathfrak{q}\equiv e^{2\pi i\tau}$ (the parameter $y$ comes from analytically continuing the relation $y = \mathfrak{q}^n$ in $SL(n, \mathbb{C})$ Chern-Simons theory \cite{Beem:2012mb}).  From \cite{Closset:2018ghr}, we have that
\begin{equation}
\lim_{\tau\to 0} B^\alpha(\nu, -\tau)B^\alpha(\nu, \tau) = \mathcal{H}^\alpha(\nu)^{-1}.
\label{blockstoH}
\end{equation}
With $2\pi i\tau = \hbar$, \eqref{CStoblock} gives
\begin{equation}
\lim_{\tau\to 0} B^\alpha(\nu, -\tau)B^\alpha(\nu, \tau) = \lim_{\tau\to 0} Z_\text{CS}^\alpha(-2\pi i\tau)Z_\text{CS}^\alpha(2\pi i\tau) = \exp(2S_1^\alpha),
\end{equation}
so we obtain from \eqref{blockstoH} the first equality in \eqref{dictionary}.  From \cite{Closset:2018ghr}, we also have that
\begin{equation}
\lim_{\tau\to 0} \frac{B^\alpha\left(\frac{\nu}{p\tau + q}, \frac{s\tau - t}{p\tau + q}\right)_{\mathfrak{m}}}{B^\alpha(\nu, \tau)} = \mathcal{G}_{q, p}^\alpha(\nu)_{\mathfrak{m}}
\label{blockstoG}
\end{equation}
where $q > 0$ by assumption and $B^\alpha(\nu, \tau)_{\mathfrak{m}}\equiv B^\alpha(\nu + \mathfrak{m}\tau, \tau)$.  The numerator in \eqref{blockstoG} is the result of applying a modular transformation by
\begin{equation}
\tilde{g} = \left(\begin{array}{cc} s & -t \\ p & q \end{array}\right)\in SL(2, \mathbb{Z})
\label{gtilde}
\end{equation}
to $B^\alpha(\nu, \tau)_{\mathfrak{m}}$.  We need only the $\mathfrak{m} = 0$ case of \eqref{blockstoG}: for $q = p = 1$, it gives
\begin{equation}
\lim_{\tau\to 0} \frac{B^\alpha\left(\frac{\nu}{\tau + 1}, \frac{\tau}{\tau + 1}\right)}{B^\alpha(\nu, \tau)} = \mathcal{F}^\alpha(\nu).
\end{equation}
Note that we have chosen $t = 0$ to ensure a finite limit,\footnote{Geometrically, the holomorphic block is a partition function on a disk fibered over a circle, and when this fibration approaches a rotation by a rational angle $t/q$, the block becomes the local model of a $(q, p)$ exceptional fiber.  We have used the freedom to define $t$ in \eqref{gtilde} to eliminate any residual twisting as $\tau\to 0$.} in which case
\begin{equation}
\lim_{\tau\to 0} \frac{B^\alpha\left(\frac{\nu}{\tau + 1}, \frac{\tau}{\tau + 1}\right)}{B^\alpha(\nu, \tau)} = \lim_{\tau\to 0} \frac{Z_\text{CS}^\alpha\left(\frac{2\pi i\tau}{\tau + 1}\right)}{Z_\text{CS}^\alpha(2\pi i\tau)} = \exp\left(-\frac{iS_0^\alpha}{2\pi}\right).
\end{equation}
This result differs by a sign from \eqref{dictionary}, but such a mismatch is easily fixed by invoking the parity symmetry $\mathcal{F}(\nu, k) = \mathcal{F}(-\nu, -k)^{-1}$ where $\nu$ and $k$ denote real masses and Chern-Simons levels, respectively \cite{Closset:2017zgf}.

These manipulations are simple enough, but trying to extend them to $(q, p)$-fibering operators in the most obvious way leads to trouble.  Consider that
\begin{align}
\mathcal{G}_{q, p}^\alpha(\nu) &= \lim_{\tau\to 0} \frac{Z_\text{CS}^\alpha\left(\frac{2\pi i(s\tau - t)}{p\tau + q}\right)}{Z_\text{CS}^\alpha(2\pi i\tau)} \\
&= \lim_{\tau\to 0} \exp\left[\left(\frac{p\tau + q}{s\tau - t} - \frac{1}{\tau}\right)\frac{S_0^\alpha}{2\pi i} + \sum_{j=2}^\infty \left(\frac{2\pi i(s\tau - t)}{p\tau + q}\right)^{j-1}S_j^\alpha\right].
\end{align}
In this expression, we are free to shift $(t, s)\to (t + mq, s - mp)$ for any $m\in \mathbb{Z}$ while preserving the condition $pt + qs = 1$.  The pole in the coefficient of the $S_0^\alpha$ term must cancel for the limit to be well-defined, which requires $t = 0$.  Moreover, if $t$ were nonzero, then the limit would contain an infinite sum of perturbative Chern-Simons invariants $S_{j\geq 2}^\alpha$, and evaluating it would come down to evaluating the full Chern-Simons partition function with a particular $\hbar$.  Since we may label nontrivial $\mathcal{G}_{q, p}^\alpha$ by $q > 0$ and $0\leq p < q$ with $(q, p)$ coprime, this further requires $q = s = 1$, so we can only handle the trivial case $\mathcal{F}^\alpha$ in this way.

From this point of view, there is something special about the $\mathcal{H}^\alpha$ and $\mathcal{F}^\alpha$ operators as compared to the generic $(q, p)$-fibering operators.  The fact that only the $SL(2, \mathbb{Z})$ transformations that lead to the handle-gluing operator and (powers of) the fibering operator have straightforward interpretations in terms of perturbative Chern-Simons invariants suggests that the other operators give interesting new invariants.  One way of phrasing this specialness is that there exists a simple asymptotic formula for the blocks as $\mathfrak{q}\to 1$, expressed in terms of the twisted superpotential, but not as $\mathfrak{q}\to$ root of unity.  An exception to this statement is the case of the free chiral multiplet, where the former limit can be used to take the latter limit of a block, yielding an expression in terms of powers of the ordinary fibering and flux operators (see \eqref{qidentity}).  However, for more general gauge theories, there appears to be no simple way to write the $(q, p)$-fibering operators in terms of the twisted superpotential alone.

The partial dictionary is already explicit enough for computing some observables, but we wish to understand the general case.  We therefore pursue a different tack.

\section{State-Integral Model} \label{stateint}

Our starting point is \cite{Dimofte:2014zga}, which uses the $L(k, p)_b$ partition functions of $T_n[M]$ to construct, via the DGG algorithm, a state-integral model for $SL(n, \mathbb{C})$ Chern-Simons theory at level $k$, or (for $p > 1$) a deformation thereof.  Likewise, we implicitly define a state-integral model for an $\mathcal{M}_3$ invariant of $M$ on the field theory side of the 3D-3D correspondence, in terms of operations on 3D $\mathcal{N} = 2$ gauge theories.  On the geometry side, dualities among these theories express invariance of the triangulation of $M$ under relabeling of the tetrahedra and under local 2-3 moves, ensuring that $T[M]$ (and hence its $\mathcal{M}_3$ partition function) is a topological invariant of $M$.  This is an indirect definition of the TQFT that computes this topological invariant.  We comment later on some more explicit properties of this TQFT.

We focus on $\mathfrak{g} = \mathfrak{su}(2)$, hence $T_2[M]$, and in particular on the building block for such theories, namely the tetrahedron theory $T_\Delta$.\footnote{It turns out that our tetrahedron theory differs from that in the literature by its polarization, i.e., by a $T$-transformation.  It is presented in the $U(1)_{-1/2}$ quantization.}  As in \cite{Dimofte:2014zga}, the state-integral model consists of composing the DGG algorithm with the evaluation of an $\mathcal{M}_3$ partition function:
\begin{equation}
T_2[M] = \left(\left.\bigotimes_{i=1}^N T_{\Delta_i}\right/\sim\right) \quad \rightsquigarrow \quad Z_{\mathcal{M}_3}[M] = \left(\left.\prod_{i=1}^N Z_{\mathcal{M}_3}[\Delta_i]\right/\sim\right),
\end{equation}
where we have suppressed polarization-dependence and written $Z_{\mathcal{M}_3}[M]\equiv Z_{\mathcal{M}_3}[T_2[M]]$ to emphasize its interpretation as a topological invariant of $M$.  The symplectic reduction operation $\sim$ is implemented on the left by standard Lagrangian operations on 3D $\mathcal{N} = 2$ gauge theories, and on the right by the corresponding actions on their $\mathcal{M}_3$ partition functions.  The latter operations can be read off from the results of \cite{Closset:2018ghr}.  The precise affine symplectic action on partition functions depends on the Seifert geometry, which we comment on in Appendix \ref{DGGcomparison}.

In other words, the gluing construction of $T_2[M]$ translates to a prescription for calculating the partition function of $T_2[M]$ on a Seifert manifold, given that of $T_\Delta$.  In this section, we show invariance of $Z_{\mathcal{M}_3}[\text{tetrahedron}]$ under the affine $ST$-transformation $\rho$ described in Section \ref{DGGreview} and invariance of $Z_{\mathcal{M}_3}[\text{bipyramid}]$ under the 2-3 move.

Because the DGG algorithm constructs only a subsector of the theory $T_n[M]$, the $L(k, p)_b$ state-integral models of \cite{Dimofte:2014zga} capture only a subsector of $SL(n, \mathbb{C})$ Chern-Simons theory on $M$.  As such, there are a number of subtleties in defining their Hilbert spaces, which admit an $ISp(2N, \mathbb{Z})$ action as a consequence of the corresponding action on the theories $T_n[M]$ and their $L(k, p)_b$ partition functions.  The machinery of \cite{Dimofte:2014zga}, which we do not repeat here, essentially carries through to our setting.  We simply focus on presenting the necessary ingredients for this setup.

\subsection{Linearity}

Gauging is a linear operation for the lens space partition functions, but not for general Seifert manifolds or the fibering operators from which they are constructed.  There are two ways to think about this nonlinearity.  First, the gauging operation can be implemented by a superficially linear JK contour integral prescription:\footnote{For more on the JK contour prescription, see Sections 6 and 7.2 of \cite{Closset:2018ghr}.  See also \cite{Closset:2017zgf}, particularly Section 4, Appendix D, and references therein.} however, this contour depends on the integrand (i.e., the theory).  Alternatively, solving the Bethe equations is a nonlinear operation.  Linearity was important in \cite{Dimofte:2011ju, Dimofte:2011py}: for instance, it implies that the $Sp(2N, \mathbb{Z})$ action that changes the polarization acts on the space of partition functions as a linear representation.

We would like to ask: which spaces are exceptions to this nonlinearity?  Namely, on which Seifert manifolds can partition functions be computed via a linear $\sigma$-contour integral formula?  Suppose $c_1(\mathcal{L}_0)\neq 0$:
\begin{enumerate}
\item If the base is smooth and has genus zero, then the $\sigma$-contour formula is conjecturally valid for any 3D $\mathcal{N} = 2$ gauge theory.
\item If the base is not smooth or does not have genus zero, then the $\sigma$-contour formula is conjecturally valid for abelian 3D $\mathcal{N} = 2$ gauge theories.
\end{enumerate}
Since class-$\mathcal{R}$ theories are abelian, we may focus on the second case.  The significance of spaces with $c_1(\mathcal{L}_0)\neq 0$ is that they lack fermion zero modes, which allows one to use the same theory-independent contour as in the $S^3$ and lens space cases, making gauging a linear operation.\footnote{More precisely, there may still be a contribution from gaugino zero modes, as discussed in Section 6.3 of \cite{Closset:2018ghr}, but this is only an issue for nonabelian theories.  Namely, as described in Section 4.1 of \cite{Closset:2017zgf}, there are always $g + 1$ complex fermion zero modes.  For $g > 0$, $g$ of these are associated to cycles of the Riemann surface, and pair up with bosonic zero modes associated to the holonomies of the gauge field along these cycles; integrating these out gives the Hessian determinant.  The remaining fermion zero mode is more subtle.  Proceeding as in Appendix D.1 of \cite{Closset:2017zgf} leads to the JK contour formula.  In some cases, by a different choice of gauge fixing, we may instead use the $\sigma$-contour formula.  This was not carefully derived by localization other than in some simple cases, like $S^3$, but the form of the answer suggests that there is no longer any fermion zero mode in this gauge, at least in the absence of special fibers.  However, with special fibers, it seems that there are fermion zero modes coming from nonabelian gauginos at the special fibers, which modify the na\"ive $\sigma$-contour for nonabelian theories.  Moreover, if $g > 0$, then the additional fermion zero modes can modify the $\sigma$-contour for nonabelian theories, even for smooth base.}  Therefore, such spaces have a chance at allowing for a state-sum TQFT interpretation similar to that in \cite{Dimofte:2014zga}.  One can write a ``Coulomb branch formula'' for the partition function of a class-$\mathcal{R}$ theory on these spaces analogous to that for $S^3$ and lens spaces.

For simplicity, we further focus on the case that $\cM_3$ is a rational homology sphere ($\mathbb{Q}$HS).  A Seifert manifold is a rational homology sphere if and only if $g = 0$ and $c_1(\mathcal{L}_0)\neq 0$.\footnote{Indeed, from Section 2.2.1 of \cite{Closset:2018ghr}, we see that these are precisely the conditions for the free parts of the integral homology groups $H_1(\mathcal{M}_3, \mathbb{Z})$ and $H_2(\mathcal{M}_3, \mathbb{Z})$ to vanish, and hence for the corresponding rational homology groups to vanish.}  For instance, all of the spherical manifolds given by quotients of $S^3$ by finite subgroups $\Gamma\subset SU(2)$ (which we denote by $\mathcal{S}^3[ADE]\cong S^3/\Gamma_{ADE}$, following \cite{Closset:2018ghr}) are rational homology spheres, including the exceptional cases $\mathcal{S}^3[E_{m+3}]\cong S^3/\Gamma_{E_{m+3}}$ with $m > 5$, with the exception of the $E_9$ case ($m = 6$), which has $c_1(\mathcal{L}_0) = 0$.  Note that the more stringent conditions characterizing a Seifert integral homology sphere ($\mathbb{Z}$HS) are \cite{Closset:2018ghr}:
\begin{equation}
\mathcal{M}_3\cong [0; 0; (q_1, p_1), \ldots, (q_n, p_n)], \quad c_1(\mathcal{L}_0) = \pm\frac{1}{\prod_{i=1}^n q_i}.
\label{zhsconditions}
\end{equation}
In particular, the $q$-values must be mutually coprime.  Examples include the Poincar\'e homology sphere ($m = 5$) and the $E_{10}$ integral homology sphere ($m = 7$).

One might expect that if the Seifert rational homology sphere partition functions of $T[M]$ are indeed Chern-Simons invariants of $M$, then $k, \sigma$ of the complex Chern-Simons theory should be specified by the Seifert geometry, the intuition being that $k$ comes from its degree as a fibration over $S^2$ \cite{Cordova:2013cea, Dimofte:2014zga}.  But note that a $\mathbb{Q}$HS, unlike a lens space, admits no squashing parameter that could appear in the continuous level.  What seems more likely is that whatever may be the Chern-Simons interpretation of the TQFT whose state-integral model arises from a $\mathbb{Q}$HS, the ``levels'' depend on the homology of the space, with different manifolds with the same homology leading to twisted versions of the theory, as in the $L(k, p)$ case.

We check some of the basic dualities that are necessary for interpreting the $\cM_3$ partition function of the $T[M]$ theory as computing a TQFT observable on $M$.  While such dualities hold at the level of $(q, p)$-fibering operators in the Bethe-sum formula \cite{Closset:2018ghr}, our goal is to show this directly from the $\sigma$-contour integral formula, reflecting the fact that rational homology spheres are special in that the mirror relations are linear integral relations.

\subsection{\texorpdfstring{$\sigma$}{Sigma}-Contour Identities}

Given a simple $\sigma$-contour integral expression for $Z_{\cM_3}$, we state the integral identities corresponding to two examples of 3D $\cN = 2$ mirror symmetry: the duality between $U(1)_{1/2}$ gauge theory with a charge-one chiral and a free chiral, as well as the duality between $U(1)$ gauge theory with one flavor (two chirals of charges $\pm 1$) and the XYZ model.  These are the basic ingredients necessary to demonstrate that the corresponding partition functions $Z_{\cM_3}(T[M])$ lead, via the 3D-3D correspondence, to a state-integral model for a 3D TQFT.

Our conventions are as follows.  We choose $c\equiv c_1(\mathcal{L}_0) > 0$ by reversing orientation if necessary, which in general takes $\mathbf{d}\to -\mathbf{d}$ and $p_i\to -p_i$.  We always use ``unnormalized'' Seifert invariants with $\mathbf{d} = 0$.  Exceptional fibers $(q_i, p_i)$ are indexed by $i = 1, \ldots, n$, while $(q_0, p_0) = (1, 0)$ is an ordinary fiber.  In our conventions, the effective dilaton is independent of the $R$-charge:
\begin{equation}
e^{2\pi i\Omega} = \Pi^\Phi(u + \nu_A q_A)^{-1}
\end{equation}
for a matter field of charge $q_A$, with $\Pi^\Phi(u) = (1 - e^{2\pi iu})^{-1}$ being the ordinary flux operator for a chiral multiplet.  This then contributes a factor of $\Pi^\Phi(u + \nu_A q_A)^{1-g}$ to the partition function, as one sees by putting \eqref{handlehessrelation} into \eqref{bethesum} and \eqref{sigmacontour}.  The ordinary $R$-symmetry flux contributes an additional power of $\Pi^\Phi(u + \nu_A q_A)^{r\mathfrak{n}_{R, 0}}$.  There are also contributions from the special fibers that depend on the corresponding fractional $R$-symmetry fluxes $\mathfrak{n}_{R, i}$.

To simplify the $\sigma$-contour analysis, we specialize in this subsection to the case that $\cM_3$ is a $\mathbb{Z}$HS.  These examples are nice for two reasons.

First, on such spaces, there exists an $R$-symmetry background with $\mathbf{L}_R$ topologically trivial, and hence no quantization condition on the $R$-charge.\footnote{This is also the case for all of the spherical manifolds.}  In particular, one can couple theories with arbitrary real $R$-charges to this background, including SCFTs.  Indeed, recall that the $R$-symmetry background is determined by \eqref{rsymbackground}.  For fixed $\mathfrak{n}_0^R$ and $\mathfrak{n}_i^R$, these equations determine $\ell_0^R$, $\ell_i^R$, $\nu_R$.  Hence we see that a given $\cM_3$ with $c_1(\cL_0)\neq 0$ admits trivial $\mathbf{L}_R$ ($\mathfrak{n}_0^R = \mathfrak{n}_i^R = 0$) if and only if $2\nu_R\in \mathbb{Z}$ and $2\nu_R p_i\equiv 1\text{ (mod $q_i$)}$ for $i = 1, \ldots, n$, where $\nu_R = -c_1(\cK)/2c_1(\cL_0)$.  One can check using \eqref{zhsconditions} that these conditions are satisfied for a $\mathbb{Z}$HS.\footnote{Whereas we choose the $R$-symmetry fluxes to vanish, the standard A-twist entails $\mathfrak{n}_0^R = g - 1$ and $\mathfrak{n}_i^R = (q_i - 1)/2$.  The two choices agree for even $R$-charges, but may disagree (and indeed, the A-twist may not be well-defined) for more general $R$-charges.}

Second, these spaces have no integer homology (by definition) and therefore trivial 3D Picard group \eqref{3dpicard}.  Then without loss of generality, we may set all gauge fluxes $\mathfrak{n}_i$ to zero, as general fluxes can be obtained through large gauge transformations, i.e., by shifting $u$ and $\nu$ by appropriate integers.  This means that the sum over discrete fluxes in \eqref{sigmacontour} (in which $\mathfrak{n}_i$ ranges from 0 to $q_i - 1$) is trivial on a $\mathbb{Z}$HS, unlike for a general lens space.  Consequently, the $\sigma$-contour integral expression is particularly simple.

Three good examples to keep in mind are the (rationally) squashed sphere
\begin{equation}
S^3_b\cong [0; 0; (1, 0), (q_1, p_1), (q_2, p_2)], \quad b^2 = q_1/q_2, \quad q_1 p_2 + q_2 p_1 = 1,
\end{equation}
the Poincar\'e homology sphere (PHS)
\begin{equation}
\mathcal{M}_\text{PHS}\cong [0; 0; (1, 0), (2, -1), (3, 1), (5, 1)],
\end{equation}
and the $E_{10}$ integral homology sphere
\begin{equation}
\mathcal{M}_{E_{10}}\cong [0; 0; (1, 0), (2, 1), (3, -1), (7, -1)].
\end{equation}
All of these examples admit an $R$-symmetry background with $\n_0^R = \n_i^R = 0$, where
\begin{equation}
\nu_R = \frac{q_1 + q_2}{2}, \quad (\ell_0^R, \ell_1^R, \ell_2^R) = (2, p_2 - p_1 - 1, p_1 - p_2 - 1)
\end{equation}
in the first case and
\begin{equation}
\nu_R = \pm\frac{1}{2}, \quad (\ell_0^R, \ell_1^R, \ell_2^R, \ell_3^R) = (2, 0, -1, -1)
\end{equation}
in the latter two cases, respectively.

Finally, it is important that the $\sigma$-contour should separate all poles coming from positively and negatively charged chirals (as described briefly in \cite{Closset:2018ghr} and in more detail in Section 4.6.1 of \cite{Closset:2017zgf}).

\subsubsection{Gauged/Free Chiral Duality} \label{duality1}

Our conventions for this elementary mirror symmetry, relating $U(1)_{1/2}$ with one chiral multiplet of charge $+1$ (``theory $\mathcal{T}$'') and a free chiral multiplet (``theory $\mathcal{T}^D$''), are as in Section 5.2 of \cite{Closset:2018ghr}.  The chiral multiplet of $\mathcal{T}^D$ is identified with the monopole of $\mathcal{T}$, and carries charge $+1$ under their shared $U(1)_T$ flavor symmetry.  Moreover, the $R$-charges of the chiral multiplets in $\mathcal{T}$ and $\mathcal{T}^D$ are identified as $r$ and $1 - r$, respectively, and the duality incurs some relative Chern-Simons contact terms for the $R$-symmetry.  To begin, we keep $g$ and the sum over gauge fluxes general.

First consider the theory $\mathcal{T} = U(1)_{1/2} + \Phi$.  The gauge and flavor symmetries, and their corresponding parameters, are
\begin{equation}
U(1)_G\times U(1)_T\times U(1)_R, \qquad u, \zeta, \nu_R, \qquad \mathfrak{n}, \mathfrak{n}_T, \mathfrak{n}_R.
\end{equation}
We often leave the $R$-symmetry fluxes implicit.  By \eqref{gaugeflux} and \eqref{effectivedilaton}, the Hessian and effective dilaton are given by\footnote{It follows that $\cH = e^{2\pi i\Omega}H$ for this theory is trivial.  By contrast, the standard A-twist would give $e^{2\pi i\Omega} = \Pi^\Phi(u + \nu_R r)^{r-1}$ and $\mathcal{H} = \Pi^\Phi(u + \nu_R r)^r$.}
\begin{equation}
H = e^{-2\pi i\Omega} = \Pi^\Phi(u + \nu_R r).
\end{equation}
The $(q, p)$-fibering operator is
\begin{align}
\mathcal{G}_{q, p}(u, \zeta; \nu_R)_{\mathfrak{n}_T} &= \sum_{\mathfrak{n} = 0}^{q-1} \mathcal{G}_{q, p}(u, \zeta; \nu_R)_{\mathfrak{n}, \mathfrak{n}_T}, \\
\mathcal{G}_{q, p}(u, \zeta; \nu_R)_{\mathfrak{n}, \mathfrak{n}_T} &\equiv \frac{1}{\sqrt{q}}\mathcal{G}_{q, p}^{(0)}\mathcal{G}_{q, p}^\Phi(u + \nu_R r)_{\mathfrak{n} + \mathfrak{n}_R r}\mathcal{G}_{q, p}^{GG}(u)_{\mathfrak{n}}\mathcal{G}_{q, p}^{GT}(u, \zeta)_{\mathfrak{n}, \mathfrak{n}_T},
\end{align}
which includes vector multiplet, chiral multiplet, CS, and FI contributions.  Since all of the fibering operators are given in the $U(1)_{-1/2}$ quantization (see Appendix \ref{explicit}), the effect of the CS contribution is (as per the definition of $\mathcal{T}$) to adjust the quantization of the chiral multiplet to $U(1)_{1/2}$.  There is a single Bethe vacuum $\hat{u}$.  We set
\begin{equation}
Z_G(u, \zeta; \nu_R)_{\mathfrak{n}, \mathfrak{n}_T}\equiv \Pi^\Phi(u + \nu_R r)^{1-g}\prod_{i=0}^n \mathcal{G}_{q_i, p_i}(u, \zeta; \nu_R)_{\mathfrak{n}_i, \mathfrak{n}_{T, i}}
\end{equation}
where
\begin{equation}
\mathcal{G}_{1, 0}(u, \zeta; \nu_R)_{\mathfrak{n}_0, \mathfrak{n}_{T, 0}} = \mathcal{G}_{1, 0}(u, \zeta; \nu_R)_{0, \mathfrak{n}_{T, 0}} = e^{2\pi i\mathfrak{n}_{T, 0}u}\Pi^\Phi(u + \nu_R r)^{\mathfrak{n}_{R, 0}r},
\end{equation}
and then compute using \eqref{bethesum} that
\begin{align}
Z^{\mathcal{T}}_\text{Bethe}(\zeta; \nu_R)_{\mathfrak{n}_T} &= H(\hat{u}; \nu_R)^{g-1}\sum_{\mathfrak{n}} Z_G(\hat{u}, \zeta; \nu_R)_{\mathfrak{n}, \mathfrak{n}_T} \label{newformula} \\
&= e^{2\pi i\mathfrak{n}_{T, 0}\hat{u}}\Pi^\Phi(\hat{u} + \nu_R r)^{\mathfrak{n}_{R, 0}r}\sum_{\mathfrak{n}} \prod_{i=1}^n \mathcal{G}_{q_i, p_i}(\hat{u}, \zeta; \nu_R)_{\mathfrak{n}_i, \mathfrak{n}_{T, i}}.
\end{align}
From \eqref{sigmacontour}, an alternative representation of the partition function for $\mathcal{T}$ is
\begin{align}
Z^{\mathcal{T}}_\text{$\sigma$-contour} &= -\int_{\mathcal{C}_\sigma} du\, H(u; \nu_R)^g\sum_{\mathfrak{n}} Z_G(u, \zeta; \nu_R)_{\mathfrak{n}, \mathfrak{n}_T} \label{newformulasigma} \\
&= -\int_{\mathcal{C}_\sigma} du\, e^{2\pi i\mathfrak{n}_{T, 0}u}\Pi^\Phi(u + \nu_R r)^{1 + \mathfrak{n}_{R, 0}r}\sum_{\mathfrak{n}} \prod_{i=1}^n \mathcal{G}_{q_i, p_i}(u, \zeta; \nu_R)_{\mathfrak{n}_i, \mathfrak{n}_{T, i}}.
\end{align}
On a $\mathbb{Z}$HS, with vanishing genus $g$ and gauge fluxes $\mathfrak{n}$, the integrand of \eqref{newformulasigma} is simply $Z_G(u, \zeta; \nu_R)_{0, \mathfrak{n}_T}$.

Now consider the dual theory $\mathcal{T}^D$, which has $k_{TR} = -r$ and $k_{RR} = r^2$.  There are no Bethe vacua.  Since $H$ is trivial, we have from \eqref{bethesum} that
\begin{align}
Z^{\mathcal{T}^D} &= \Pi^\Phi(\zeta + (1 - r)\nu_R)^{1-g} \nonumber \\
&\phantom{==} \times \prod_{i=0}^n \mathcal{G}_{q_i, p_i}^{RR}(\nu_R)^{r^2}\mathcal{G}_{q_i, p_i}^{TR}(\zeta; \nu_R)_{\mathfrak{n}_{T, i}}^{-r}\mathcal{G}_{q_i, p_i}^\Phi(\zeta + (1 - r)\nu_R)_{\mathfrak{n}_{T, i} + (1 - r)\mathfrak{n}_{R, i}}.
\end{align}
Again, this differs from the result obtained using the A-twist.\footnote{Since the results for the fibering operators are stated in the $U(1)_{-1/2}$ quantization, we have for the free chiral theory that up to $\zeta$-independent factors,
\begin{equation}
\lim_{\zeta\to i\infty} Z^{\mathcal{T}^D}\sim e^{-2\pi i\kappa r\zeta}, \quad \lim_{\zeta\to -i\infty} Z^{\mathcal{T}^D}\sim e^{2\pi i(\frac{1}{2}c\zeta^2 - \mathfrak{n}_{T, \text{tot}}\zeta)}.
\end{equation}
These asymptotics follow from the relation $\mathcal{G}_{1, 0}^\Phi(u)_{\mathfrak{n}} = \Pi^\Phi(u)^{\mathfrak{n}}$ and the asymptotics for $\mathcal{G}_{q_i, p_i}^\Phi$, to be stated momentarily.}

To determine the $\sigma$-contour, it is helpful to know the analytic properties of the integrand.  For theory $\mathcal{T}$, we can write
\begin{equation}
Z_G(u, \zeta; \nu_R)_{\mathfrak{n}, \mathfrak{n}_T} = Z_\Phi(u)_{\mathfrak{n}}Z_\text{CS}(u)_{\mathfrak{n}}Z_\text{FI}(u, \zeta)_{\mathfrak{n}, \mathfrak{n}_T}\prod_{i=0}^n \frac{1}{\sqrt{q_i}}\mathcal{G}_{q_i, p_i}^{(0)}
\end{equation}
where
\begin{align}
Z_\Phi(u)_{\mathfrak{n}} &\equiv \Pi^\Phi(u + \nu_R r)^{1-g}\prod_{i=0}^n \mathcal{G}_{q_i, p_i}^\Phi(u + \nu_R r)_{\mathfrak{n}_i + \mathfrak{n}_{R, i}r} \nonumber \\
&\to \begin{cases} 1 & \operatorname{Im}(u)\to +\infty, \\ (\text{phase})e^{2\pi i(cu^2/2 + (\kappa(1 - r) - \mathfrak{n}_\text{tot})u)} & \operatorname{Im}(u)\to -\infty, \end{cases} \label{chiralasymptotic} \\
Z_\text{CS}(u)_{\mathfrak{n}} &\equiv \prod_{i=0}^n \mathcal{G}_{q_i, p_i}^{GG}(u)_{\mathfrak{n}_i} \xrightarrow{\operatorname{Im}(u)\to \pm\infty} (\text{phase})e^{-2\pi i(cu^2/2 - \mathfrak{n}_\text{tot}u)}, \label{CSasymptotic} \\
Z_\text{FI}(u, \zeta)_{\mathfrak{n}, \mathfrak{n}_T} &\equiv \prod_{i=0}^n \mathcal{G}_{q_i, p_i}^{GT}(u, \zeta)_{\mathfrak{n}_i, \mathfrak{n}_{T, i}} \xrightarrow{\operatorname{Im}(u)\to \pm\infty} (\text{phase})e^{-2\pi i(cu\zeta - \mathfrak{n}_\text{tot}\zeta - \mathfrak{n}_{T, \text{tot}}u)}. \label{FIasymptotic}
\end{align}
We have defined\footnote{The parameter $\kappa$ is simply related to other topological invariants of $\hat{\Sigma}_g$, such as the orbifold Euler characteristic: $c_1(\cK) = -\chi(\hat{\Sigma}_g) = 2\kappa$.}
\begin{equation}
\kappa\equiv g - 1 + \frac{1}{2}\sum_{i=1}^n \frac{q_i - 1}{q_i}, \quad \mathfrak{n}_\text{tot}\equiv \sum_{i=0}^n \frac{\mathfrak{n}_i}{q_i} = \mathfrak{n}_0 + \sum_{i=1}^n \frac{\mathfrak{n}_i}{q_i},
\end{equation}
etc.  The asymptotic \eqref{chiralasymptotic} follows from the fact that up to a $u$-independent phase,
\begin{equation}
\lim_{u\to i\infty} \mathcal{G}_{q, p}^\Phi(u)_{\mathfrak{n}} = 1, \quad \lim_{u\to -i\infty} \mathcal{G}_{q, p}^\Phi(u)_{\mathfrak{n}}\sim e^{-\frac{2\pi i}{q}\mathfrak{n}u}e^{i\pi\frac{p}{q}u^2}e^{i\pi\frac{q - 1}{q}u}
\end{equation}
(see Appendix D.3 of \cite{Closset:2018ghr}) and the fact that when $\mathbf{d} = 0$, \eqref{rsymbackground} implies that
\begin{equation}
\mathfrak{n}_{R, \text{tot}} = \kappa + c\nu_R.
\end{equation}
The other asymptotics follow straightforwardly from the explicit formulas in Appendix \ref{explicit}.  Combining \eqref{chiralasymptotic}--\eqref{FIasymptotic} and omitting $u$-independent constants, we have
\begin{equation}
Z_G(u, \zeta; \nu_R)_{\mathfrak{n}, \mathfrak{n}_T}\sim \begin{cases} e^{2\pi i(-cu^2/2 + (-c\zeta + \mathfrak{n}_\text{tot} + \mathfrak{n}_{T, \text{tot}})u)} & \operatorname{Im}(u)\to +\infty, \\ e^{2\pi i(\kappa(1 - r) - c\zeta + \mathfrak{n}_{T, \text{tot}})u} & \operatorname{Im}(u)\to -\infty. \end{cases}
\label{ZGasymptotics}
\end{equation}
Finally, there exists an upper bound on the real part of any pole $u_\ast$ of $Z_\Phi(u)_{\mathfrak{n}}$,
\begin{equation}
c\operatorname{Re}(u_\ast)\leq \mathfrak{n}_\text{tot} + \kappa r - g
\label{upperboundpole}
\end{equation}
($Z_\text{CS}$ and $Z_\text{FI}$ introduce no additional poles into the integrand), so we impose that all poles lie to the left of the integration contour.

We now specialize to a $\mathbb{Z}$HS with $\mathfrak{n}_\text{tot} = g = 0$ and consider a straight $\sigma$-contour parallel to the imaginary axis: $u\in \delta + i\mathbb{R}$.  In light of \eqref{upperboundpole}, imposing that all poles lie to the left of the contour gives
\begin{equation}
\kappa r < c\delta.
\end{equation}
In addition, \eqref{ZGasymptotics} gives the condition for the integrand to converge in both directions:
\begin{equation}
\mathfrak{n}_{T, \text{tot}} - c\operatorname{Re}(\zeta) - c\delta > 0 > \mathfrak{n}_{T, \text{tot}} - c\operatorname{Re}(\zeta) + \kappa(1 - r).
\end{equation}
These inequalities restrict the range of the flavor parameters, in the combination $\mathfrak{n}_{T, \text{tot}} - c\operatorname{Re}(\zeta)$, to a compact interval, where the condition for the interval to be nonempty is
\begin{equation}
c\delta < \kappa(r - 1).
\end{equation}
Thus we must have $\kappa < 0$.  If this condition does not hold, then a suitable contour may still exist, but it cannot be taken to be straight: it must remain on the correct side of the poles at finite $u$ and approach regions where the integral decays at large $u$.  The contour can be deformed for better convergence, using the fact that for $\operatorname{Im}(u)\to +\infty$, the integrand decays more quickly for larger negative $\operatorname{Re}(u)$.

\subsubsection{\texorpdfstring{SQED$_1$}{SQED1}/XYZ Duality} \label{duality2}

For the SQED$_1$/XYZ ($\mathcal{T}$/$\mathcal{T}^D$) duality, our conventions are as in Section 3.2 of \cite{Closset:2016arn} and Section 6.2 of \cite{Closset:2017zgf} (see also Appendix D.4.1 of \cite{Closset:2018ghr} for the relevant integral formula on $S_b^3$).  Theory $\mathcal{T}$ is a $U(1)$ gauge theory with two chiral multiplets $Q, \tilde{Q}$ of charges $\pm 1$ and $R$-charge $r$.  This theory has an axial symmetry $U(1)_A$ and a topological symmetry $U(1)_T$.  Theory $\mathcal{T}^D$ consists of three chiral multiplets $M, T^+, T^-$, the latter two being identified with monopoles in $\mathcal{T}$, and a cubic superpotential $MT^+ T^-$.  We list the charges of the various fields in the following table:
\vspace{0.1\baselineskip}
\begin{center}
\renewcommand{\arraystretch}{1.1}
\begin{tabular}{c|cc|ccc}
& $Q$ & $\tilde{Q}$ & $M$ & $T^+$ & $T^-$ \\ \hline
$U(1)_G$ & 1 & $-1$ & 0 & 0 & 0 \\ \hline
$U(1)_A$ & 1 & 1 & 2 & $-1$ & $-1$ \\
$U(1)_T$ & 0 & 0 & 0 & 1 & $-1$ \\
$U(1)_R$ & $r$ & $r$ & $2r$ & $1 - r$ & $1 - r$
\end{tabular}
\end{center}
\vspace{0.1\baselineskip}
Again, we begin by keeping $g$ and the sum over gauge fluxes general.

First consider theory $\mathcal{T}$.  The gauge and flavor symmetries, along with their chemical potentials and fluxes, are
\begin{equation}
U(1)_G\times U(1)_A\times U(1)_T\times U(1)_R, \qquad u, \nu_A, \zeta, \nu_R, \qquad \mathfrak{n}, \mathfrak{n}_A, \mathfrak{n}_T, \mathfrak{n}_R.
\end{equation}
Below, we omit $\mathbf{u}$-independent terms, where $\mathbf{u}$ includes all gauge and flavor parameters except for $\nu_R$.  With $k_{GT} = k_{GG} = 1$, the twisted superpotential is $\mathcal{W} = \mathcal{W}_\text{CS} + \mathcal{W}_\text{matter}$ where 
\begin{align}
\mathcal{W}_\text{CS} &= \frac{1}{2}(u^2 + (1 + 2\nu_R)u) + \zeta u, \\[5 pt]
\mathcal{W}_\text{matter} &= \sum_\pm \mathcal{W}^\Phi(\pm u + \nu_A + \nu_R r), \quad \mathcal{W}^\Phi(u)\equiv \frac{1}{(2\pi i)^2}\operatorname{Li}_2(e^{2\pi iu}).
\end{align}
The gauge flux operator is
\begin{equation}
\Pi = (-1)^{1 + 2\nu_R}e^{2\pi i(\zeta + u)}\left[\frac{1 - e^{2\pi i(-u + \nu_A + \nu_R r)}}{1 - e^{2\pi i(u + \nu_A + \nu_R r)}}\right],
\end{equation}
and there is a single Bethe vacuum $\hat{u}$:
\begin{equation}
e^{2\pi i\hat{u}} = \frac{(-1)^{2\nu_R}e^{2\pi i(\zeta + \nu_A + \nu_R r)} - 1}{(-1)^{2\nu_R}e^{2\pi i\zeta} - e^{2\pi i(\nu_A + \nu_R r)}}.
\end{equation}
The Hessian is
\begin{equation}
H = \frac{1}{2\pi i}\partial_u\log\Pi = \frac{1}{1 - e^{2\pi i(u + \nu_A + \nu_R r)}} - \frac{1}{1 - e^{2\pi i(u - \nu_A - \nu_R r)}}.
\end{equation}
The $(q, p)$-fibering operator is
\begin{align}
\mathcal{G}_{q, p}(u, \zeta, \nu_A; \nu_R)_{\mathfrak{n}_T, \mathfrak{n}_A} &= \sum_{\mathfrak{n} = 0}^{q-1} \mathcal{G}_{q, p}(u, \zeta, \nu_A; \nu_R)_{\mathfrak{n}, \mathfrak{n}_T, \mathfrak{n}_A}, \\
\mathcal{G}_{q, p}(u, \zeta, \nu_A; \nu_R)_{\mathfrak{n}, \mathfrak{n}_T, \mathfrak{n}_A} &\equiv \resizebox{0.95\width}{!}{$\displaystyle \frac{1}{\sqrt{q}}\mathcal{G}_{q, p}^{(0)}\mathcal{G}_{q, p}^{GG}(u)_{\mathfrak{n}}\mathcal{G}_{q, p}^{GT}(u, \zeta)_{\mathfrak{n}, \mathfrak{n}_T}\prod_\pm \mathcal{G}_{q, p}^\Phi(\pm u + \nu_A + \nu_R r)_{\pm\mathfrak{n} + \mathfrak{n}_A + \mathfrak{n}_R r}$}.
\end{align}
We set
\begin{align}
Z_G(u, \zeta, \nu_A; \nu_R)_{\mathfrak{n}, \mathfrak{n}_T, \mathfrak{n}_A} &\equiv \prod_\pm \Pi^\Phi(\pm u + \nu_A + \nu_R r)^{1-g}\prod_{i=0}^n \mathcal{G}_{q_i, p_i}(u, \zeta, \nu_A; \nu_R)_{\mathfrak{n}_i, \mathfrak{n}_{T, i}, \mathfrak{n}_{A, i}}
\end{align}
and write, using \eqref{bethesum} and \eqref{sigmacontour},
\begin{align}
Z^{\mathcal{T}}_\text{Bethe}(\zeta, \nu_A; \nu_R)_{\mathfrak{n}_T, \mathfrak{n}_A} &= H(\hat{u}, \nu_A; \nu_R)^{g-1}\sum_{\mathfrak{n}} Z_G(\hat{u}, \zeta, \nu_A; \nu_R)_{\mathfrak{n}, \mathfrak{n}_T, \mathfrak{n}_A}, \label{sqedbethe} \\
Z^{\mathcal{T}}_\text{$\sigma$-contour} &= -\int_{\mathcal{C}_\sigma} du\, H(u, \nu_A; \nu_R)^g\sum_{\mathfrak{n}} Z_G(u, \zeta, \nu_A; \nu_R)_{\mathfrak{n}, \mathfrak{n}_T, \mathfrak{n}_A}, \label{sqedsigma}
\end{align}
which can be checked to be equivalent.  On a $\mathbb{Z}$HS, the integrand of \eqref{sqedsigma} simplifies to $Z_G(u, \zeta, \nu_A; \nu_R)_{0, \mathfrak{n}_T, \mathfrak{n}_A}$.

Now consider the theory $\mathcal{T}^D$.  There are no Bethe vacua and the twisted superpotential has no $u$-dependence, so $H = 1$.  We have the contact terms
\begin{equation}
k_{TT} = 1, \quad k_{AA} = 2, \quad k_{RR} = -\frac{1}{2}, \quad k_g = -1.
\end{equation}
The $(q, p)$-fibering operator is
\begin{align}
\mathcal{G}_{q, p}(\zeta, \nu_A; \nu_R)_{\mathfrak{n}_T, \mathfrak{n}_A} &= \mathcal{G}_{q, p}^{TT}(\zeta)_{\mathfrak{n}_T}\mathcal{G}_{q, p}^{AA}(\nu_A)_{\mathfrak{n}_A}^2\mathcal{G}_{q, p}^{AR}(\nu_A; \nu_R)_{\mathfrak{n}_A}^{2r}\mathcal{G}_{q, p}^{RR}(\nu_R)^{2r^2}(\mathcal{G}_{q, p}^{(0)})^{-1} \nonumber \\
&\phantom{==} \times \mathcal{G}_{q, p}^\Phi(2\nu_A + 2\nu_R r)_{2\mathfrak{n}_A + 2\mathfrak{n}_R r} \nonumber \\
&\phantom{==} \times \mathcal{G}_{q, p}^\Phi(\zeta - \nu_A + \nu_R(1 - r))_{\mathfrak{n}_T - \mathfrak{n}_A + \mathfrak{n}_R(1 - r)} \nonumber \\
&\phantom{==} \times \mathcal{G}_{q, p}^\Phi(-\zeta - \nu_A + \nu_R(1 - r))_{-\mathfrak{n}_T - \mathfrak{n}_A + \mathfrak{n}_R(1 - r)}. \label{qpxyz}
\end{align}
To justify the contact terms in \eqref{qpxyz}, imagine separately giving the fields $M, T^\pm$ large positive real masses (for the contact terms generated by integrating out chiral multiplets in our quantization conventions, see Section 4.3.2 of \cite{Closset:2017zgf}):
\begin{itemize}
\item With $r = 0$, we get $AA$ contact terms from $M, T^\pm$ and $TT$ contact terms from $T^\pm$.  There are no $AT$ contact terms since the contributions from $T^\pm$ cancel.
\item With $r\neq 0$, we get additional $AR$ contact terms from $M, T^\pm$ that are linear in $r$ and $RR$ contact terms from $M, T^\pm$ that are quadratic in $r$.  There are no $TR$ contact terms since the contributions from $T^\pm$ cancel.
\end{itemize}
We then have that
\begin{align}
Z^{\mathcal{T}^D} &= e^{i\pi/4}\Pi^\Phi(2\nu_A + 2\nu_R r)^{1-g}\Pi^\Phi(\zeta - \nu_A + \nu_R(1 - r))^{1-g} \nonumber \\
&\phantom{==} \times \Pi^\Phi(-\zeta - \nu_A + \nu_R(1 - r))^{1-g}\prod_{i=0}^n \mathcal{G}_{q_i, p_i}(\zeta, \nu_A; \nu_R)_{\mathfrak{n}_{T, i}, \mathfrak{n}_{A, i}},
\end{align}
where we find empirically that we must include an extra phase factor of $e^{i\pi/4}$ for $Z^{\mathcal{T}^D}$ to match the dual expressions $Z^{\mathcal{T}}_\text{Bethe}$ and $Z^{\mathcal{T}}_\text{$\sigma$-contour}$.

Now we summarize the analytic properties of the integrand for theory $\mathcal{T}$.  We can write
\begin{equation}
Z_G(u, \zeta, \nu_A; \nu_R)_{\mathfrak{n}, \mathfrak{n}_T, \mathfrak{n}_A} = Z_\Phi(u)_{\mathfrak{n}, \mathfrak{n}_A}Z_\Phi(-u)_{-\mathfrak{n}, \mathfrak{n}_A}Z_\text{CS}(u)_{\mathfrak{n}}Z_\text{FI}(u, \zeta)_{\mathfrak{n}, \mathfrak{n}_T}\prod_{i=0}^n \frac{1}{\sqrt{q_i}}\mathcal{G}_{q_i, p_i}^{(0)}
\end{equation}
where, using \eqref{chiralasymptotic},
\begin{align}
Z_\Phi(u)_{\mathfrak{n}, \mathfrak{n}_A} &\equiv \Pi^\Phi(u + \nu_A + \nu_R r)^{1-g}\prod_{i=0}^n \mathcal{G}_{q_i, p_i}^\Phi(u + \nu_A + \nu_R r)_{\mathfrak{n}_i + \mathfrak{n}_{A, i} + \mathfrak{n}_{R, i}r} \nonumber \\
&\to \begin{cases} 1 & \operatorname{Im}(u)\to +\infty, \\ (\text{phase})e^{2\pi i(c(u + \nu_A)^2/2 + (\kappa(1 - r) - \mathfrak{n}_\text{tot} - \mathfrak{n}_{A, \text{tot}})(u + \nu_A))} & \operatorname{Im}(u)\to -\infty. \end{cases}
\end{align}
Note that we have not absorbed all $u$-independent constants into the phase because $\nu_A$ may have nonzero imaginary part.  Using \eqref{CSasymptotic} and \eqref{FIasymptotic}, this then gives
\begin{align}
\lim_{u\to \pm i\infty} Z_G(u, \zeta, \nu_A; \nu_R)_{\mathfrak{n}, \mathfrak{n}_T, \mathfrak{n}_A} &\sim (\text{phase})\prod_{i=0}^n \frac{1}{\sqrt{q_i}}\times e^{2\pi i\mathfrak{n}_\text{tot}\zeta}e^{2\pi i(c\nu_A^2/2 + (\kappa(1 - r)\pm \mathfrak{n}_\text{tot} - \mathfrak{n}_{A, \text{tot}})\nu_A)} \nonumber \\
&\phantom{==} \times e^{2\pi i(\mp\kappa(1 - r) - c(\zeta\pm \nu_A) + \mathfrak{n}_{T, \text{tot}}\pm \mathfrak{n}_{A, \text{tot}})u} \label{sqedasymptotics}
\end{align}
(note that $e^{2\pi i\mathfrak{n}_\text{tot}\zeta}$ is not a phase if $\zeta$ has nonzero imaginary part).  We further have by \eqref{upperboundpole} that any pole $u_\ast$ of $Z_\Phi(u)_{\mathfrak{n}, \mathfrak{n}_A} = Z_\Phi(u + \nu_A)_{\mathfrak{n} + \mathfrak{n}_A}$ satisfies
\begin{equation}
c\operatorname{Re}(u_\ast + \nu_A)\leq \mathfrak{n}_\text{tot} + \mathfrak{n}_{A, \text{tot}} + \kappa r - g,
\end{equation}
and any pole of $Z_\Phi(-u)_{-\mathfrak{n}, \mathfrak{n}_A}$ satisfies
\begin{equation}
c\operatorname{Re}(-u_\ast + \nu_A)\leq -\mathfrak{n}_\text{tot} + \mathfrak{n}_{A, \text{tot}} + \kappa r - g.
\end{equation}
The $\sigma$-contour should separate all poles from positively and negatively charged chirals.

Finally, we specialize to the case of a $\mathbb{Z}$HS.  On a $\mathbb{Z}$HS, with $\mathfrak{n} = 0$, \eqref{sqedasymptotics} becomes
\begin{align}
\lim_{u\to \pm i\infty} Z_G(u, \zeta, \nu_A; \nu_R)_{\mathfrak{n}, \mathfrak{n}_T, \mathfrak{n}_A} &\sim (\text{phase})\prod_{i=0}^n \frac{1}{\sqrt{q_i}}\times e^{2\pi i(c\nu_A^2/2 + (\kappa(1 - r) - \mathfrak{n}_{A, \text{tot}})\nu_A)} \nonumber \\
&\phantom{==} \times e^{2\pi i(\mp\kappa(1 - r) - c(\zeta\pm \nu_A) + \mathfrak{n}_{T, \text{tot}}\pm \mathfrak{n}_{A, \text{tot}})u}.
\end{align}
Let us take $u\in \delta + i\mathbb{R}$.  The integrand decays in both directions (irrespective of $\delta$) if
\begin{equation}
-\kappa(1 - r) - c\operatorname{Re}(\zeta + \nu_A) + \mathfrak{n}_{T, \text{tot}} + \mathfrak{n}_{A, \text{tot}} > 0 > \kappa(1 - r) - c\operatorname{Re}(\zeta - \nu_A) + \mathfrak{n}_{T, \text{tot}} - \mathfrak{n}_{A, \text{tot}}. \hspace{0.4 mm}
\end{equation}
This is equivalent to
\begin{equation}
\mathfrak{n}_{A, \text{tot}} - c\operatorname{Re}(\nu_A) - \kappa(1 - r) > |\mathfrak{n}_{T, \text{tot}} - c\operatorname{Re}(\zeta)|
\end{equation}
(and thus, \emph{a fortiori}, $\mathfrak{n}_{A, \text{tot}} - c\operatorname{Re}(\nu_A) > \kappa(1 - r)$).  Given the pole structure of the integrand and the requirements on the $\sigma$-contour, we further want (with $\mathfrak{n}_\text{tot} = g = 0$)
\begin{equation}
\mathfrak{n}_{A, \text{tot}} + \kappa r - c\operatorname{Re}(\nu_A) < c\delta < -\mathfrak{n}_{A, \text{tot}} - \kappa r + c\operatorname{Re}(\nu_A).
\end{equation}
This is equivalent to
\begin{equation}
-\kappa r - [\mathfrak{n}_{A, \text{tot}} - c\operatorname{Re}(\nu_A)] > |c\delta|
\end{equation}
(and thus, \emph{a fortiori}, $\mathfrak{n}_{A, \text{tot}} - c\operatorname{Re}(\nu_A) < -\kappa r$).  To summarize, we must have $\kappa < 0$ for a straight $\sigma$-contour, in which case we must choose
\begin{equation}
f_A\in (\kappa(1 - r), -\kappa r), \quad |f_T| < f_A - \kappa(1 - r), \quad |c\delta| < -\kappa r - f_A,
\end{equation}
where
\begin{equation}
f_A\equiv \mathfrak{n}_{A, \text{tot}} - c\operatorname{Re}(\nu_A), \quad f_T\equiv \mathfrak{n}_{T, \text{tot}} - c\operatorname{Re}(\zeta).
\end{equation}
Recall that $c > 0$ by convention.

\subsubsection{Comments on Straight Contour}

In general, the choice of $\sigma$-contour depends on both the theory and the Seifert geometry.  The rank-one theories $U(1)_{1/2} + \Phi$ and SQED$_1$ are distinguished in that their corresponding mirror symmetries express consistency under gluing in the 3D-3D correspondence.  Hence both theories should have unambiguously defined $\sigma$-contours, for any geometry.

In SQED$_1$, unlike in $U(1)_{1/2} + \Phi$, the condition for the integrand $Z_G$ to decay as $\operatorname{Im}(u)\to \pm\infty$ is symmetric in the $\pm$ directions and independent of $\operatorname{Re}(u)$.  If this condition is satisfied and if $\kappa > 0$, then there exists no straight line parallel to the imaginary axis in the $u$-plane that separates the poles coming from the two chirals (the poles would be interleaved, so that the $\sigma$-contour would need to zigzag over a horizontal interval in Mellin-Barnes style for finite $u$ before going to $\operatorname{Im}(u) = \pm\infty$: see Section 6.2 of \cite{Closset:2018ghr}, as well as \cite{Closset:2017zgf}).  Hence we restrict our attention to geometries with $\kappa < 0$ (e.g., lens spaces and the PHS, but not the $E_{10}$ integral homology sphere).

In this case, the best convergence for SQED$_1$ is achieved by taking $f_A$ as large as possible, i.e., as close as possible to $-\kappa r$ from below; moreover, taking $f_T = 0$ ensures equally good convergence in both directions.  But since $0 < -\kappa(1 - r) + f_A < -\kappa$, the rate of convergence is bounded from above by $-\kappa$, no matter how we deform the contour and regardless of the value of $r$ (for instance, the PHS has $\kappa = -1/60$).  This situation should be contrasted with that of $U(1)_{1/2} + \Phi$, where it is possible to deform the contour to improve convergence.

Suppose we wish to use the \emph{same} contour in both theories $U(1)_{1/2} + \Phi$ and SQED$_1$.  Let $r_\Phi$ denote the $R$-charge of the chiral multiplet in the theory $U(1)_{1/2} + \Phi$.  A straight $\sigma$-contour with $\operatorname{Re}(u) = \delta$ for $U(1)_{1/2} + \Phi$ must satisfy
\begin{equation}
\kappa r_\Phi < c\delta < f_{T, \Phi} < \kappa(r_\Phi - 1).
\end{equation}
On the other hand, for such a contour to work for SQED$_1$, we choose $f_A = -\kappa(r_\text{SQED} - x)$ with $0 < x < 1$ and then demand that
\begin{equation}
|f_{T, \text{SQED}}| < -\kappa(1 - x), \quad |c\delta| < -\kappa x.
\end{equation}
So we need the interval $(\kappa r_\Phi, \kappa(r_\Phi - 1))$ to overlap with the interval $(\kappa x, -\kappa x)$ for some $0 < x < 1$, which means that we need
\begin{equation}
-1 < r_\Phi < 2.
\end{equation}
To summarize, whenever $\kappa < 0$, there exists a straight $\sigma$-contour that works for both $U(1)_{1/2} + \Phi$ and SQED$_1$ (with some range of flavor parameters for both theories) as long as the $R$-charge for the first theory satisfies the above inequality.  This situation can be compared to that in \cite{Dimofte:2014zga}, where it is shown that convergence of the state integrals defining $Z_{L(k, p)_b}[M]$ imposes positivity conditions on the angles in a triangulation of $M$.  Physically, these positivity conditions constrain the $U(1)_R$ charges of operators in $T_n[M]$.  Under these conditions, $Z_{L(k, p)_b}[M]$ is invariant under 2-3 moves relating positive ideal triangulations.

More generally, one might worry that convergence of the $\sigma$-contour integrals at intermediate stages of the DGG construction requires that the flavor parameters lie in certain windows, but that these flavor parameters must be integrated over at later stages.  In practice, such conflicts can always be resolved by suitable analytic continuation \cite{Dimofte:2014zga}.

The apparent requirement that $\kappa < 0$ could either be a limitation of our approach, which relies crucially on the formalism in \cite{Closset:2018ghr}, or a fundamental constraint on which Seifert geometries admit TQFT duals under the 3D-3D correspondence --- e.g., via the Gauss-Bonnet theorem, an obstruction to the hyperbolicity of $\hat{\Sigma}_g$.  Conservatively (and optimistically), we favor the first possibility.

\subsubsection{Example: Poincar\'e Homology Sphere} \label{PHSexample}

We pause to illustrate the preceding discussion with a quick example.  For the PHS, we have:
\begin{equation}
Z_\text{PHS}(\nu) = \frac{(-1)^{\operatorname{rank}(G)}}{|\cW|}\int_{\cC_\sigma} d^{\operatorname{rank}(G)}u\, e^{-2\pi i\Omega(u, \nu)}\cG_{2, -1}(u, \nu)_{0, 0}\cG_{3, 1}(u, \nu)_{0, 0}\cG_{5, 1}(u, \nu)_{0, 0}.
\label{PHSint}
\end{equation}
According to our conventions, we should also include a $(1, 0)$ fibering operator, but it contributes trivially in this case since we set all fluxes to zero and $\n_0^R = 0$.  In particular, the $R$-charges appear only in the contributions of the chiral multiplets via shifts of their arguments, $u\to u + \nu_R r = u + \frac{1}{2}r$, where $r\in \R$.

We can use the integral formula \eqref{PHSint} to set up checks of some simple 3D dualities.  We define, for later convenience, the partition function of a chiral multiplet of $R$-charge zero on the PHS:
\begin{equation}
Z_\text{PHS}^\Phi(u) = \Pi^\Phi(u)\cG^\Phi_{2, -1}(u)_0\cG^\Phi_{3, 1}(u)_0\cG^\Phi_{5, 1}(u)_0.
\end{equation}
It is convenient to understand the pole structure and asymptotic behavior of $Z_\text{PHS}^\Phi(u)$.  We first note that it can be written as the following infinite product, up to regularization:
\begin{equation}
Z_\text{PHS}^\Phi(u) = \prod_{n\in \Z} \left(\frac{1}{n + u}\right)^{d_n}, \quad d_n\equiv 1 + \left\lfloor -\frac{n}{2}\right\rfloor + \left\lfloor\frac{n}{3}\right\rfloor + \left\lfloor\frac{n}{5}\right\rfloor.
\end{equation}
There are poles of generally increasing order as we move toward large negative $u$, and zeros of increasing order toward large positive $u$.  The rightmost pole occurs at $u = 0$.  Therefore, a chiral multiplet with positive $R$-charge has all poles lying in the negative half-plane, $\operatorname{Re}(u) < 0$.  As for the asymptotic behavior, one computes that
\begin{equation}
\lim_{u\to i\infty} Z_\text{PHS}^\Phi(u) = 1, \quad \lim_{u\to -i\infty} Z_\text{PHS}^\Phi(u) = -i\exp\left[\frac{i\pi}{30}\left(u^2 - u + \frac{1}{6}\right)\right],
\label{PHSasymptotics}
\end{equation}
where the constant phase can be deduced from Appendix D.3 of \cite{Closset:2018ghr}.

Now let us consider the duality between the theory $U(1)_{1/2} + \Phi$ and a free chiral multiplet.  Omitting constant factors and including both CS and FI contributions (in the absence of flux), the partition function of the gauge theory is given by
\begin{equation}
Z_\text{PHS}^{U(1)_{1/2} + \Phi}(\zeta)\propto \int du\, e^{-\frac{i\pi}{30}u^2}e^{-\frac{2\pi i}{30}u\zeta}Z_\text{PHS}^\Phi\left(u + \frac{1}{2}r\right),
\end{equation}
where the integration is over a cycle homologous to the imaginary $u$-axis.  Specifically, if we take $u = \delta + ix$, then the asymptotic behavior of the integrand is:
\begin{equation}
\left|e^{-\frac{i\pi}{30}u^2}e^{-\frac{2\pi i}{30}u\zeta}Z_\text{PHS}^\Phi\left(u + \frac{1}{2}r\right)\right|\sim \begin{cases} e^{\frac{\pi}{15}(\delta + \operatorname{Re}(\zeta))x} & x\to \infty, \\ e^{\frac{\pi}{15}(\frac{1}{2}(1 - r) + \operatorname{Re}(\zeta))x} & x\to -\infty. \end{cases}
\end{equation}
In particular, to decay in both directions, we must have
\begin{equation}
\operatorname{Re}(\zeta) + \delta < 0 < \operatorname{Re}(\zeta) + \frac{1}{2}(1 - r).
\end{equation}
For the contour to lie to the right of all poles, we must also take $\delta > -\frac{1}{2}r$, so we see that $\delta + \frac{1}{2}r$ must belong to the interval $(0, \frac{1}{2})$, and then the domain for $\zeta$ is restricted as above.  Then the duality implies the relation
\begin{equation}
\boxed{Z_\text{PHS}^{U(1)_{1/2} + \Phi}(\zeta) = Z_\text{contact}(\zeta)Z_\text{PHS}^\Phi\left(\zeta + \frac{1}{2}(1 - r)\right),}
\end{equation}
where the first factor includes $FR$, $RR$, and gravitational Chern-Simons contact terms.  This relation can be checked numerically, using the exact formulas from Section \ref{duality1}.

Next, consider the duality between the $U(1)$ gauge theory with a pair of chirals of charges $\pm 1$ and the XYZ model.  The relevant partition functions are
\begin{align}
Z_\text{PHS}^\text{SQED$_1$}(\zeta, \nu) &\propto \int du\, e^{-\frac{i\pi}{30}u^2}e^{-\frac{2\pi i}{30}u\zeta}\prod_\pm Z_\text{PHS}^\Phi\left(\pm u + \nu + \frac{1}{2}r\right), \\
Z_\text{PHS}^\text{XYZ}(\zeta, \nu) &= Z_\text{PHS}^\Phi(2\nu + r)\prod_\pm Z_\text{PHS}^\Phi\left(\pm\zeta - \nu + \frac{1}{2}(1 - r)\right).
\end{align}
Then we expect a relation of the form
\begin{equation}
\boxed{Z_\text{PHS}^\text{SQED$_1$}(\zeta, \nu) = Z_\text{contact}(\zeta, \nu)Z_\text{PHS}^\text{XYZ}(\zeta, \nu).}
\end{equation}
For a straight contour, such a relation holds as long as
\begin{equation}
f\equiv \operatorname{Re}(\nu) + \frac{1}{2}r\in \left(0, \frac{1}{2}\right), \quad \frac{1}{2} - f > |{\operatorname{Re}(\zeta)}|, \quad f > |\delta|.
\end{equation}
It can also be checked numerically using the exact formulas from Section \ref{duality2}.

In Appendix \ref{DGGcomparison}, we make some further remarks on the geometrical interpretation of identities such as those above.

\section{Difference Equations} \label{diffeqs}

The gluing procedure of DGG says little about the TQFT interpretation of the fibering operators, such as what kinds of Chern-Simons invariants they might correspond to.  Indeed, properties such as affine $ST$-invariance of the tetrahedron wavefunction of the dual TQFT and independence of triangulation are in principle guaranteed by mirror symmetry, regardless of $\mathcal{M}_3$.  To gain a better handle on the TQFT dual to $\mathcal{M}_3$, we need to examine in more detail the partition functions $Z_{\mathcal{M}_3}$ themselves rather than their composition under gluing.

We begin by recalling that the lens space partition functions of the theories $T_2[M]$ obey a set of difference equations that can be derived from those of the fundamental theory $T_\Delta$.  This can be understood by noting that $\mathcal{P}_{\partial M}$ is the phase space of $SL(2, \mathbb{C})$ Chern-Simons theory on $M$, and $\mathcal{L}_M$ is the space of classical solutions (i.e., the moduli space of supersymmetric vacua of the theory $T_2[M]$ on a circle) \cite{Dimofte:2011py}.  The classical defining equations of $\mathcal{L}_M$ are promoted to quantum operators that annihilate the wavefunctions of Chern-Simons theory on $M$.  There are holomorphic and antiholomorphic versions of these operators, hence two sets of difference equations.

From the SCFT side of the 3D-3D correspondence, the existence of holomorphic and antiholomorphic difference equations can be understood by factorizing the lens space partition functions into holomorphic blocks \cite{Beem:2012mb}.  The holomorphic block $B_\alpha(x; \mathfrak{q})$ of a 3D $\mathcal{N} = 2$ SCFT is essentially the partition function on a twisted product $D^2\times_{\mathfrak{q}} S^1$, labeled by a choice of vacuum $\alpha$ for the massive theory on $\mathbb{R}^2\times S^1$, with $U(1)$ flavor symmetry fugacities denoted by $x$.  The vacuum $\alpha$ is determined by boundary conditions on the $D^2$, viewed as a semi-infinite cigar.

From this point of view, the holomorphic blocks for a given theory satisfy a set of $\mathfrak{q}$-difference equations that follow from the algebra of half-BPS line operators for background flavor gauge fields wrapping the $S^1$ and inserted at the tip of the cigar (to preserve supersymmetry).  They satisfy as many difference equations as the number of $U(1)$ flavor symmetries $N$.  The operators that annihilate the blocks are polynomials in the Wilson and 't Hooft lines $\hat{x}_i$ and $\hat{p}_i$ ($i = 1, \ldots, N$).  The blocks can be characterized as providing a basis for the vector space of solutions to these difference equations, with suitable analyticity properties.

In this section, we derive the difference equations obeyed by the tetrahedron partition functions on $\mathcal{M}_3$ from properties of the Seifert fibering operators.  We find a richer structure than in the case of the holomorphic-antiholomorphic factorization for lens spaces.  Having derived the difference equations for $T_\Delta$, the difference equations for a general theory of class $\mathcal{R}$ follow from the standard operations of changing polarization and gluing.  This is accomplished via appropriate eliminations in tensor products of the tetrahedron operator algebra \cite{Dimofte:2011py, Beem:2012mb, Dimofte:2014zga}.  We will see in Section \ref{quantization} that these difference equations contain important physical information about the TQFT dual to $\mathcal{M}_3$.

In what follows, we make heavy use of various $\mathfrak{q}$-deformed Pochhammer symbols, which we define here for reference.  The $\mathfrak{q}$-Pochhammer symbol is defined by
\begin{equation}
(x; \mathfrak{q})_n\equiv \begin{cases} \prod_{j=0}^{n-1} (1 - x\mathfrak{q}^j) & \text{for $n > 0$}, \\[2 pt] \prod_{j=1}^{|n|} (1 - x\mathfrak{q}^{-j})^{-1} & \text{for $n < 0$}, \end{cases}
\label{finiteqPoch}
\end{equation}
and $(x; \mathfrak{q})_0\equiv 1$.  The extended $\mathfrak{q}$-Pochhammer symbol is defined by
\begin{equation}
(x; \mathfrak{q})_\infty\equiv \begin{cases} \prod_{j=0}^\infty (1 - x\mathfrak{q}^j) & \text{for $|\mathfrak{q}| < 1$}, \\[2 pt] \prod_{j=0}^\infty (1 - x\mathfrak{q}^{-j - 1})^{-1} & \text{for $|\mathfrak{q}| > 1$}. \end{cases}
\label{qPoch}
\end{equation}
It is analytic for $|\mathfrak{q}| < 1$ and $|\mathfrak{q}| > 1$ but diverges for $|\mathfrak{q}| = 1$.  It satisfies $(x; \mathfrak{q}^{-1})_\infty = (\mathfrak{q}x; \mathfrak{q})_\infty^{-1}$.

\subsection{Lens Spaces} \label{lensspaces}

We first recall the difference equations satisfied by the partition function of the tetrahedron theory on (squashed) lens spaces.

On $L(k, p)_b$ with arbitrary holonomy $m\in \mathbb{Z}/k\mathbb{Z}$ turned on, we have\footnote{On $S_b^3$, this becomes $Z_b^{(1, 1)}[\Delta](\nu, 0) = e^{\frac{i\pi}{2}(iQ/2 - \nu)^2 + \frac{i\pi}{24}(b^2 + b^{-2})}s_b(iQ/2 - \nu)$ \cite{Dimofte:2011ju, Dimofte:2014zga}.  On $L(k, 1)_b$ with holonomy turned on, the answer can be written as a lattice product \cite{Dimofte:2014zga}.} 
\begin{equation}
Z_b^{(k, p)}[\Delta](\nu, m) = (\mathfrak{q}x^{-1}; \mathfrak{q})_\infty(\tilde{\mathfrak{q}}\tilde{x}^{-1}; \tilde{\mathfrak{q}})_\infty
\label{Lkppartition}
\end{equation}
in the $U(1)_{1/2}$ quantization \cite{Dimofte:2014zga}, where
\begin{equation}
\mathfrak{q}\equiv e^{\frac{2\pi i}{k}(b^2 + p)}, \quad \tilde{\mathfrak{q}}\equiv e^{\frac{2\pi i}{k}(b^{-2} + r)}, \quad x\equiv e^{\frac{2\pi i}{k}(-ib\nu - pm)}, \quad \tilde{x}\equiv e^{\frac{2\pi i}{k}(-ib^{-1}\nu + m)},
\end{equation}
and $pr\equiv 1\text{ (mod $k$)}$.  The shift operators
\begin{equation}
\mathbf{x} = x, \quad \tilde{\mathbf{x}} = \tilde{x}, \quad \mathbf{y} = e^{ib\partial_\nu - \partial_m}, \quad \tilde{\mathbf{y}} = e^{ib^{-1}\partial_\nu + r\partial_m}
\end{equation}
satisfy two commuting algebras:
\begin{equation}
\mathbf{y}\mathbf{x} = \mathfrak{q}\mathbf{x}\mathbf{y}, \quad \tilde{\mathbf{y}}\tilde{\mathbf{x}} = \tilde{\mathfrak{q}}\tilde{\mathbf{x}}\tilde{\mathbf{y}}, \quad \mathbf{y}\tilde{\mathbf{x}} = \tilde{\mathbf{x}}\mathbf{y}, \quad \tilde{\mathbf{y}}\mathbf{x} = \mathbf{x}\tilde{\mathbf{y}}.
\end{equation}
From \eqref{qPoch}, we have $(\mathfrak{q}x; \mathfrak{q})_\infty = (1 - x)^{-1}(x; \mathfrak{q})_\infty$, so that
\begin{equation}
\mathbf{y}(\mathfrak{q}x^{-1}; \mathfrak{q})_\infty = (x^{-1}; \mathfrak{q})_\infty = (1 - x^{-1})(\mathfrak{q}x^{-1}; \mathfrak{q})_\infty
\end{equation}
and similarly for $\tilde{\mathbf{y}}$, which implies that
\begin{equation}
(\mathbf{y} + \mathbf{x}^{-1} - 1)Z_b^{(k, p)}[\Delta] = (\tilde{\mathbf{y}} + \tilde{\mathbf{x}}^{-1} - 1)Z_b^{(k, p)}[\Delta] = 0.
\label{Lkpdifference}
\end{equation}
As anticipated, we find a ``holomorphic'' difference equation and an ``antiholomorphic'' counterpart.

\subsection{Properties of Fibering Operators}

To derive the analogous difference equations obeyed by $Z_{\mathcal{M}_3}[\Delta]$, we take an empirical approach that we retroactively justify in Section \ref{lineops}.  The difference equations that we deduce in this section will also be shown in the next section to be uniquely fixed by higher-dimensional considerations.

As a first step, we observe that the difference equations for $Z_b^{(k, p)}[\Delta]\equiv Z_{L(k, p)_b}[\Delta]$ stem from simple multiplicative properties of the $\mathfrak{q}$-Pochhammer symbols (holomorphic blocks).  What similarly useful properties might the Seifert fibering operators satisfy?  For a free chiral multiplet, the $(q, p)$-fibering operator is related to the holomorphic block by \cite{Closset:2018ghr}
\begin{equation}
\mathcal{G}_{q, p}^\Phi(\nu) = \lim_{\tau\to 0}\frac{B_{\tilde{g}}^\Phi(\nu, \tau)}{B^\Phi(\nu, \tau)} = \lim_{\tau\to 0}\frac{(\tilde{\mathfrak{q}}\tilde{y}; \tilde{\mathfrak{q}})_\infty}{(\mathfrak{q}y; \mathfrak{q})_\infty},
\label{fiberingfromblock}
\end{equation}
where we have defined the variables
\begin{equation}
y = e^{2\pi i\nu}, \quad \mathfrak{q} = e^{2\pi i\tau}, \quad \tilde{y} = \exp\left(\frac{2\pi i\nu}{p\tau + q}\right), \quad \tilde{\mathfrak{q}} = \exp\left(\frac{2\pi i(s\tau - t)}{p\tau + q}\right)
\label{definitions}
\end{equation}
and introduced integers $t, s$ associated to $q, p$ satisfying
\begin{equation}
pt + qs = 1.
\end{equation}
Given \eqref{fiberingfromblock}, the finite-$\tau$ difference equations satisfied by the $\mathfrak{q}$-Pochhammer symbols naturally give rise to difference equations satisfied by the fibering operators (we indicate the relevant operators in this case with hats to distinguish them from those in the lens space case).  For example, by analogy with the lens space case, we have
\begin{equation}
\hat{y} = y, \mbox{ } \hat{\pi} = e^{-\tau\partial_\nu} \implies \hat{\pi}\hat{y} = \mathfrak{q}^{-1}\hat{y}\hat{\pi}.
\label{badalgebra}
\end{equation}
Of course, this is not quite what we want, both because the algebra \eqref{badalgebra} trivializes in the $\tau\to 0$ limit and because the $(\mathfrak{q}y; \mathfrak{q})_\infty$ in \eqref{fiberingfromblock} is just a normalization factor, so the difference operators should really act on $(\tilde{\mathfrak{q}}\tilde{y}; \tilde{\mathfrak{q}})_\infty$.  So at finite $\tau$, the relevant shift operator should take $\tilde{y}\mapsto \tilde{\mathfrak{q}}^{-1}\tilde{y}$, which is equivalent to $\nu\mapsto \nu - s\tau + t$ in view of \eqref{definitions}.  This operation does have a nontrivial limit as $\tau\to 0$, namely $\nu\mapsto \nu + t$.  Therefore, we might define the operators
\begin{equation}
\hat{y} = e^{2\pi i\nu/q}, \quad \hat{\pi} : \nu\mapsto \nu + t
\label{goodalgebra}
\end{equation}
acting on $\mathcal{G}_{q, p}^\Phi(\nu)$.  Using the explicit formulas in Appendix \ref{explicit}, we compute that
\begin{equation}
\mathcal{G}_{q, p}^\Phi(\nu + t) = (1 - e^{2\pi i\nu})^{-s}(1 - e^{2\pi i\nu/q})\mathcal{G}_{q, p}^\Phi(\nu).
\label{Gqpdifference}
\end{equation}
Actually, because $\mathcal{G}_{q, p}^\Phi(\nu)$ is not purely a function of $e^{2\pi i\nu/q}$, the action of $\hat{\pi}$ in \eqref{goodalgebra} is ambiguous: we could equally well have defined it as
\begin{equation}
\hat{\pi} : \nu\mapsto \nu + t + mq
\label{alternative}
\end{equation}
for $m\in \mathbb{Z}$ (note that the operator $\hat{\sigma} : \nu\mapsto \nu + q$ commutes with $\hat{\pi}$).  Correspondingly, we are free to redefine $(t, s)\to (t + mq, s - mp)$ in \eqref{Gqpdifference}.

Of course, understanding the action of certain difference operators on the fibering operators is not the whole story: our real interest lies in how these fibering operators combine into full $\mathcal{M}_3$ partition functions, and the action of the appropriate difference operators thereon.  But at the very least, the above discussion suggests that we need to understand the transformation properties of the Seifert fibering operators under arbitrary shifts of their arguments and fluxes.  To this end, we recall that\footnote{At an operational level, the fact that the flux operators are well-defined only when the total flux $\mathfrak{n}$ is an integer makes clear how the $R$-symmetry fluxes constrain the matter $R$-charges.}
\begin{equation}
\mathcal{G}_{q, p}^\Phi(\nu)_\mathfrak{n}\equiv \Pi_{q, p}^\Phi(\nu)_\mathfrak{n}\mathcal{G}_{q, p}^\Phi(\nu), \quad \Pi_{q, p}^\Phi(\nu)_\mathfrak{n}\equiv (e^{2\pi i\nu/q}; e^{2\pi it/q})_{-\mathfrak{n}}
\label{withfluxes}
\end{equation}
and consider the properties
\begin{equation}
\mathcal{G}_{q, p}^\Phi(\nu)_{\mathfrak{n} + q} = \Pi^\Phi(\nu)\mathcal{G}_{q, p}^\Phi(\nu)_\mathfrak{n}, \quad \mathcal{G}_{q, p}^\Phi(\nu + 1)_{\mathfrak{n} + p} = \mathcal{G}_{q, p}^\Phi(\nu)_\mathfrak{n}
\label{equivalencerelations}
\end{equation}
(as spelled out in Section 4.3.2 of \cite{Closset:2018ghr}).  The relations \eqref{equivalencerelations} collectively give
\begin{equation}
\mathcal{G}_{q, p}^\Phi(\nu + a)_{\mathfrak{n} + qb + pa} = \Pi^\Phi(\nu)^b\mathcal{G}_{q, p}^\Phi(\nu)_\mathfrak{n}.
\end{equation}
The first relation in \eqref{equivalencerelations} expresses the fact that $q$ units of fractional flux at a special fiber is equivalent to one unit of ordinary flux, while the second comes from invariance under large gauge transformations.  Note that the property
\begin{equation}
\mathcal{G}_{q, p + q}^\Phi(\nu)_\mathfrak{n} = \mathcal{G}_{1, 1}^\Phi(\nu)\mathcal{G}_{q, p}^\Phi(\nu)_\mathfrak{n}
\end{equation}
is a special case of the statement that $q$ units of fractional flux equate to one unit of ordinary flux, applied to the flux describing the $S^1$ fibration.

To be more explicit, it is helpful to use the identity
\begin{equation}
\prod_{\ell=0}^{q-1} (1 - e^{2\pi i(\nu + \ell)/q}) = 1 - e^{2\pi i\nu}.
\label{completeresidueproperty}
\end{equation}
This formula holds whenever $\ell$ ranges over a complete set of residues (mod $q$), and for instance, after rescaling $\ell\to t\ell$ (because $q$ and $t$ are coprime).  Note that the LHS of \eqref{completeresidueproperty} can be written as $(e^{2\pi i\nu/q}; e^{2\pi i/q})_q$.  Then the first equation in \eqref{equivalencerelations} is seen to be a property of the flux operator, namely
\begin{equation}
\Pi_{q, p}^\Phi(\nu)_{\mathfrak{n} + q} = \Pi^\Phi(\nu)\Pi_{q, p}^\Phi(\nu)_\mathfrak{n},
\end{equation}
as follows from \eqref{finiteqPoch}, \eqref{withfluxes}, and \eqref{completeresidueproperty}.  Now write $p = [p]_q' q + [p]_q$ with $0\leq [p]_q < q$ (we always assume that $q > 0$; for normalized Seifert invariants, we have $[p]_q' = 0$ and $[p]_q = p$).  The second equation in \eqref{equivalencerelations} states that
\begin{equation}
\mathcal{G}_{q, p}^\Phi(\nu + 1) = \frac{\Pi_{q, p}^\Phi(\nu)_\mathfrak{n}}{\Pi_{q, p}^\Phi(\nu + 1)_{\mathfrak{n} + p}}\mathcal{G}_{q, p}^\Phi(\nu) = (1 - e^{2\pi i\nu})^{[p]_q'}\left[\prod_{j=0}^{[p]_q-1} (1 - e^{2\pi i(\nu + tj)/q})\right]\mathcal{G}_{q, p}^\Phi(\nu),
\label{Gqpshift1}
\end{equation}
the quantity in square brackets being $(e^{2\pi i\nu/q}; e^{2\pi it/q})_{[p]_q}$.  The prefactor on the RHS can be written as $\Pi^\Phi(\nu)^{-[p]_q'}\Pi_{q, p}^\Phi(\nu)_{-[p]_q}$, which reduces to $\Pi_{q, p}^\Phi(\nu)_{-p}$ for normalized $(q, p)$.  Indeed, we can infer this property directly from \eqref{Gqpchiral}, using \eqref{completeresidueproperty} and the fact that $p$ is the inverse of $t$ (mod $q$).  Hence the second equation in \eqref{equivalencerelations} specifies that the fibering operator transforms by a multiplicative factor under a shift of $\nu$ by 1.

In light of these facts, we collect some useful formulas below.

\paragraph{Shifts of the flux operator.}

Combining the properties $(x; \mathfrak{q})_{-n-1} = (1 - x\mathfrak{q}^{-(n + 1)})^{-1}\allowbreak (x; \mathfrak{q})_{-n}$ and $(\mathfrak{q}x; \mathfrak{q})_n = (1 - x)^{-1}(x; \mathfrak{q})_{n+1}$ gives
\begin{align}
\Pi_{q, p}^\Phi(\nu + kt)_{\mathfrak{m} + m} &= \frac{(e^{2\pi i(\nu - t(\mathfrak{m} + m))/q}; e^{2\pi it/q})_k}{(e^{2\pi i\nu/q}; e^{2\pi it/q})_k(e^{2\pi i(\nu - t(\mathfrak{m} + 1))/q}; e^{-2\pi it/q})_m}\Pi_{q, p}^\Phi(\nu)_{\mathfrak{m}}
\label{Piqpshift}
\end{align}
for $k, m\in \mathbb{Z}$.

\paragraph{Generic shifts of the fibering operator.}

Using the identity \eqref{completeresidueproperty} and the explicit formula \eqref{Gqpchiral} (or \eqref{Gqpshift1} and induction), we get
\begin{equation}
\mathcal{G}_{q, p}^\Phi(\nu + n) = (1 - e^{2\pi i\nu})^{n[p]_q'}\left[\prod_{j=0}^{[p]_q-1} (e^{2\pi i(\nu + tj)/q}; e^{2\pi i/q})_n\right]\mathcal{G}_{q, p}^\Phi(\nu)
\label{Gqpshiftn}
\end{equation}
for $n\in \mathbb{Z}$.

\paragraph{Shifts of the fibering operator by multiples of $t$.}

When $t | n$, shifts by $n$ simplify substantially.  Using \eqref{completeresidueproperty}, we deduce from \eqref{Gqpshiftn} that
\begin{equation}
\mathcal{G}_{q, p}^\Phi(\nu + mt) = (1 - e^{2\pi i\nu})^{-ms}(e^{2\pi i\nu/q}; e^{2\pi it/q})_m\mathcal{G}_{q, p}^\Phi(\nu) \hspace{0.6 mm}
\label{Gqpshiftmt}
\end{equation}
for $m\in \mathbb{Z}$.

\paragraph{Shifts of the fibering operator by multiples of $q$.}

When $q | n$, shifts by $n$ simplify even further.  Again using \eqref{completeresidueproperty}, we deduce from \eqref{Gqpshiftn} that
\begin{equation}
\mathcal{G}_{q, p}^\Phi(\nu + mq) = (1 - e^{2\pi i\nu})^{mp}\mathcal{G}_{q, p}^\Phi(\nu)
\label{Gqpshiftmq}
\end{equation}
for $m\in \mathbb{Z}$.

\subsection{Spaces with \texorpdfstring{$H_1 = 0$}{H1 = 0}} \label{H1is0}

Let us see whether we can upgrade the above observations to a prescription for writing down the difference equations for $Z_{\mathcal{M}_3}[\Delta]$.  We do so by looking at a series of examples.  The most tractable examples are those for which fractional fluxes play no essential role, namely those for which $H_1(\mathcal{M}_3, \mathbb{Z}) = 0$, so we start with these.  For these examples, it helps to know that on a $\mathbb{Q}$HS, the partition function of a chiral multiplet of vanishing $R$-charge in the absence of flux is given by
\begin{equation}
Z^\Phi(\nu) = \Pi^\Phi(\nu)\widehat{Z}^\Phi(\nu), \quad \widehat{Z}^\Phi(\nu)\equiv \prod_{i=1}^n \mathcal{G}_{q_i, p_i}^\Phi(\nu).
\label{zchiralnofluxes}
\end{equation}
We have stripped off a factor of $\Pi^\Phi(\nu)$, which is insensitive to integer shifts of $\nu$.

The general structure that we find is as follows.  For spaces with no holonomies (and with ordinary and fractional fluxes set to zero), we can associate to each special fiber a pair of operators \eqref{goodalgebra}.  The operators collectively obey
\begin{equation}
\hat{\pi}_q\hat{y}_q = e^{2\pi it/q}\hat{y}_q\hat{\pi}_q, \quad \hat{\pi}_q\hat{y}_{q'} = \hat{y}_{q'}\hat{\pi}_q \quad (q\neq q'),
\label{commutationrelations}
\end{equation}
and each pair gives rise to a separate difference equation
\begin{equation}
(\hat{\pi}_q + \hat{y}_q - 1)\widehat{Z}^\Phi(\nu) = 0 \Longleftrightarrow (\hat{\pi}_q + \hat{y}_q - 1)Z^\Phi(\nu) = 0
\end{equation}
satisfied by the partition function.

\paragraph{Example.}

On $S^3 = L(1, 1)_{b=1}$, we have $\mathfrak{q} = \tilde{\mathfrak{q}} = 1$ and \eqref{Gqpdifference} becomes
\begin{equation}
\mathcal{G}_{1, 1}^\Phi(\nu + 1) = (1 - e^{2\pi i\nu})\mathcal{G}_{1, 1}^\Phi(\nu) \Longleftrightarrow (\hat{\pi} + \hat{y} - 1)\mathcal{G}_{1, 1}^\Phi(\nu) = 0,
\end{equation}
as expected from \eqref{Lkpdifference}.  To be precise, the Pochhammer symbols in \eqref{Lkppartition} diverge when $\mathfrak{q} = \tilde{\mathfrak{q}} = 1$, so we cannot directly compare to \eqref{Lkpdifference} in this case; however, we do expect that in the absence of squashing, the two difference equations in \eqref{Lkpdifference} collapse into one.

\paragraph{Example.}

On $S_b^3 = L(1, 1)_b = L(1, 0)_b$, there are no holonomies and we have
\begin{equation}
\widehat{Z}^\Phi(\nu) = \mathcal{G}_{q_1, p_1}^\Phi(\nu)\mathcal{G}_{q_2, p_2}^\Phi(\nu), \quad b^2 = q_1/q_2, \quad q_1 p_2 + q_2 p_1 = 1.
\label{znofluxsb3}
\end{equation}
Using \eqref{Gqpdifference} and \eqref{Gqpshiftmq}, we get
\begin{equation}
\widehat{Z}^\Phi(\nu + q_2) = (1 - e^{2\pi i\nu/q_1})\widehat{Z}^\Phi(\nu), \quad \widehat{Z}^\Phi(\nu + q_1) = (1 - e^{2\pi i\nu/q_2})\widehat{Z}^\Phi(\nu),
\end{equation}
so that with $\hat{y}_1 = e^{2\pi i\nu/q_1}$, $\hat{y}_2 = e^{2\pi i\nu/q_2}$, and $\hat{\pi}_1$ and $\hat{\pi}_2$ defined as in \eqref{goodalgebra},
\begin{equation}
(\hat{\pi}_1 + \hat{y}_1 - 1)\widehat{Z}^\Phi(\nu) = (\hat{\pi}_2 + \hat{y}_2 - 1)\widehat{Z}^\Phi(\nu) = 0
\end{equation}
as expected.  What if we had used \eqref{alternative} instead?  Define
\begin{align}
\hat{\pi}_1 : \nu &\mapsto \nu + t_1 + m_1 q_1 = \nu + q_2 + m_1 q_1, \\
\hat{\pi}_2 : \nu &\mapsto \nu + t_2 + m_2 q_2 = \nu + q_1 + m_2 q_2.
\end{align}
Regardless of $m_1$ and $m_2$, we have
\begin{equation}
\hat{\pi}_1\hat{y}_1 = e^{2\pi ib^{-2}}\hat{y}_1\hat{\pi}_1, \quad \hat{\pi}_2\hat{y}_2 = e^{2\pi ib^2}\hat{y}_2\hat{\pi}_2.
\end{equation}
To ensure that $\hat{\pi}_1\hat{y}_2 = \hat{y}_2\hat{\pi}_1$ and $\hat{\pi}_2\hat{y}_1 = \hat{y}_1\hat{\pi}_2$, we must have $m_1 = q_2 m_1'$ and $m_2 = q_1 m_2'$ for $m_1', m_2'\in \mathbb{Z}$, so that
\begin{equation}
\hat{\pi}_1\widehat{Z}^\Phi(\nu) = (1 - \hat{y}_1)(1 - \hat{y}_2)^{m_1'}\widehat{Z}^\Phi(\nu), \quad \hat{\pi}_2\widehat{Z}^\Phi(\nu) = (1 - \hat{y}_2)(1 - \hat{y}_1)^{m_2'}\widehat{Z}^\Phi(\nu).
\end{equation}
We may set $m_1' = m_2' = 0$ by acting with shifts of $\nu$ by $q_1 q_2$, which commute with both $\hat{\pi}_1$ and $\hat{\pi}_2$.  Hence our original prescription was the correct one.

\paragraph{Example.}

On the PHS, we have
\begin{equation}
\widehat{Z}^\Phi(\nu) = \mathcal{G}_{2, -1}^\Phi(\nu)\mathcal{G}_{3, 1}^\Phi(\nu)\mathcal{G}_{5, 1}^\Phi(\nu)
\label{znofluxphs}
\end{equation}
where for $(q, p) = (2, -1), (3, 1), (5, 1)$, we have
\begin{equation}
(t, s)\in (1, 1) + (2, 1)\mathbb{Z}, \mbox{ } (1, 0) + (3, -1)\mathbb{Z}, \mbox{ } (1, 0) + (5, -1)\mathbb{Z},
\end{equation}
respectively.  Set $\hat{y}_q = e^{2\pi i\nu/q}$ for $q = 2, 3, 5$.  Letting $[n]_q\equiv n\text{ (mod $q$)}\in \{0, \ldots, q - 1\}$, we have from \eqref{completeresidueproperty} and \eqref{Gqpshiftmt} (with $t = 1$) that for $n\in \mathbb{Z}$,
\begin{equation}
\widehat{Z}^\Phi(\nu + n) = (1 - e^{2\pi i\nu})^{n/30 - [n]_2/2 - [n]_3/3 - [n]_5/5}\prod_{q = 2, 3, 5}\prod_{\ell_q = 0}^{[n]_q - 1} (1 - e^{2\pi i(\nu + \ell_q)/q})\widehat{Z}^\Phi(\nu).
\label{PHSshift}
\end{equation}
Clearly, shifting $\nu$ by 30 shifts the ordinary flux by $-1$: $Z^\Phi(\nu + 30) = (1 - e^{2\pi i\nu})Z^\Phi(\nu)$.  We expect one difference equation for each special fiber, since each $\hat{\pi}$ should act nontrivially on only one $\hat{y}$.  Hence we choose
\begin{align}
(q, p) = (2, -1) &\implies (t, s) = (1, 1) + 7(2, 1) = (15, 8), \nonumber \\
(q, p) = (3, 1) &\implies (t, s) = (1, 0) + 3(3, -1) = (10, -3), \\
(q, p) = (5, 1) &\implies (t, s) = (1, 0) + (5, -1) = (6, -1), \nonumber
\end{align}
and accounting for the ambiguity (mod $30$), we define
\begin{equation}
\hat{\pi}_2 : \nu\mapsto \nu + 15 + 30m_2, \quad \hat{\pi}_3 : \nu\mapsto \nu + 10 + 30m_3, \quad \hat{\pi}_5 : \nu\mapsto \nu + 6 + 30m_5
\end{equation}
with $m_q\in \mathbb{Z}$.  We find that \eqref{commutationrelations} is satisfied, with $t = 1$ in all cases.  Using \eqref{PHSshift}, we have
\begin{equation}
\hat{\pi}_q\widehat{Z}^\Phi(\nu) = (1 - e^{2\pi i\nu})^{m_q}(1 - \hat{y}_q)\widehat{Z}^\Phi(\nu)
\end{equation}
for $q = 2, 3, 5$.  Shifts by $\operatorname{lcm}(2, 3, 5) = 30$ are redundant and commute with $\hat{\pi}_{2, 3, 5}$, so by making such shifts, we may set $m_2 = m_3 = m_5 = 0$ to obtain the ``elementary'' operations $\hat{\pi}_{2, 3, 5}$.

\paragraph{Comment.}

Let us try to extend these considerations to a rationally squashed lens space $L(p, q)_b$, with
\begin{equation}
\widehat{Z}^\Phi(\nu) = \mathcal{G}_{q_1, p_1}^\Phi(\nu)\mathcal{G}_{q_2, p_2}^\Phi(\nu), \quad b^2 = q_1/q_2, \quad p = p_1 q_2 + p_2 q_1, \quad q = q_1 s_2 - p_1 t_2
\end{equation}
and, as usual, $p_1 t_1 + q_1 s_1 = 1$ and $p_2 t_2 + q_2 s_2 = 1$ \cite{Closset:2018ghr}.  Define
\begin{equation}
\hat{y}_1 = e^{2\pi i\nu/q_1}, \quad \hat{y}_2 = e^{2\pi i\nu/q_2}, \quad \hat{\pi}_1 : \nu\mapsto \nu + t_1 + m_1 q_1, \quad \hat{\pi}_2 : \nu\mapsto \nu + t_2 + m_2 q_2.
\end{equation}
Then we have
\begin{equation}
\hat{\pi}_1\hat{y}_1 = e^{2\pi it_1/q_1}\hat{y}_1\hat{\pi}_1, \quad \hat{\pi}_2\hat{y}_2 = e^{2\pi it_2/q_2}\hat{y}_2\hat{\pi}_2.
\label{commutationlpq}
\end{equation}
We require that
\begin{equation}
\hat{\pi}_i\hat{y}_j = \hat{y}_j\hat{\pi}_i \quad (i\neq j),
\end{equation}
which means that $m_1, m_2$ must be chosen such that $q_2 | (t_1 + m_1 q_1)$ and $q_1 | (t_2 + m_2 q_2)$:
\begin{equation}
t_1 + m_1 q_1 = r_1 q_2, \quad t_2 + m_2 q_2 = r_2 q_1.
\label{mrequations1}
\end{equation}
Using \eqref{Gqpshiftn}, we then have
\begin{align}
\widehat{Z}^\Phi(\nu + t_1 + m_1 q_1) &= (1 - e^{2\pi i\nu})^{m_1 p_1 - s_1 + r_1 p_2}(1 - e^{2\pi i\nu/q_1})\widehat{Z}^\Phi(\nu), \\
\widehat{Z}^\Phi(\nu + t_2 + m_2 q_2) &= (1 - e^{2\pi i\nu})^{m_2 p_2 - s_2 + r_2 p_1}(1 - e^{2\pi i\nu/q_2})\widehat{Z}^\Phi(\nu).
\end{align}
So we want
\begin{equation}
m_1 p_1 - s_1 + r_1 p_2 = 0, \quad m_2 p_2 - s_2 + r_2 p_1 = 0.
\label{mrequations2}
\end{equation}
Solving the equations \eqref{mrequations1} and \eqref{mrequations2} for $m_i, r_i$ gives
\begin{equation}
m_1 = \frac{q_2 s_1 - p_2 t_1}{p}, \quad m_2 = \frac{q}{p}.
\end{equation}
So this procedure works only for $p = 1$, in which case $L(1, q)_b\cong S_b^3$.  In this case, the commutation relations \eqref{commutationlpq} coincide with the known relations for lens spaces:
\begin{equation}
\hat{\pi}_1\hat{y}_1 = e^{\frac{2\pi i}{p}b^{-2}}\hat{y}_1\hat{\pi}_1 = e^{\frac{2\pi iq_2}{q_1}}\hat{y}_1\hat{\pi}_1, \quad \hat{\pi}_2\hat{y}_2 = e^{\frac{2\pi i}{p}b^2}\hat{y}_2\hat{\pi}_2 = e^{\frac{2\pi iq_1}{q_2}}\hat{y}_2\hat{\pi}_2,
\end{equation}
by virtue of $p_1 q_2\equiv 1 \text{ (mod $q_1$)}$, $p_2 q_1\equiv 1 \text{ (mod $q_2$)}$, and $p_i t_i\equiv 1 \text{ (mod $q_i$)}$.

\subsection{Including Holonomies} \label{holonomies}

We now describe how to incorporate holonomies into the difference equations arising from the Seifert formalism, and how they depend only on the ``global'' flux rather than the individual fractional fluxes.  Including fluxes, the $\mathbb{Q}$HS partition function of a chiral multiplet is
\begin{equation}
Z^\Phi(\nu)_{\mathfrak{m}} = \Pi^\Phi(\nu)\widehat{Z}^\Phi(\nu)_{\mathfrak{m}}, \quad \widehat{Z}^\Phi(\nu)_{\mathfrak{m}}\equiv \prod_{i=0}^n \mathcal{G}_{q_i, p_i}^\Phi(\nu)_{\mathfrak{m}_i} = \left[\Pi^\Phi(\nu)^{\mathfrak{m}_0}\prod_{i=1}^n \Pi_{q_i, p_i}^\Phi(\nu)_{\mathfrak{m}_i}\right]\widehat{Z}^\Phi(\nu),
\label{zchiralwithfluxes}
\end{equation}
with $\widehat{Z}^\Phi(\nu)$ as in \eqref{zchiralnofluxes}.

Regardless of geometry, there is locally a notion of fractional flux, which can be globally redefined away for homology spheres.  For lens spaces, our difference equations (which depend on $\nu, \mathfrak{m}_0, \mathfrak{m}_1, \mathfrak{m}_2$) admit many redundancies in description relative to those of \cite{Dimofte:2014zga} (which depend only on $\nu, \mathfrak{m}$).  Indeed, the difference equations of \cite{Dimofte:2014zga} have many different representations in terms of $\mathfrak{m}_1, \mathfrak{m}_2$.  These redundancies are encoded in the fact that the large gauge transformation
\begin{equation}
\nu\to \nu + 1, \quad \mathfrak{m}_0\to \mathfrak{m}_0 + \mathbf{d}, \quad \mathfrak{m}_i\to \mathfrak{m}_i + p_i
\label{largegauge}
\end{equation}
is a trivial operation in the 3D Picard group \cite{Closset:2018ghr}.  Likewise, the equivalence relations
\begin{equation}
[\mathbf{d}; g; (q_i, b_i)]\cong \left[\mathbf{d} - \sum_i \mathfrak{m}_i; g; (q_i, b_i + \mathfrak{m}_i q_i)\right]
\end{equation}
are trivial operations in the 2D Picard group \cite{Closset:2018ghr}.  The partition function should be invariant under the former for any choice of Seifert invariants in the defining line bundle related by the latter.

As examples, we derive the difference equations obeyed by the tetrahedron theory on all of the spherical manifolds $S^3/\Gamma_{ADE}$.  The general picture that emerges is that the definitions of the variables differ from those for lens spaces with continuous squashing, but the difference equations take exactly the same form (one for each exceptional fiber).  For example, for spaces with $H_1 = 0$, we find for each constituent fibering operator $\mathcal{G}_{q, p}^\Phi(\nu)_{\mathfrak{m}}$ that
\begin{equation}
\hat{y} = e^{2\pi i(\nu - t\mathfrak{m})/q}, \quad \hat{\pi} : (\nu, \mathfrak{m})\mapsto (\nu + t, \mathfrak{m}),
\end{equation}
where $\hat{\pi}$ can be defined \emph{not to act} on any of the fluxes (due to the equivalence relations in the 2D and 3D Picard groups).  However, when $H_1\neq 0$, at least one of the $\hat{\pi}$'s \emph{must} act on the holonomies.  One consistency check that these equations are correct is that by demanding that these equations hold while completely accounting for large gauge transformations and other equivalences between fractional fluxes, one can reproduce the known homology groups for these manifolds.  In Section \ref{lineops}, we offer an alternative derivation that removes all doubts as to the correctness of these equations.

As a warmup, we begin by corroborating and generalizing our analysis for spaces with $H_1 = 0$ by turning on fractional fluxes.  Below, we often use $\hat{\pi}_\ast : \bullet$ as shorthand for $\hat{\pi}_\ast : (\nu, \mathfrak{m}_0, \mathfrak{m}_1, \ldots, \mathfrak{m}_n)\mapsto \bullet$.

On $S^3$, we have $\mathcal{G}_{1, 1}^\Phi(\nu)_{\mathfrak{m}} = \Pi^\Phi(\nu)^{\mathfrak{m}}\mathcal{G}_{1, 1}^\Phi(\nu)$ where
\begin{equation}
\mathcal{G}_{1, 1}^\Phi(\nu + n)_{\mathfrak{m} + n'} = \Pi^\Phi(\nu)^{\mathfrak{m} + n'}\mathcal{G}_{1, 1}^\Phi(\nu + n) = (1 - e^{2\pi i\nu})^{n - n'}\mathcal{G}_{1, 1}^\Phi(\nu)_{\mathfrak{m}}
\label{s3operation}
\end{equation}
for $n, n'\in \mathbb{Z}$.  The partition function is invariant under large gauge transformations:
\begin{equation}
\widehat{Z}^\Phi(\nu)_{\mathfrak{m}_0, \mathfrak{m}_1} = \Pi^\Phi(\nu)^{\mathfrak{m}_0}\mathcal{G}_{1, 1}^\Phi(\nu)_{\mathfrak{m}_1} = \Pi^\Phi(\nu + 1)^{\mathfrak{m}_0 + 0}\mathcal{G}_{1, 1}^\Phi(\nu + 1)_{\mathfrak{m}_1 + 1}.
\end{equation}
We see that for any operation of the form \eqref{s3operation}, one can use a large gauge transformation to set the shift in the flux $n'$ to zero; then we are left with $n = t$ and we can set $n = 1$ by redefining $(t, s)\to (t + m, s - m)$, giving the desired difference equation.

On $S_b^3$, we have
\begin{equation}
\widehat{Z}^\Phi(\nu)_{\mathfrak{m}_0, \mathfrak{m}_1, \mathfrak{m}_2} = \Pi^\Phi(\nu)^{\mathfrak{m}_0}\Pi_{q_1, p_1}^\Phi(\nu)_{\mathfrak{m}_1}\Pi_{q_2, p_2}^\Phi(\nu)_{\mathfrak{m}_2}\widehat{Z}^\Phi(\nu)
\end{equation}
with $\widehat{Z}^\Phi(\nu)$ as in \eqref{znofluxsb3}.  We have using \eqref{Piqpshift} that
\begin{align}
\widehat{Z}^\Phi(\nu + q_2)_{\mathfrak{m}_0, \mathfrak{m}_1, \mathfrak{m}_2} &= \frac{\Pi_{q_1, p_1}^\Phi(\nu + q_2)_{\mathfrak{m}_1}}{\Pi_{q_1, p_1}^\Phi(\nu)_{\mathfrak{m}_1}}(1 - e^{2\pi i\nu/q_1})\widehat{Z}^\Phi(\nu)_{\mathfrak{m}_0, \mathfrak{m}_1, \mathfrak{m}_2} \nonumber \\
&= (1 - e^{2\pi i(\nu - q_2\mathfrak{m}_1)/q_1})\widehat{Z}^\Phi(\nu)_{\mathfrak{m}_0, \mathfrak{m}_1, \mathfrak{m}_2}, \\
\widehat{Z}^\Phi(\nu + q_1)_{\mathfrak{m}_0, \mathfrak{m}_1, \mathfrak{m}_2} &= \frac{\Pi_{q_2, p_2}^\Phi(\nu + q_1)_{\mathfrak{m}_2}}{\Pi_{q_2, p_2}^\Phi(\nu)_{\mathfrak{m}_2}}(1 - e^{2\pi i\nu/q_2})\widehat{Z}^\Phi(\nu)_{\mathfrak{m}_0, \mathfrak{m}_1, \mathfrak{m}_2} \nonumber \\
&= (1 - e^{2\pi i(\nu - q_1\mathfrak{m}_2)/q_2})\widehat{Z}^\Phi(\nu)_{\mathfrak{m}_0, \mathfrak{m}_1, \mathfrak{m}_2},
\end{align}
which shows that the difference equations can be defined without acting on the ho\-lo\-no\-mies, as consistent with $H_1 = 0$ in this case.  To show that the action on the holonomies can \emph{always} be gauged away, consider the most general parametrization
\begin{align}
\hat{y}_1 &= e^{2\pi i(\nu - q_2\mathfrak{m}_1)/q_1}, & \hat{\pi}_1 &: (\nu + m_1, \mathfrak{m}_0 + m_{10}, \mathfrak{m}_1 + m_{11}, \mathfrak{m}_2 + m_{12}), \\
\hat{y}_2 &= e^{2\pi i(\nu - q_1\mathfrak{m}_2)/q_2}, & \hat{\pi}_2 &: (\nu + m_2, \mathfrak{m}_0 + m_{20}, \mathfrak{m}_1 + m_{21}, \mathfrak{m}_2 + m_{22}),
\end{align}
with arbitrary integer shifts $m_q$ and $m_{qq'}$.  We must have
\begin{align}
m_1 - m_{11}q_2 &\equiv q_2\text{ (mod $q_1$)}, & m_1 - m_{12}q_1 &\equiv 0\text{ (mod $q_2$)}, \\
m_2 - m_{22}q_1 &\equiv q_1\text{ (mod $q_2$)}, & m_2 - m_{21}q_2 &\equiv 0\text{ (mod $q_1$)}
\end{align}
for the actions of $\hat{\pi}_1, \hat{\pi}_2$ on $\hat{y}_1, \hat{y}_2$ to be correct.  Using the coprimality of $q_1$ and $q_2$, we can therefore write
\begin{align}
\hat{\pi}_1 &: (\nu + q_2 + m_{11}q_2 + m_{12}q_1 + k_1 q_1 q_2, \mathfrak{m}_0 + m_{10}, \mathfrak{m}_1 + m_{11}, \mathfrak{m}_2 + m_{12}), \\
\hat{\pi}_2 &: (\nu + q_1 + m_{22}q_1 + m_{21}q_2 + k_2 q_1 q_2, \mathfrak{m}_0 + m_{20}, \mathfrak{m}_1 + m_{21}, \mathfrak{m}_2 + m_{22}).
\end{align}
Using \eqref{completeresidueproperty} for simplification, we compute that
\begin{equation}
\hat{\pi}_q\widehat{Z}^\Phi(\nu)_{\mathfrak{m}_0, \mathfrak{m}_1, \mathfrak{m}_2} = (1 - e^{2\pi i\nu})^{k_q - m_{q0}}(1 - \hat{y}_q)\widehat{Z}^\Phi(\nu)_{\mathfrak{m}_0, \mathfrak{m}_1, \mathfrak{m}_2},
\end{equation}
which requires $k_q = m_{q0}$ to obtain the desired difference equations ($q = 1, 2$).  Under these conditions, we readily see from the fact that shifts
\begin{equation}
\nu + 1, \mathfrak{m}_1 + p_1, \mathfrak{m}_2 + p_2
\end{equation}
act trivially on the partition function (since they comprise a large gauge transformation), the flux equivalences
\begin{equation}
(\nu, \mathfrak{m}_0, \mathfrak{m}_1, \mathfrak{m}_2)\sim (\nu, \mathfrak{m}_0 - N_1 - N_2, \mathfrak{m}_1 + N_1 q_1, \mathfrak{m}_2 + N_2 q_2),
\end{equation}
and $q_1 p_2 + q_2 p_1 = 1$ that these operators are equivalent to
\begin{equation}
\hat{\pi}_1 : \nu\mapsto \nu + q_2, \quad \hat{\pi}_2 : \nu\mapsto \nu + q_1.
\label{sb3naive}
\end{equation}
So our na\"ive definitions of $\hat{\pi}_1, \hat{\pi}_2$ were correct.  Note in particular that
\begin{align}
(\nu, \mathfrak{m}_0, \mathfrak{m}_1, \mathfrak{m}_2) &\sim (\nu + q_1 q_2, \mathfrak{m}_0 + 1, \mathfrak{m}_1, \mathfrak{m}_2) \nonumber \\
&\sim (\nu + q_2, \mathfrak{m}_0, \mathfrak{m}_1 + 1, \mathfrak{m}_2) \nonumber \\
&\sim (\nu + q_1, \mathfrak{m}_0, \mathfrak{m}_1, \mathfrak{m}_2 + 1), \label{sb3naivebasis}
\end{align}
meaning that any shifts of the fluxes can be absorbed into shifts of $\nu$.

On the PHS with $(q, p) = (2, -1), (3, 1), (5, 1)$, we have
\begin{align}
\widehat{Z}^\Phi(\nu + t)_{\mathfrak{m}_0, \mathfrak{m}_2, \mathfrak{m}_3, \mathfrak{m}_5} &= \frac{\Pi_{q, p}^\Phi(\nu + 1)_{\mathfrak{m}_q}}{\Pi_{q, p}^\Phi(\nu)_{\mathfrak{m}_q}}(1 - e^{2\pi i\nu/q})\widehat{Z}^\Phi(\nu)_{\mathfrak{m}_0, \mathfrak{m}_2, \mathfrak{m}_3, \mathfrak{m}_5} \nonumber \\
&= (1 - e^{2\pi i(\nu - \mathfrak{m}_q)/q})\widehat{Z}^\Phi(\nu)_{\mathfrak{m}_0, \mathfrak{m}_2, \mathfrak{m}_3, \mathfrak{m}_5}
\end{align}
for $t = 15, 10, 6$.  More generally, consider
\begin{equation}
\hat{\pi}_q : (\nu + t(q) + 15m_{q2} + 10m_{q3} + 6m_{q5} + 30k_q, \{\mathfrak{m}_{q'} + m_{qq'} \,|\, q' = 0, 2, 3, 5\}) \label{PHSshifts}
\end{equation}
where $t(2) = 15$, $t(3) = 10$, $t(5) = 6$.  This is the most general parametrization such that $\hat{\pi}_q$ and $\hat{y}_q = e^{2\pi i(\nu - \mathfrak{m}_q)/q}$ obey the correct commutation relations for $q = 2, 3, 5$ (as follows from solving the required system of congruences using the Chinese remainder theorem).  To compute the action of the $\hat{\pi}_q$ on
\begin{equation}
\widehat{Z}^\Phi(\nu)_{\mathfrak{m}_0, \mathfrak{m}_2, \mathfrak{m}_3, \mathfrak{m}_5} = \Pi^\Phi(\nu)^{\mathfrak{m}_0}\Pi_{2, -1}^\Phi(\nu)_{\mathfrak{m}_2}\Pi_{3, 1}^\Phi(\nu)_{\mathfrak{m}_3}\Pi_{5, 1}^\Phi(\nu)_{\mathfrak{m}_5}\widehat{Z}^\Phi(\nu)
\end{equation}
with $\widehat{Z}^\Phi(\nu)$ as in \eqref{znofluxphs}, it suffices to use \eqref{Piqpshift} and \eqref{Gqpshiftmt} for arbitrary shifts of the flux and fibering operators because we can choose $t = 1$ for all exceptional fibers, and all shifts are multiples of 1.  We compute that
\begin{equation}
\hat{\pi}_q\widehat{Z}^\Phi(\nu)_{\mathfrak{m}_0, \mathfrak{m}_2, \mathfrak{m}_3, \mathfrak{m}_5} = (1 - e^{2\pi i\nu})^{k_q - m_{q0}}(1 - \hat{y}_q)\widehat{Z}^\Phi(\nu)_{\mathfrak{m}_0, \mathfrak{m}_2, \mathfrak{m}_3, \mathfrak{m}_5}
\label{PHSalmostdifference}
\end{equation}
for $q = 2, 3, 5$.  Hence the conditions that we need to impose are again
\begin{equation}
k_q = m_{q0} \quad (q = 2, 3, 5).
\label{PHSconditions}
\end{equation}
Using the conditions \eqref{PHSconditions}, combined with the fact that large gauge transformations
\begin{equation}
(\nu + 1, \mathfrak{m}_0, \mathfrak{m}_2 - 1, \mathfrak{m}_3 + 1, \mathfrak{m}_5 + 1)
\end{equation}
and shifts of the form
\begin{equation}
(\nu, \mathfrak{m}_0 - N_2 - N_3 - N_5, \mathfrak{m}_2 + 2N_2, \mathfrak{m}_3 + 3N_3, \mathfrak{m}_5 + 5N_5)
\end{equation}
leave $\widehat{Z}^\Phi(\nu)_{\mathfrak{m}_0, \mathfrak{m}_2, \mathfrak{m}_3, \mathfrak{m}_5}$ invariant, shows that the shifts \eqref{PHSshifts} are equivalent to\footnote{Indeed, applying the equivalence relations directly to \eqref{PHSshifts} reduces them to $\hat{\pi}_q : (\nu + t(q), \mathfrak{m}_0 + m_{q0} - k_q, \mathfrak{m}_2, \mathfrak{m}_3, \mathfrak{m}_5)$, which facilitates the computation of \eqref{PHSalmostdifference}.}
\begin{equation}
\hat{\pi}_2 : \nu\mapsto \nu + 15, \quad \hat{\pi}_3 : \nu\mapsto \nu + 10, \quad \hat{\pi}_5 : \nu\mapsto \nu + 6.
\label{PHSnaive}
\end{equation}
Similarly, we find that
\begin{align}
(\nu, \mathfrak{m}_0, \mathfrak{m}_2, \mathfrak{m}_3, \mathfrak{m}_5) &\sim (\nu + 30, \mathfrak{m}_0 + 1, \mathfrak{m}_2, \mathfrak{m}_3, \mathfrak{m}_5) \nonumber \\
&\sim (\nu + 15, \mathfrak{m}_0, \mathfrak{m}_2 + 1, \mathfrak{m}_3, \mathfrak{m}_5) \nonumber \\
&\sim (\nu + 10, \mathfrak{m}_0, \mathfrak{m}_2, \mathfrak{m}_3 + 1, \mathfrak{m}_5) \nonumber \\
&\sim (\nu + 6, \mathfrak{m}_0, \mathfrak{m}_2, \mathfrak{m}_3, \mathfrak{m}_5 + 1). \label{PHSnaivebasis}
\end{align}
We again see that any shifts of the fluxes can be absorbed into shifts of $\nu$.  Hence we can define the $\hat{\pi}_q$ so that all fluxes remain inert, meaning there is no global flux.

We have seen that we can gauge away holonomies on geometries with $H_1 = 0$.  Let us take stock of the examples considered so far.  On $S^3$, we have
\begin{equation}
Z^\Phi(\nu)_{\mathfrak{m}_0, \mathfrak{m}_1} = \Pi^\Phi(\nu)^{1 + \mathfrak{m}_0}\mathcal{G}_{1, 1}^\Phi(\nu)_{\mathfrak{m}_1}
\end{equation}
where
\begin{align}
\hat{y} &= e^{2\pi i\nu}, & \hat{\pi} : \nu &\mapsto \nu + 1, & (\hat{\pi} + \hat{y} - 1)Z^\Phi(\nu)_{\mathfrak{m}_0, \mathfrak{m}_1} &= 0.
\end{align}
Shifting $\nu\to \nu + 1$ is equivalent to shifting the ordinary flux by $-1$, and we also have that $(\nu, \mathfrak{m}_1)\sim (\nu + 1, \mathfrak{m}_1 + 1)$.  On $S_b^3$ with $b^2 = q_1/q_2$ and $q_1 p_2 + q_2 p_1 = 1$, we have
\begin{equation}
Z^\Phi(\nu)_{\mathfrak{m}_0, \mathfrak{m}_1, \mathfrak{m}_2} = \Pi^\Phi(\nu)^{1 + \mathfrak{m}_0}\mathcal{G}_{q_1, p_1}^\Phi(\nu)_{\mathfrak{m}_1}\mathcal{G}_{q_2, p_2}^\Phi(\nu)_{\mathfrak{m}_2}
\end{equation}
where
\begin{align}
\hat{y}_1 &= e^{2\pi i(\nu - q_2\mathfrak{m}_1)/q_1}, & \hat{\pi}_1 : \nu &\mapsto \nu + q_2, & (\hat{\pi}_1 + \hat{y}_1 - 1)Z^\Phi(\nu)_{\mathfrak{m}_0, \mathfrak{m}_1, \mathfrak{m}_2} &= 0, \\
\hat{y}_2 &= e^{2\pi i(\nu - q_1\mathfrak{m}_2)/q_2}, & \hat{\pi}_2 : \nu &\mapsto \nu + q_1, & (\hat{\pi}_2 + \hat{y}_2 - 1)Z^\Phi(\nu)_{\mathfrak{m}_0, \mathfrak{m}_1, \mathfrak{m}_2} &= 0.
\end{align}
Shifting $\nu$ by $\operatorname{lcm}(q_1, q_2) = q_1 q_2$ is equivalent to shifting the ordinary flux by $-1$.  On the PHS, we have
\begin{equation}
Z^\Phi(\nu)_{\mathfrak{m}_0, \mathfrak{m}_2, \mathfrak{m}_3, \mathfrak{m}_5} = \Pi^\Phi(\nu)^{1 + \mathfrak{m}_0}\mathcal{G}_{2, -1}^\Phi(\nu)_{\mathfrak{m}_2}\mathcal{G}_{3, 1}^\Phi(\nu)_{\mathfrak{m}_3}\mathcal{G}_{5, 1}^\Phi(\nu)_{\mathfrak{m}_5}
\end{equation}
where
\begin{align}
\hat{y}_2 &= e^{2\pi i(\nu - \mathfrak{m}_2)/2}, & \hat{\pi}_2 : \nu &\mapsto \nu + 15, & (\hat{\pi}_2 + \hat{y}_2 - 1)Z^\Phi(\nu)_{\mathfrak{m}_0, \mathfrak{m}_2, \mathfrak{m}_3, \mathfrak{m}_5} &= 0, \\
\hat{y}_3 &= e^{2\pi i(\nu - \mathfrak{m}_3)/3}, & \hat{\pi}_3 : \nu &\mapsto \nu + 10, & (\hat{\pi}_3 + \hat{y}_3 - 1)Z^\Phi(\nu)_{\mathfrak{m}_0, \mathfrak{m}_2, \mathfrak{m}_3, \mathfrak{m}_5} &= 0, \\
\hat{y}_5 &= e^{2\pi i(\nu - \mathfrak{m}_5)/5}, & \hat{\pi}_5 : \nu &\mapsto \nu + 6, & (\hat{\pi}_5 + \hat{y}_5 - 1)Z^\Phi(\nu)_{\mathfrak{m}_0, \mathfrak{m}_2, \mathfrak{m}_3, \mathfrak{m}_5} &= 0.
\end{align}
Shifting $\nu$ by $\operatorname{lcm}(2, 3, 5) = 30$ is equivalent to shifting the ordinary flux by $-1$.

We now move on to spaces \emph{with} holonomies, for which the action of the $\hat{\pi}$'s on the fractional fluxes can no longer be completely gauged away.  In Section \ref{H1is0}, we argued that as far as lens spaces are concerned, we can only handle cases with no holonomies.  Apart from lens spaces, some good examples with holonomies are the spherical three-manifolds considered in \cite{Closset:2018ghr}:
\begin{align}
\mathcal{S}^3[A_{p-1}] &\cong [0; 0; (1, 0), (q_1, 1), (q_2, 1)], & p &= q_1 + q_2\geq 2, \nonumber \\
\mathcal{S}^3[D_{n+2}] &\cong [0; 0; (1, 0), (2, -1), (2, 1), (n, 1)], & n &\geq 1, \\
\mathcal{S}^3[E_{m+3}] &\cong [0; 0; (1, 0), (2, -1), (3, 1), (m, 1)], & m &= 3, 4, 5. \nonumber
\end{align}
All of these examples have $\kappa < 0$; the only one without holonomies in this class is the PHS, since the compact group $E_8$ is unique among simple, compact Lie groups in being simply connected and having trivial center (extending the range of $m$, the $E_{10}$ case has no fractional flux, but $\kappa > 0$).  Since the $A$-series is
\begin{equation}
L(p, p - 1), \quad p = q_1 + q_2\geq 2, \quad b^2 = q_1/q_2,
\end{equation}
we restrict our attention to the $D$- and $E$-series, for which
\begin{equation}
\kappa_D = -\frac{1}{2n} < 0, \quad \kappa_E = \frac{1}{2}\left(\frac{1}{6} - \frac{1}{m}\right) < 0
\end{equation}
for $n\geq 1$ and $m = 3, 4, 5$.  We consider $n\geq 2$, since $D_3 = A_3$.  By requiring that the expected difference equations hold for these geometries, we verify that $\widetilde{\operatorname{Pic}}(\mathcal{S}^3[D_{n+2}])\cong \mathbb{Z}_2\times \mathbb{Z}_2$ for $n$ even and $\mathbb{Z}_4$ for $n$ odd, as well as $\widetilde{\operatorname{Pic}}(\mathcal{S}^3[E_{m+3}])\cong \mathbb{Z}_{6-m}$.

On $\mathcal{S}^3[D_{n+2}]$ with $n\geq 2$, we have $(q, p) = (2, -1), (2, 1), (n, 1)$, so
\begin{equation}
(t, s)\in (1, 1) + (2, 1)\mathbb{Z}, \mbox{ } (1, 0) + (2, -1)\mathbb{Z}, \mbox{ } (1, 0) + (n, -1)\mathbb{Z}
\end{equation}
and
\begin{equation}
\widehat{Z}^\Phi(\nu)_{\mathfrak{m}_0, \mathfrak{m}_-, \mathfrak{m}_+, \mathfrak{m}_n} = \Pi^\Phi(\nu)^{\mathfrak{m}_0}\Pi_{2, -1}^\Phi(\nu)_{\mathfrak{m}_-}\Pi_{2, 1}^\Phi(\nu)_{\mathfrak{m}_+}\Pi_{n, 1}^\Phi(\nu)_{\mathfrak{m}_n}\widehat{Z}^\Phi(\nu)
\end{equation}
where
\begin{equation}
\widehat{Z}^\Phi(\nu)\equiv \mathcal{G}_{2, -1}^\Phi(\nu)\mathcal{G}_{2, 1}^\Phi(\nu)\mathcal{G}_{n, 1}^\Phi(\nu).
\end{equation}
At the level of the partition function, we have the equivalences
\begin{align}
(\nu, \mathfrak{m}_0, \mathfrak{m}_-, \mathfrak{m}_+, \mathfrak{m}_n) &\sim (\nu + N, \mathfrak{m}_0, \mathfrak{m}_- - N, \mathfrak{m}_+ + N, \mathfrak{m}_n + N) \label{dseriesequivalences} \\
&\sim (\nu, \mathfrak{m}_0 - N_- - N_+ - N_n, \mathfrak{m}_- + 2N_-, \mathfrak{m}_+ + 2N_+, \mathfrak{m}_n + nN_n). \nonumber
\end{align}
For $n$ even, these equivalence relations imply that
\begin{align}
(\nu, \mathfrak{m}_0, \mathfrak{m}_-, \mathfrak{m}_+, \mathfrak{m}_n) &\sim (\nu + n, \mathfrak{m}_0 + 1, \mathfrak{m}_-, \mathfrak{m}_+, \mathfrak{m}_n) \nonumber \\
&\sim (\nu + n + 1, \mathfrak{m}_0, \mathfrak{m}_- + 1, \mathfrak{m}_+ + 1, \mathfrak{m}_n + 1) \nonumber \\
&\sim (\nu + n, \mathfrak{m}_0, \mathfrak{m}_-, \mathfrak{m}_+ + 2, \mathfrak{m}_n) \nonumber \\
&\sim (\nu + 2, \mathfrak{m}_0, \mathfrak{m}_-, \mathfrak{m}_+, \mathfrak{m}_n + 2). \label{devennaivebasis}
\end{align}
From \eqref{devennaivebasis}, we infer $(\nu, \mathfrak{m}_-)\sim (\nu + n, \mathfrak{m}_- + 2)$.  So in addition to the shift $\nu\to \nu + n$ (where $n$ is the LCM of the $q$'s) being equivalent to a shift of the ordinary flux by $-1$, all three fractional fluxes are effectively valued in $\mathbb{Z}_2$ modulo shifts of $\nu$.  For $n$ odd, we derive instead that
\begin{align}
(\nu, \mathfrak{m}_0, \mathfrak{m}_-, \mathfrak{m}_+, \mathfrak{m}_n) &\sim (\nu + n + 1, \mathfrak{m}_0 + 1, \mathfrak{m}_-, \mathfrak{m}_+, \mathfrak{m}_n + 1) \nonumber \\
&\sim (\nu + n, \mathfrak{m}_0, \mathfrak{m}_- + 1, \mathfrak{m}_+ + 1, \mathfrak{m}_n) \nonumber \\
&\sim (\nu + 2n, \mathfrak{m}_0, \mathfrak{m}_-, \mathfrak{m}_+ + 4, \mathfrak{m}_n) \nonumber \\
&\sim (\nu + n + 1, \mathfrak{m}_0, \mathfrak{m}_-, \mathfrak{m}_+ + 2, \mathfrak{m}_n + 1). \label{doddnaivebasis}
\end{align}
The equivalences \eqref{doddnaivebasis} imply that shifts of the form $(\nu + 2n, \mathfrak{m}_0 + 2)$, $(\nu + 2n, \mathfrak{m}_- + 4)$, $(\nu + 2, \mathfrak{m}_n + 2)$ are trivial.  Therefore, a shift of $\nu$ by the LCM of the $q$'s ($2n$) is equivalent to a shift of the ordinary flux by $-2$, and the fractional fluxes $\mathfrak{m}_-, \mathfrak{m}_+, \mathfrak{m}_n$ are effectively valued in $\mathbb{Z}_4, \mathbb{Z}_4, \mathbb{Z}_2$ modulo shifts of $\nu$ (this example is the only one considered in this paper for which arbitrary shifts of the ordinary flux $\mathfrak{m}_0$ cannot be absorbed into shifts of $\nu$).  Now suppose that we have three mutually commuting pairs $(\hat{y}, \hat{\pi})$.  We find that we must define the $\hat{\pi}$ to shift fractional fluxes nontrivially to achieve this, and that we may need to shift more fluxes than \emph{a priori} necessary to get the expected difference equations.  The result of our analysis is that for $\mathcal{S}^3[D_{n+2}]$, (the torsion part of) $H_1$ is $\mathbb{Z}_2\times \mathbb{Z}_2$ for $n$ even and $\mathbb{Z}_4$ for $n$ odd, as expected.  We first write down the most general shift operators that satisfy the correct commutation relations with
\begin{equation}
\hat{y}_- = e^{2\pi i(\nu - \mathfrak{m}_-)/2}, \quad \hat{y}_+ = e^{2\pi i(\nu - \mathfrak{m}_+)/2}, \quad \hat{y}_n = e^{2\pi i(\nu - \mathfrak{m}_n)/n}.
\end{equation}
Using, e.g., the Chinese remainder theorem for non-coprime moduli and the equivalences \eqref{dseriesequivalences} for simplification, we find that these operators can be parametrized as
\begin{align}
\hat{\pi}_- &: (\nu, \mathfrak{m}_0 + m_{-0}, \mathfrak{m}_- + 1, \mathfrak{m}_+, \mathfrak{m}_n), \nonumber \\
\hat{\pi}_+ &: (\nu, \mathfrak{m}_0 + m_{+0}, \mathfrak{m}_-, \mathfrak{m}_+ + 1, \mathfrak{m}_n), \\
\hat{\pi}_n &: (\nu + 1, \mathfrak{m}_0 + m_{n0}, \mathfrak{m}_- + 1, \mathfrak{m}_+ + 1, \mathfrak{m}_n) \nonumber
\end{align}
for some integers $m_{\ast 0}$, regardless of whether $n$ is even or odd.  We now use the formulas \eqref{Piqpshift} and \eqref{Gqpshiftmt} for shifts of $\nu$ by multiples of $t = 1$ and hence take $(t, s) = (1, 1)$, $(1, 0)$, $(1, 0)$ for the three exceptional fibers.  We compute that
\begin{equation}
\hat{\pi}_\ast\widehat{Z}^\Phi(\nu)_{\mathfrak{m}_0, \mathfrak{m}_-, \mathfrak{m}_+, \mathfrak{m}_n} = (1 - e^{2\pi i\nu})^{-1 - m_{\ast 0}}(1 - \hat{y}_\ast)\widehat{Z}^\Phi(\nu)_{\mathfrak{m}_0, \mathfrak{m}_-, \mathfrak{m}_+, \mathfrak{m}_n}
\end{equation}
for $\ast = -, +, n$.  So the conditions that we want to impose are
\begin{equation}
m_{-0} = m_{+0} = m_{n0} = -1.
\end{equation}
We may use these conditions, as well as the equivalences \eqref{devennaivebasis} and \eqref{doddnaivebasis}, to simplify the shift operators so that they act on as few fluxes as possible.  For instance, we can write
\begin{align}
\hat{\pi}_- &: (\nu + n, \mathfrak{m}_0, \mathfrak{m}_- + 1, \mathfrak{m}_+, \mathfrak{m}_n), \nonumber \\
\hat{\pi}_+ &: (\nu + n, \mathfrak{m}_0, \mathfrak{m}_-, \mathfrak{m}_+ + 1, \mathfrak{m}_n), \\
\hat{\pi}_n &: (\nu + n + 1, \mathfrak{m}_0, \mathfrak{m}_- + 1, \mathfrak{m}_+ + 1, \mathfrak{m}_n) \nonumber
\end{align}
for $n$ even and
\begin{align}
\hat{\pi}_- &: (\nu + n, \mathfrak{m}_0, \mathfrak{m}_-, \mathfrak{m}_+ + 1, \mathfrak{m}_n), \nonumber \\
\hat{\pi}_+ &: (\nu + n, \mathfrak{m}_0, \mathfrak{m}_- + 1, \mathfrak{m}_+, \mathfrak{m}_n), \\
\hat{\pi}_n &: (\nu + 1, \mathfrak{m}_0, \mathfrak{m}_-\pm 1, \mathfrak{m}_+\mp 1, \mathfrak{m}_n) \nonumber
\end{align}
for $n$ odd, where we have chosen $\mathfrak{m}_n$ to remain inert in all cases.  For $n$ even, it is easy to see from \eqref{devennaivebasis} that all $(\mathfrak{m}_-, \mathfrak{m}_+)\in \mathbb{Z}_2\times \mathbb{Z}_2$ are independent.  But for $n$ odd, we can further use \eqref{doddnaivebasis} to make both $\mathfrak{m}_+$ and $\mathfrak{m}_n$ (or $\mathfrak{m}_-$ and $\mathfrak{m}_n$) inert by redefining $\hat{\pi}_-$ and $\hat{\pi}_n$ (or $\hat{\pi}_+$ and $\hat{\pi}_n$) appropriately.  Hence the only non-redundant flux in this case is valued in $\mathbb{Z}_4$, as desired.

On $\mathcal{S}^3[E_6]$, we have $(q, p) = (2, -1), (3, 1), (3, 1)$, so
\begin{equation}
(t, s)\in (1, 1) + (2, 1)\mathbb{Z}, \mbox{ } (1, 0) + (3, -1)\mathbb{Z}, \mbox{ } (1, 0) + (3, -1)\mathbb{Z}
\end{equation}
and
\begin{equation}
\widehat{Z}^\Phi(\nu)_{\mathfrak{m}_0, \mathfrak{m}_2, \mathfrak{m}_3, \mathfrak{m}_{3'}} = \Pi^\Phi(\nu)^{\mathfrak{m}_0}\Pi_{2, -1}^\Phi(\nu)_{\mathfrak{m}_2}\Pi_{3, 1}^\Phi(\nu)_{\mathfrak{m}_3}\Pi_{3, 1}^\Phi(\nu)_{\mathfrak{m}_{3'}}\widehat{Z}^\Phi(\nu)
\end{equation}
where
\begin{equation}
\widehat{Z}^\Phi(\nu)\equiv \mathcal{G}_{2, -1}^\Phi(\nu)\mathcal{G}_{3, 1}^\Phi(\nu)^2.
\end{equation}
The equivalences
\begin{align}
(\nu, \mathfrak{m}_0, \mathfrak{m}_2, \mathfrak{m}_3, \mathfrak{m}_{3'}) &\sim (\nu + N, \mathfrak{m}_0, \mathfrak{m}_2 - N, \mathfrak{m}_3 + N, \mathfrak{m}_{3'} + N) \label{e6equivalences} \\
&\sim (\nu, \mathfrak{m}_0 - N_2 - N_3 - N_{3'}, \mathfrak{m}_2 + 2N_2, \mathfrak{m}_3 + 3N_3, \mathfrak{m}_{3'} + 3N_{3'}) \nonumber
\end{align}
imply that
\begin{align}
(\nu, \mathfrak{m}_0, \mathfrak{m}_2, \mathfrak{m}_3, \mathfrak{m}_{3'}) &\sim (\nu + 6, \mathfrak{m}_0 + 1, \mathfrak{m}_2, \mathfrak{m}_3, \mathfrak{m}_{3'}) \nonumber \\
&\sim (\nu + 3, \mathfrak{m}_0, \mathfrak{m}_2 + 1, \mathfrak{m}_3, \mathfrak{m}_{3'}) \nonumber \\
&\sim (\nu + 4, \mathfrak{m}_0, \mathfrak{m}_2, \mathfrak{m}_3 + 1, \mathfrak{m}_{3'} + 1) \nonumber \\
&\sim (\nu + 6, \mathfrak{m}_0, \mathfrak{m}_2, \mathfrak{m}_3, \mathfrak{m}_{3'} + 3). \label{e6naivebasis}
\end{align}
Hence a shift of $\nu$ by 6 (the LCM of the $q$'s) is equivalent to a shift of the ordinary flux by $-1$, and the fractional flux $\mathfrak{m}_2$ is trivial modulo shifts of $\nu$.  From \eqref{e6naivebasis}, we also see that shifts $(\nu + 6, \mathfrak{m}_3 + 3)$ are trivial, so $\mathfrak{m}_3, \mathfrak{m}_{3'}$ are both effectively valued in $\mathbb{Z}_3$.  We now determine the most general shift operators satisfying the required commutation relations with
\begin{equation}
\hat{y}_2 = e^{2\pi i(\nu - \mathfrak{m}_2)/2}, \quad \hat{y}_3 = e^{2\pi i(\nu - \mathfrak{m}_3)/3}, \quad \hat{y}_{3'} = e^{2\pi i(\nu - \mathfrak{m}_{3'})/3}.
\end{equation}
Using \eqref{e6equivalences}, these reduce to the simple expressions
\begin{equation}
\hat{\pi}_\ast : (\mathfrak{m}_0, \mathfrak{m}_\ast)\mapsto (\mathfrak{m}_0 + m_{\ast 0}, \mathfrak{m}_\ast - 1)
\end{equation}
for some undetermined integers $m_{\ast 0}$, where $\ast = 2, 3, 3'$.  We then compute that
\begin{equation}
\hat{\pi}_\ast\widehat{Z}^\Phi(\nu)_{\mathfrak{m}_0, \mathfrak{m}_2, \mathfrak{m}_3, \mathfrak{m}_{3'}} = (1 - e^{2\pi i\nu})^{-m_{\ast 0}}(1 - \hat{y}_\ast)\widehat{Z}^\Phi(\nu)_{\mathfrak{m}_0, \mathfrak{m}_2, \mathfrak{m}_3, \mathfrak{m}_{3'}}
\end{equation}
for $\ast = 2, 3, 3'$.  Hence we must impose the conditions $m_{\ast 0} = 0$.  Using these conditions as well as the equivalences \eqref{e6equivalences}, it is possible to rewrite the shift operators such that $\mathfrak{m}_2$ and one of $\mathfrak{m}_3$ or $\mathfrak{m}_{3'}$ remain inert under their action.  For instance, they can be taken to act only on the fractional flux $\mathfrak{m}_{3'}$:
\begin{align}
\hat{\pi}_2 &: (\nu + 3, \mathfrak{m}_0, \mathfrak{m}_2, \mathfrak{m}_3, \mathfrak{m}_{3'}), \nonumber \\
\hat{\pi}_3 &: (\nu + 4, \mathfrak{m}_0, \mathfrak{m}_2, \mathfrak{m}_3, \mathfrak{m}_{3'} + 1), \\
\hat{\pi}_{3'} &: (\nu, \mathfrak{m}_0, \mathfrak{m}_2, \mathfrak{m}_3, \mathfrak{m}_{3'} - 1). \nonumber
\end{align}
The key to this rewriting is that 2 and 3 are coprime.  Hence the only non-redundant flux is valued in $\mathbb{Z}_3$, as desired.

On $\mathcal{S}^3[E_7]$, we have $(q, p) = (2, -1), (3, 1), (4, 1)$, so
\begin{equation}
(t, s)\in (1, 1) + (2, 1)\mathbb{Z}, \mbox{ } (1, 0) + (3, -1)\mathbb{Z}, \mbox{ } (1, 0) + (4, -1)\mathbb{Z}
\end{equation}
and
\begin{equation}
\widehat{Z}^\Phi(\nu)_{\mathfrak{m}_0, \mathfrak{m}_2, \mathfrak{m}_3, \mathfrak{m}_4} = \Pi^\Phi(\nu)^{\mathfrak{m}_0}\Pi_{2, -1}^\Phi(\nu)_{\mathfrak{m}_2}\Pi_{3, 1}^\Phi(\nu)_{\mathfrak{m}_3}\Pi_{4, 1}^\Phi(\nu)_{\mathfrak{m}_4}\widehat{Z}^\Phi(\nu)
\end{equation}
where
\begin{equation}
\widehat{Z}^\Phi(\nu)\equiv \mathcal{G}_{2, -1}^\Phi(\nu)\mathcal{G}_{3, 1}^\Phi(\nu)\mathcal{G}_{4, 1}^\Phi(\nu).
\end{equation}
The equivalences
\begin{align}
(\nu, \mathfrak{m}_0, \mathfrak{m}_2, \mathfrak{m}_3, \mathfrak{m}_4) &\sim (\nu + N, \mathfrak{m}_0, \mathfrak{m}_2 - N, \mathfrak{m}_3 + N, \mathfrak{m}_4 + N) \label{e7equivalences} \\
&\sim (\nu, \mathfrak{m}_0 - N_2 - N_3 - N_4, \mathfrak{m}_2 + 2N_2, \mathfrak{m}_3 + 3N_3, \mathfrak{m}_4 + 4N_4) \nonumber
\end{align}
imply that
\begin{align}
(\nu, \mathfrak{m}_0, \mathfrak{m}_2, \mathfrak{m}_3, \mathfrak{m}_4) &\sim (\nu + 12, \mathfrak{m}_0 + 1, \mathfrak{m}_2, \mathfrak{m}_3, \mathfrak{m}_4) \nonumber \\
&\sim (\nu + 9, \mathfrak{m}_0, \mathfrak{m}_2 + 1, \mathfrak{m}_3, \mathfrak{m}_4 + 1) \nonumber \\
&\sim (\nu + 4, \mathfrak{m}_0, \mathfrak{m}_2, \mathfrak{m}_3 + 1, \mathfrak{m}_4) \nonumber \\
&\sim (\nu + 6, \mathfrak{m}_0, \mathfrak{m}_2, \mathfrak{m}_3, \mathfrak{m}_4 + 2). \label{e7naivebasis}
\end{align}
Hence a shift of $\nu$ by 12 (the LCM of the $q$'s) is equivalent to a shift of the ordinary flux by $-1$, and the fractional flux $\mathfrak{m}_3$ is trivial modulo shifts of $\nu$.  From \eqref{e7naivebasis}, we also deduce that shifts $(\nu + 12, \mathfrak{m}_2 + 2)$ are trivial, so $\mathfrak{m}_2$ and $\mathfrak{m}_4$ are effectively valued in $\mathbb{Z}_2$.  Using these equivalences for simplification, we again find that the most general shift operators satisfying the required commutation relations with $\hat{y}_q = e^{2\pi i(\nu - \mathfrak{m}_q)/q}$ for $q = 2, 3, 4$ take the simple form
\begin{equation}
\hat{\pi}_q : (\mathfrak{m}_0, \mathfrak{m}_q)\mapsto (\mathfrak{m}_0 + m_{q0}, \mathfrak{m}_q - 1),
\end{equation}
and that they act on the partition function as
\begin{equation}
\hat{\pi}_q\widehat{Z}^\Phi(\nu)_{\mathfrak{m}_0, \mathfrak{m}_2, \mathfrak{m}_3, \mathfrak{m}_4} = (1 - e^{2\pi i\nu})^{-m_{q0}}(1 - \hat{y}_q)\widehat{Z}^\Phi(\nu)_{\mathfrak{m}_0, \mathfrak{m}_2, \mathfrak{m}_3, \mathfrak{m}_4}.
\end{equation}
Hence we must impose that $m_{q0} = 0$.  Under these conditions, the equivalences \eqref{e7naivebasis} allow us to write $\hat{\pi}_3$ so that it acts on no fluxes,
\begin{equation}
\hat{\pi}_3 : (\nu + 4, \mathfrak{m}_0, \mathfrak{m}_2, \mathfrak{m}_3, \mathfrak{m}_4),
\end{equation}
as well as to choose $\mathfrak{m}_3$ and $\mathfrak{m}_4$ to be inert by taking
\begin{align}
\hat{\pi}_2 &: (\nu, \mathfrak{m}_0, \mathfrak{m}_2 - 1, \mathfrak{m}_3, \mathfrak{m}_4), \nonumber \\
\hat{\pi}_4 &: (\nu - 3, \mathfrak{m}_0, \mathfrak{m}_2 - 1, \mathfrak{m}_3, \mathfrak{m}_4),
\end{align}
or to choose $\mathfrak{m}_2$ and $\mathfrak{m}_3$ to be inert by taking
\begin{align}
\hat{\pi}_2 &: (\nu + 3, \mathfrak{m}_0, \mathfrak{m}_2, \mathfrak{m}_3, \mathfrak{m}_4 - 1), \nonumber \\
\hat{\pi}_4 &: (\nu, \mathfrak{m}_0, \mathfrak{m}_2, \mathfrak{m}_3, \mathfrak{m}_4 - 1).
\end{align}
We see in any case that the only non-redundant flux is valued in $\mathbb{Z}_2$, as desired.

\subsection{Eliminating Redundancies}

We now describe a systematic, and far less effortful, approach to the above computations.  Specifically, for spaces with holonomies, we show how to write the difference equations explicitly in a canonical form, using a natural basis of fluxes with no redundancy.\footnote{I thank Brian Willett for discussions on this point.}  This relies on using the Smith normal form of the matrix of Picard group relations to eliminate redundant fluxes in the partition function (which works for any theory, not just the tetrahedron theory). 

Consider, for a general theory and general $\mathcal{M}_3$,
\begin{equation}
\widehat{Z}(\nu)_{\mathfrak{m}_0, \mathfrak{m}_1, \ldots, \mathfrak{m}_n} = \prod_{i=0}^n \mathcal{G}_{q_i, p_i}(\nu)_{\mathfrak{m}_i}.
\end{equation}
The dependence of $\widehat{Z}$ on the given variables is redundant.  We wish to eliminate this redundancy.  The 3D Picard group is abelian and admits an additive presentation
\begin{equation}
\widetilde{\operatorname{Pic}}(\mathcal{M}_3) = \langle x_0, x_1, \ldots, x_n | Ax = 0\rangle, \quad x\equiv \left(\begin{array}{c} x_0 \\ x_1 \\ \vdots \\ x_n \end{array}\right)
\end{equation}
where we have defined the $(n + 1)\times (n + 1)$ integer matrix
\begin{equation}
A\equiv \left(\begin{array}{ccccc}
-1 & q_1 & 0 & \cdots & 0 \\
-1 & 0 & q_2 & \cdots & 0 \\
\vdots & \vdots & \vdots & \ddots & \vdots \\
-1 & 0 & 0 & \cdots & q_n \\
d & p_1 & p_2 & \cdots & p_n
\end{array}\right),
\end{equation}
whose first $n$ rows encode the relations in the 2D Picard group and whose last row encodes the additional relation in the 3D Picard group.  The Smith normal form of $A$ can be written as
\begin{equation}
D = SAT, \quad D = \operatorname{diag}(\alpha_0, \alpha_1, \ldots, \alpha_n)\geq 0, \quad \alpha_0 = 1,
\end{equation}
where $S$ and $T$ are unimodular $(n + 1)\times (n + 1)$ matrices.  Since
\begin{equation}
\det A = (-1)^n c_1(\mathcal{L}_0)\prod_{i=1}^n q_i,
\end{equation}
the rank of $A$ is either $n + 1$ or $n$ (i.e., $\alpha_n > 0$ or $\alpha_n = 0$) depending on whether $c_1(\mathcal{L}_0)\neq 0$ or $c_1(\mathcal{L}_0) = 0$:
\begin{equation}
\widetilde{\operatorname{Pic}}(\mathcal{M}_3) = \begin{cases} \mathbb{Z}_{\alpha_1}\oplus \cdots\oplus \mathbb{Z}_{\alpha_n} & \text{if $c_1(\mathcal{L}_0)\neq 0$}, \\ \mathbb{Z}_{\alpha_1}\oplus \cdots\oplus \mathbb{Z}_{\alpha_{n-1}}\oplus \mathbb{Z} & \text{if $c_1(\mathcal{L}_0) = 0$}. \end{cases}
\end{equation}
At the level of fluxes, the natural basis is found by writing
\begin{equation}
Ax = 0 \Longleftrightarrow D(T^{-1}x) = 0.
\end{equation}
At the level of the partition function $\widehat{Z}$, the 3D Picard group relation also involves a shift in $\nu$:
\begin{equation}
(\nu, \mathfrak{m}_0, \mathfrak{m}_1, \ldots, \mathfrak{m}_n)\sim (\nu + 1, \mathfrak{m}_0 + d, \mathfrak{m}_1 + p_1, \ldots, \mathfrak{m}_n + p_n).
\end{equation}
Therefore, it is convenient to augment $A$ to an $(n + 1)\times (n + 2)$ matrix and write
\begin{equation}
\left(\begin{array}{c|c} {\begin{array}{c} \vec{0} \\ 1 \end{array}} & A \end{array}\right)\left(\begin{array}{c} x_\nu \vphantom{\Big(\Big)} \\ \hline x \vphantom{\Big(\Big)} \end{array}\right) = 0 \Longleftrightarrow S\left(\begin{array}{c|c} {\begin{array}{c} \vec{0} \\ 1 \end{array}} & AT \end{array}\right)\left(\begin{array}{c} x_\nu \vphantom{\Big(\Big)} \\ \hline T^{-1}x \vphantom{\Big(\Big)} \end{array}\right) = \left(\begin{array}{c|c} \beta & D \end{array}\right)\left(\begin{array}{c} x_\nu \vphantom{\Big(\Big)} \\ \hline T^{-1}x \vphantom{\Big(\Big)} \end{array}\right) = 0
\end{equation}
where $\beta$ is the rightmost column of $S$.  Hence the $n + 1$ relations in the 3D Picard group can be written as
\begin{equation}
(\nu, \tilde{\mathfrak{m}}_i)\sim (\nu + \beta_i, \tilde{\mathfrak{m}}_i + \alpha_i) \quad (i = 0, \ldots, n)
\end{equation}
where we have in terms of generators that
\begin{equation}
\left(\begin{array}{c} \tilde{x}_0 \\ \vdots \\ \tilde{x}_n \end{array}\right) = T^{-1}\left(\begin{array}{c} x_0 \\ \vdots \\ x_n \end{array}\right),
\end{equation}
so the new basis of fluxes is given by
\begin{equation}
\left(\begin{array}{c} \tilde{\mathfrak{m}}_0 \\ \vdots \\ \tilde{\mathfrak{m}}_n \end{array}\right) = T^T\left(\begin{array}{c} \mathfrak{m}_0 \\ \vdots \\ \mathfrak{m}_n \end{array}\right) \Longleftrightarrow \left(\begin{array}{c} \mathfrak{m}_0 \\ \vdots \\ \mathfrak{m}_n \end{array}\right) = T^{-T}\left(\begin{array}{c} \tilde{\mathfrak{m}}_0 \\ \vdots \\ \tilde{\mathfrak{m}}_n \end{array}\right)
\end{equation}
where $T^{-T}\equiv (T^{-1})^T = (T^T)^{-1}$ (passing from generators to fluxes requires distinguishing between passive and active transformations).

We further note that in the presence of flux, it is natural to define the shift operator for a given special fiber to act only on the corresponding flux and not on $\nu$, \emph{regardless} of $H_1$:
\begin{equation}
\boxed{\hat{y}_q = e^{2\pi i(\nu - t\mathfrak{m})/q}, \quad \hat{\pi}_q : (\nu, \mathfrak{m})\mapsto (\nu, \mathfrak{m} - 1).}
\label{onlyflux}
\end{equation}
Then the pairs $(\hat{y}, \hat{\pi})$ for different special fibers automatically commute, and the re\-qui\-red difference equations are manifestly satisfied because
\begin{equation}
\Pi_{q, p}^\Phi(\nu)_{\mathfrak{m} - 1} = (1 - e^{2\pi i(\nu - t\mathfrak{m})/q})\Pi_{q, p}^\Phi(\nu)_{\mathfrak{m}}
\end{equation}
(a special case of \eqref{Piqpshift}).  It can be checked that our previously obtained $\hat{\pi}$'s reduce to \eqref{onlyflux} in all cases.  As we explain in Section \ref{lineops}, this is no accident.  Below, we use the perspective \eqref{onlyflux}, which we have boxed to highlight its importance.

Let us see how these considerations work in our examples.  On $S^3$,
\begin{equation}
A = \left(\begin{array}{cc}
-1 & 1 \\
0 & 1
\end{array}\right)
\implies
S = A, \quad D = T = I_2,
\end{equation}
so $\tilde{\mathfrak{m}}_i = \mathfrak{m}_i$ ($i = 0, 1$) and we have
\begin{equation}
(\nu, \mathfrak{m}_0, \mathfrak{m}_1)\sim (\nu + 1, \mathfrak{m}_0 + 1, \mathfrak{m}_1)\sim (\nu + 1, \mathfrak{m}_0, \mathfrak{m}_1 + 1).
\end{equation}
On $S_b^3$,
\begin{equation}
A = \left(\begin{array}{ccc}
-1 & q_1 & 0 \\
-1 & 0 & q_2 \\
0 & p_1 & p_2
\end{array}\right)
\implies
S = \left(\begin{array}{ccc}
-q_2 p_1 & -q_1 p_2 & q_1 q_2 \\
p_2 & -p_2 & q_2 \\
-p_1 & p_1 & q_1
\end{array}\right), \quad D = T = I_3,
\end{equation}
so $\tilde{\mathfrak{m}}_i = \mathfrak{m}_i$ ($i = 0, 1, 2$) and we reproduce \eqref{sb3naivebasis}, which shows that we can indeed take the $\hat{\pi}$'s as in \eqref{sb3naive}.  On the PHS,
\begin{equation}
A = \left(\begin{array}{cccc}
-1 & 2 & 0 & 0 \\
-1 & 0 & 3 & 0 \\
-1 & 0 & 0 & 5 \\
0 & -1 & 1 & 1
\end{array}\right)
\implies
S = \left(\begin{array}{cccc}
15 & -10 & -6 & 30 \\
8 & -5 & -3 & 15 \\
5 & -3 & -2 & 10 \\
3 & -2 & -1 & 6
\end{array}\right), \quad D = T = I_4,
\end{equation}
so we have $\tilde{\mathfrak{m}}_q = \mathfrak{m}_q$ ($q = 0, 2, 3, 5$) and the right identifications \eqref{PHSnaivebasis} to define the $\hat{\pi}$'s simply as shifts of $\nu$, as in \eqref{PHSnaive}.  On $\mathcal{S}^3[D_{n+2}]$,
\begin{equation}
A = \left(\begin{array}{cccc}
-1 & 2 & 0 & 0 \\
-1 & 0 & 2 & 0 \\
-1 & 0 & 0 & n \\
0 & -1 & 1 & 1
\end{array}\right),
\end{equation}
so that for $n$ even,
\begin{equation}
S = \left(\begin{smallmatrix}
\frac{n}{2} & -\frac{n}{2} & -1 & n \\
1 + \frac{n}{2} & -\frac{n}{2} & -1 & n + 1 \\
\frac{n}{2} & 1 - \frac{n}{2} & -1 & n \\
1 & -1 & 0 & 2
\end{smallmatrix}\right),
\quad
D = \left(\begin{smallmatrix}
1 & 0 & 0 & 0 \\
0 & 1 & 0 & 0 \\
0 & 0 & 2 & 0 \\
0 & 0 & 0 & 2
\end{smallmatrix}\right),
\quad
T = \left(\begin{smallmatrix}
1 & 0 & 0 & 0 \\
0 & 1 & -1 & -1 \\
0 & 0 & 1 & 0 \\
0 & 0 & 0 & 1
\end{smallmatrix}\right),
\end{equation}
and for $n$ odd,
\begin{equation}
S = \left(\begin{smallmatrix}
\frac{n + 1}{2} & -\frac{n + 1}{2} & -1 & n + 1 \\
\frac{n + 1}{2} & -\frac{n - 1}{2} & -1 & n \\
\frac{n + 1}{2} & -\frac{n - 1}{2} & -1 & n + 1 \\
-n & n - 2 & 2 & -2n
\end{smallmatrix}\right),
\quad
D = \left(\begin{smallmatrix}
1 & 0 & 0 & 0 \\
0 & 1 & 0 & 0 \\
0 & 0 & 1 & 0 \\
0 & 0 & 0 & 4
\end{smallmatrix}\right),
\quad
T = \left(\begin{smallmatrix}
1 & 0 & -1 & -2 \\
0 & 1 & 0 & 1 \\
0 & 0 & 0 & -1 \\
0 & 0 & 1 & 2
\end{smallmatrix}\right).
\end{equation}
For $n$ even, we have
\begin{equation}
\tilde{\mathfrak{m}}_0 = \mathfrak{m}_0, \quad \tilde{\mathfrak{m}}_- = \mathfrak{m}_-, \quad \tilde{\mathfrak{m}}_+ = \mathfrak{m}_+ - \mathfrak{m}_-, \quad \tilde{\mathfrak{m}}_n = \mathfrak{m}_n - \mathfrak{m}_-
\label{devengoodbasis}
\end{equation}
as well as the equivalence relations
\begin{align}
(\nu, \tilde{\mathfrak{m}}_0) &\sim (\nu + n, \tilde{\mathfrak{m}}_0 + 1), \nonumber \\
(\nu, \tilde{\mathfrak{m}}_-) &\sim (\nu + n + 1, \tilde{\mathfrak{m}}_- + 1), \nonumber \\
(\nu, \tilde{\mathfrak{m}}_+) &\sim (\nu + n, \tilde{\mathfrak{m}}_+ + 2), \nonumber \\
(\nu, \tilde{\mathfrak{m}}_n) &\sim (\nu + 2, \tilde{\mathfrak{m}}_n + 2), \label{devengoodequivs}
\end{align}
which are equivalent to \eqref{devennaivebasis}.  Starting from the natural definitions
\begin{equation}
\hat{\pi}_- : \mathfrak{m}_-\mapsto \mathfrak{m}_- - 1, \quad \hat{\pi}_+ : \mathfrak{m}_+\mapsto \mathfrak{m}_+ - 1, \quad \hat{\pi}_n : \mathfrak{m}_n\mapsto \mathfrak{m}_n - 1,
\label{naturaldefinitions}
\end{equation}
we pass to the $\tilde{\mathfrak{m}}$ basis using \eqref{devengoodbasis} and eliminate $\tilde{\mathfrak{m}}_-$ using the second line of \eqref{devengoodequivs} to get that these are equivalent to
\begin{align}
\hat{\pi}_- : (\nu, \tilde{\mathfrak{m}}_+, \tilde{\mathfrak{m}}_n) &\mapsto (\nu + n + 1, \tilde{\mathfrak{m}}_+ + 1, \tilde{\mathfrak{m}}_n + 1), & \hat{\pi}_\ast : \tilde{\mathfrak{m}}_\ast &\mapsto \tilde{\mathfrak{m}}_\ast - 1 \quad (\ast = +, n)
\end{align}
acting on
\begin{equation}
\widehat{Z}^\Phi(\nu)_{\tilde{\mathfrak{m}}_0, \tilde{\mathfrak{m}}_-, \tilde{\mathfrak{m}}_+, \tilde{\mathfrak{m}}_n} = \Pi^\Phi(\nu)^{\tilde{\mathfrak{m}}_0}\mathcal{G}_{2, -1}^\Phi(\nu)_{\tilde{\mathfrak{m}}_-}\mathcal{G}_{2, 1}^\Phi(\nu)_{\tilde{\mathfrak{m}}_- + \tilde{\mathfrak{m}}_+}\mathcal{G}_{n, 1}^\Phi(\nu)_{\tilde{\mathfrak{m}}_- + \tilde{\mathfrak{m}}_n}.
\end{equation}
For $n$ odd, we have
\begin{gather}
\tilde{\mathfrak{m}}_0 = \mathfrak{m}_0, \quad \tilde{\mathfrak{m}}_- = \mathfrak{m}_-, \quad \tilde{\mathfrak{m}}_+ = -\mathfrak{m}_0 + \mathfrak{m}_n, \quad \tilde{\mathfrak{m}}_n = -2\mathfrak{m}_0 + \mathfrak{m}_- - \mathfrak{m}_+ + 2\mathfrak{m}_n
\end{gather}
as well as the equivalence relations
\begin{align}
(\nu, \tilde{\mathfrak{m}}_0) &\sim (\nu + n + 1, \tilde{\mathfrak{m}}_0 + 1), \nonumber \\
(\nu, \tilde{\mathfrak{m}}_-) &\sim (\nu + n, \tilde{\mathfrak{m}}_- + 1), \nonumber \\
(\nu, \tilde{\mathfrak{m}}_+) &\sim (\nu + n + 1, \tilde{\mathfrak{m}}_+ + 1), \nonumber \\
(\nu, \tilde{\mathfrak{m}}_n) &\sim (\nu - 2n, \tilde{\mathfrak{m}}_n + 4), \label{doddgoodequivs}
\end{align}
which are equivalent to \eqref{doddnaivebasis}.  Again writing the $\hat{\pi}$'s from \eqref{naturaldefinitions} in the $\tilde{\mathfrak{m}}$ basis, and then eliminating $\tilde{\mathfrak{m}}_-$ and $\tilde{\mathfrak{m}}_+$ using \eqref{doddgoodequivs}, we obtain the equivalent expressions
\begin{align}
\hat{\pi}_- : \nu &\mapsto \nu + 2n + 1, & \hat{\pi}_+ : \tilde{\mathfrak{m}}_n &\mapsto \tilde{\mathfrak{m}}_n + 1, & \hat{\pi}_n : (\nu, \tilde{\mathfrak{m}}_n) &\mapsto (\nu + n + 1, \tilde{\mathfrak{m}}_n - 2)
\end{align}
acting on
\begin{equation}
\widehat{Z}^\Phi(\nu)_{\tilde{\mathfrak{m}}_0, \tilde{\mathfrak{m}}_-, \tilde{\mathfrak{m}}_+, \tilde{\mathfrak{m}}_n} = \Pi^\Phi(\nu)^{\tilde{\mathfrak{m}}_0}\mathcal{G}_{2, -1}^\Phi(\nu)_{\tilde{\mathfrak{m}}_-}\mathcal{G}_{2, 1}^\Phi(\nu)_{\tilde{\mathfrak{m}}_- + 2\tilde{\mathfrak{m}}_+ - \tilde{\mathfrak{m}}_n}\mathcal{G}_{n, 1}^\Phi(\nu)_{\tilde{\mathfrak{m}}_0 + \tilde{\mathfrak{m}}_+}.
\end{equation}
On $\mathcal{S}^3[E_6]$,
\begin{gather}
A = \left(\begin{smallmatrix}
-1 & 2 & 0 & 0 \\
-1 & 0 & 3 & 0 \\
-1 & 0 & 0 & 3 \\
0 & -1 & 1 & 1
\end{smallmatrix}\right)
\implies
(S, D, T) = \left(\left(\begin{smallmatrix}
3 & -2 & -2 & 6 \\
2 & -1 & -1 & 3 \\
2 & -1 & -1 & 4 \\
3 & -2 & -1 & 6
\end{smallmatrix}\right),
\left(\begin{smallmatrix}
1 & 0 & 0 & 0 \\
0 & 1 & 0 & 0 \\
0 & 0 & 1 & 0 \\
0 & 0 & 0 & 3
\end{smallmatrix}\right),
\left(\begin{smallmatrix}
1 & 0 & 0 & 0 \\
0 & 1 & 0 & 0 \\
0 & 0 & 1 & -1 \\
0 & 0 & 0 & 1
\end{smallmatrix}\right)\right).
\end{gather}
We have
\begin{gather}
\tilde{\mathfrak{m}}_0 = \mathfrak{m}_0, \quad \tilde{\mathfrak{m}}_2 = \mathfrak{m}_2, \quad \tilde{\mathfrak{m}}_3 = \mathfrak{m}_3, \quad \tilde{\mathfrak{m}}_{3'} = -\mathfrak{m}_3 + \mathfrak{m}_{3'}
\end{gather}
as well as the equivalence relations
\begin{align}
(\nu, \tilde{\mathfrak{m}}_0) &\sim (\nu + 6, \tilde{\mathfrak{m}}_0 + 1), \nonumber \\
(\nu, \tilde{\mathfrak{m}}_2) &\sim (\nu + 3, \tilde{\mathfrak{m}}_2 + 1), \nonumber \\
(\nu, \tilde{\mathfrak{m}}_3) &\sim (\nu + 4, \tilde{\mathfrak{m}}_3 + 1), \nonumber \\
(\nu, \tilde{\mathfrak{m}}_{3'}) &\sim (\nu + 6, \tilde{\mathfrak{m}}_{3'} + 3), \label{e6goodequivs}
\end{align}
which are equivalent to \eqref{e6naivebasis}.  The $\hat{\pi}$'s from \eqref{onlyflux}, when written in the $\tilde{\mathfrak{m}}$ basis and after eliminating $\tilde{\mathfrak{m}}_2$ and $\tilde{\mathfrak{m}}_3$ using \eqref{e6goodequivs}, become
\begin{align}
\hat{\pi}_2 : \nu &\mapsto \nu + 3, & \hat{\pi}_3 : (\nu, \tilde{\mathfrak{m}}_{3'}) &\mapsto (\nu + 4, \tilde{\mathfrak{m}}_{3'} + 1), & \hat{\pi}_{3'} : \tilde{\mathfrak{m}}_{3'} &\mapsto \tilde{\mathfrak{m}}_{3'} - 1
\end{align}
acting on
\begin{equation}
\widehat{Z}^\Phi(\nu)_{\tilde{\mathfrak{m}}_0, \tilde{\mathfrak{m}}_2, \tilde{\mathfrak{m}}_3, \tilde{\mathfrak{m}}_{3'}} = \Pi^\Phi(\nu)^{\tilde{\mathfrak{m}}_0}\mathcal{G}_{2, -1}^\Phi(\nu)_{\tilde{\mathfrak{m}}_2}\mathcal{G}_{3, 1}^\Phi(\nu)_{\tilde{\mathfrak{m}}_3}\mathcal{G}_{3, 1}^\Phi(\nu)_{\tilde{\mathfrak{m}}_3 + \tilde{\mathfrak{m}}_{3'}}.
\end{equation}
On $\mathcal{S}^3[E_7]$,
\begin{gather}
A = \left(\begin{smallmatrix}
-1 & 2 & 0 & 0 \\
-1 & 0 & 3 & 0 \\
-1 & 0 & 0 & 4 \\
0 & -1 & 1 & 1
\end{smallmatrix}\right)
\implies
(S, D, T) = \left(\left(\begin{smallmatrix}
6 & -4 & -3 & 12 \\
5 & -3 & -2 & 9 \\
2 & -1 & -1 & 4 \\
3 & -2 & -1 & 6
\end{smallmatrix}\right),
\left(\begin{smallmatrix}
1 & 0 & 0 & 0 \\
0 & 1 & 0 & 0 \\
0 & 0 & 1 & 0 \\
0 & 0 & 0 & 2
\end{smallmatrix}\right),
\left(\begin{smallmatrix}
1 & 0 & 0 & 0 \\
0 & 1 & 0 & -1 \\
0 & 0 & 1 & 0 \\
0 & 0 & 0 & 1
\end{smallmatrix}\right)\right).
\end{gather}
We have
\begin{gather}
\tilde{\mathfrak{m}}_0 = \mathfrak{m}_0, \quad \tilde{\mathfrak{m}}_2 = \mathfrak{m}_2, \quad \tilde{\mathfrak{m}}_3 = \mathfrak{m}_3, \quad \tilde{\mathfrak{m}}_4 = -\mathfrak{m}_2 + \mathfrak{m}_4
\end{gather}
as well as the equivalence relations
\begin{align}
(\nu, \tilde{\mathfrak{m}}_0) &\sim (\nu + 12, \tilde{\mathfrak{m}}_0 + 1), \nonumber \\
(\nu, \tilde{\mathfrak{m}}_2) &\sim (\nu + 9, \tilde{\mathfrak{m}}_2 + 1), \nonumber \\
(\nu, \tilde{\mathfrak{m}}_3) &\sim (\nu + 4, \tilde{\mathfrak{m}}_3 + 1), \nonumber \\
(\nu, \tilde{\mathfrak{m}}_4) &\sim (\nu + 6, \tilde{\mathfrak{m}}_4 + 2), \label{e7goodequivs}
\end{align}
which are equivalent to \eqref{e7naivebasis}.  Expressing the $\hat{\pi}$'s from \eqref{onlyflux} in the $\tilde{\mathfrak{m}}$ basis and eliminating $\tilde{\mathfrak{m}}_2$ and $\tilde{\mathfrak{m}}_3$ using \eqref{e7goodequivs}, we obtain
\begin{align}
\hat{\pi}_2 : (\nu, \tilde{\mathfrak{m}}_4) &\mapsto (\nu + 9, \tilde{\mathfrak{m}}_4 + 1), & \hat{\pi}_3 : \nu &\mapsto \nu + 4, & \hat{\pi}_4 : \tilde{\mathfrak{m}}_4 &\mapsto \tilde{\mathfrak{m}}_4 - 1
\end{align}
acting on
\begin{equation}
\widehat{Z}^\Phi(\nu)_{\tilde{\mathfrak{m}}_0, \tilde{\mathfrak{m}}_2, \tilde{\mathfrak{m}}_3, \tilde{\mathfrak{m}}_4} = \Pi^\Phi(\nu)^{\tilde{\mathfrak{m}}_0}\mathcal{G}_{2, -1}^\Phi(\nu)_{\tilde{\mathfrak{m}}_2}\mathcal{G}_{3, 1}^\Phi(\nu)_{\tilde{\mathfrak{m}}_3}\mathcal{G}_{4, 1}^\Phi(\nu)_{\tilde{\mathfrak{m}}_2 + \tilde{\mathfrak{m}}_4}.
\end{equation}
Altogether, we recover the results of Section \ref{holonomies}.

\section{Quantization} \label{quantization}

We are now in a position to ask: what is the physical interpretation of these difference equations in terms of some as-yet unknown TQFT for the Seifert manifold $\mathcal{M}_3$?  In this section, we will fall well short of precisely identifying this TQFT, but we describe some properties that such a TQFT must have.

We would like to draw a parallel to Chern-Simons theory with gauge group $SL(2, \mathbb{C})$ \cite{Witten:1989ip}, about which we recall some basic facts for the sake of comparison \cite{Dimofte:2014zga}.  The action of $SL(2, \mathbb{C})$ Chern-Simons theory at level $(k, \sigma)$ contains both ho\-lo\-mor\-phic and an\-ti\-ho\-lo\-mor\-phic terms, weighted by $k\pm \sigma$:
\begin{equation}
S_{k, \sigma}(\cA, \bar{\cA}) = \frac{1}{2}(k + \sigma)S_\text{CS}(\cA) + \frac{1}{2}(k - \sigma)S_\text{CS}(\bar{\cA}).
\label{sl2cCS}
\end{equation}
Invariance under large gauge transformations requires $k\in \mathbb{Z}$, while unitarity (in the usual Hermitian structure) requires $\sigma\in i\mathbb{R}$.  The phase space of $SL(2, \mathbb{C})$ Chern-Simons theory on an ideal tetrahedron is $\mathcal{P}_{\partial\Delta}\cong (\mathbb{C}^\ast)^2$, with holomorphic symplectic form
\begin{equation}
\Omega = d\log y\wedge d\log x.
\end{equation}
The theory \eqref{sl2cCS} quantizes $\mathcal{P}_{\partial\Delta}$ with respect to the real symplectic form \cite{Dimofte:2014zga}
\begin{equation}
\omega_{k, \sigma} = \frac{(k + \sigma)\Omega + (k - \sigma)\overline{\Omega}}{4\pi} = \frac{2\pi}{k}(d\nu\wedge d\mu - dn\wedge dm),
\label{omegaksigma}
\end{equation}
where we have set $\sigma = k\big(\frac{1 - b^2}{1 + b^2}\big)$ (with $|b| = 1$ for unitarity) and passed to the more convenient coordinates
\begin{equation}
x = e^{2\pi i(-ib\mu - m)/k}, \quad y = e^{2\pi i(-ib\nu - n)/k}, \quad \bar{x} = e^{2\pi i(-ib^{-1}\mu + m)/k}, \quad \bar{y} = e^{2\pi i(-ib^{-1}\nu + n)/k}.
\end{equation}
Quantization promotes the coordinates to operators satisfying
\begin{equation}
[\boldsymbol{\nu}, \boldsymbol{\mu}] = -\frac{k}{2\pi i}, \quad [\mathbf{n}, \mathbf{m}] = \frac{k}{2\pi i}, \quad [\boldsymbol{\nu}, \mathbf{m}] = [\mathbf{n}, \boldsymbol{\mu}] = 0,
\end{equation}
or upon exponentiating, the $\mathfrak{q}$-commutation relations
\begin{equation}
\mathbf{yx} = \mathfrak{q}\mathbf{xy}, \quad \bar{\mathbf{y}}\bar{\mathbf{x}} = \bar{\mathfrak{q}}^{-1}\bar{\mathbf{x}}\bar{\mathbf{y}}, \quad \mathbf{y}\bar{\mathbf{x}} = \bar{\mathbf{x}}\mathbf{y}, \quad \bar{\mathbf{y}}\mathbf{x} = \mathbf{x}\bar{\mathbf{y}},
\end{equation}
where $\mathfrak{q} = e^{2\pi i(1 + b^2)/k}$ and $\bar{\mathfrak{q}}^{-1} = e^{2\pi i(1 + b^{-2})/k}$.  Since $m, n$ correspond to compact directions in phase space, the eigenvalues of $\mathbf{m}, \mathbf{n}$ are valued in $\mathbb{Z}_k$, hence trivial when $k = 1$.\footnote{In particular, quantization of $SL(2, \mathbb{C})$ Chern-Simons at $k = 1$ looks like quantization of $SL(2, \mathbb{R})$, as if we drop the quantized holonomies and keep only the noncompact part of the phase space $(\mathbb{C}^\ast)^2\cong \mathbb{R}^2\times (S^1)^2$ \cite{Dimofte:2011py}.}  We obtain a Hilbert space of wavefunctions depending on $\mu\in \mathbb{R}$ and $m\in \mathbb{Z}_k$.  It is useful to analytically continue the theory away from $|b| = 1$, in particular to real $b$; in this case, we write $\tilde{x}, \tilde{y}$ (which are no longer related to $x, y$ by complex conjugation) and write $\tilde{\mathfrak{q}}$ in place of $\bar{\mathfrak{q}}^{-1}$.

This discussion should be compared to that of lens spaces $L(k, p)_b$ in Section \ref{lensspaces}, with $p = 1$.  The difference equations for lens space partition functions of $T_2[M]$ have been studied exhaustively.  The associated difference operators are quantizations of classical ``A-polynomials'' that cut out Lagrangian submanifolds in the phase space of $SL(2, \mathbb{C})$ Chern-Simons theory on $M$ \cite{Dimofte:2011gm, Beem:2012mb, Dimofte:2013iv}, which in the simplest case $M = \Delta$ is the phase space $\mathcal{P}_{\partial\Delta}$ described above.

\subsection{Phase Space}

Our object of interest is the ``$\mathcal{M}_3$-TQFT'' described by the $\mathcal{M}_3$ state-integral model for the theory $T_2[M]$.  We have learned that the Seifert partition function can be refined by both masses and insertions of flavor Wilson lines.  Hence the wavefunction of the $\mathcal{M}_3$-TQFT on $M$ is a function of (exponentiated) real mass parameters $\nu_i$ and background holonomies $m_i$ for the $U(1)$ factors in the maximal torus of the flavor symmetry group of $T_2[M]$, the latter being valued in $\pi_1$ of the Seifert geometry.  For triangulated $M$, the number of pairs of variables $(\nu_i, m_i)$ should depend on the topology of $\partial M$ \cite{Dimofte:2014zga}.

More precisely, it is not $\pi_1$ of $\mathcal{M}_3$ that is parametrized by phases in the partition function, but rather the torsion part of its abelianization $H_1$ (which is a finite abelian group).  For example, $\pi_1$ of the PHS is the binary icosahedral group $\hat{I}$, and the discrete parameters that refine the PHS partition function can all be ``gauged away'' because the abelianization of $\hat{I}$ is trivial.  In general, recall that the 3D Picard group $\widetilde{\operatorname{Pic}}(\mathcal{M}_3)$ given in \eqref{3dpicard} is the pullback of the 2D orbifold Picard group $\operatorname{Pic}(\hat{\Sigma}_g)$ along the Seifert fibration (tensoring with $\mathcal{L}_0$ corresponds to a large gauge transformation in the 3D A-model, as in \eqref{largegauge}).  For $c_1(\mathcal{L}_0)\neq 0$, we have
\begin{equation}
\widetilde{\operatorname{Pic}}(\mathcal{M}_3)\cong \operatorname{Tor} H_1(\mathcal{M}_3, \mathbb{Z}).
\end{equation}
The holonomies are encoded in the partition function as fractional fluxes, which pa\-ra\-met\-rize $\operatorname{Pic}(\hat{\Sigma}_g)$.

For simplicity, consider the tetrahedron theory $T_\Delta = T_2[\Delta]$, i.e., a free chiral multiplet.  From the above discussion, we conclude that for a given Seifert geometry $\mathcal{M}_3$, the Hilbert space of its TQFT dual on a four-punctured sphere (i.e., the boundary of an ideal tetrahedron) is given by quantizing the classical phase space
\begin{equation}
\mathbb{R}^2\times (S^1)^{2\#}
\end{equation}
where $\#$ is the number of nontrivial fluxes, i.e., the rank of the 3D Picard group (when $c_1(\mathcal{L}_0)\neq 0$, $\#$ is the number of generators of $\operatorname{Tor} H_1(\mathcal{M}_3, \mathbb{Z})$ as a finite abelian group).  This is true regardless of the number of special fibers.  For rationally squashed lens spaces (those for which the squashing parameter satisfies $b^2 = q_1/q_2$ where $q_1$ and $q_2$ are coprime positive integers), $\# = 1$.

The partition function of the tetrahedron theory on $\mathcal{M}_3$ (i.e., the TQFT wavefunction on the four-punctured sphere) is a function of a single real mass parameter $\nu$ and $\#$ discrete holonomies: these variables and their momenta are interpreted as the coordinates on the aforementioned phase space, which has a canonical symplectic form and can be quantized to yield operators acting on the wavefunction.  By studying the properties of the tetrahedron partition function on $\mathcal{M}_3$, we have shown that these wavefunctions are annihilated by a set of difference operators determined by the Seifert geometry.  The number of difference operators is the number of exceptional fibers.  For example, rationally squashed lens spaces are Seifert manifolds with $S^2$ base and two exceptional fibers, and indeed, the analysis of \cite{Dimofte:2014zga} gives two difference equations in this case.  In our examples, we can write the difference operators as polynomials in the exponentiated phase space coordinates in a canonical way (thanks to Smith normal form).  These difference operators are extra data that are not determined by the symplectic form on phase space.

To illustrate this last point, we again proceed by way of example.  We consider a few simple examples that involve no flux and are therefore all ``$SL(2, \mathbb{R})$-like,'' but whose tetrahedron wavefunctions are all annihilated by different sets of difference operators.

\subsubsection*{Squashed Three-Sphere}

The rationally squashed three-sphere $S^3_b$ with $b^2 = q_1/q_2$ is a Seifert manifold whose exceptional fibers are characterized by the pairs of integers $(q_1, p_1)$ and $(q_2, p_2)$ where $p_1$ and $p_2$ satisfy $q_1 p_2 + q_2 p_1 = 1$.  It has no holonomies ($\# = 0$), so we can set $\mathfrak{m}_1 = \mathfrak{m}_2 = 0$ and write $\hat{y}_1 = e^{2\pi i\nu/q_1}$ and $\hat{y}_2 = e^{2\pi i\nu/q_2}$.  Then we have the representations
\begin{equation}
\hat{\pi}_1 = e^{q_2\partial_\nu}, \quad \hat{\pi}_2 = e^{q_1\partial_\nu},
\end{equation}
which are consistent with the Weyl algebras
\begin{align}
\hat{\pi}_1\hat{y}_1 &= \mathfrak{q}_1\hat{y}_1\hat{\pi}_1, & \mathfrak{q}_1 &= e^{2\pi iq_2/q_1} = e^{2\pi ib^{-2}}, \\
\hat{\pi}_2\hat{y}_2 &= \mathfrak{q}_2\hat{y}_2\hat{\pi}_2, & \mathfrak{q}_2 &= e^{2\pi iq_1/q_2} = e^{2\pi ib^2}.
\end{align}
The $S^3_b$ partition function of $T_2[\Delta]$ is annihilated by the difference operators
\begin{equation}
\hat{\pi}_1 + \hat{y}_1 - 1, \quad \hat{\pi}_2 + \hat{y}_2 - 1.
\label{s3bdifference1}
\end{equation}
The difference operators can also be viewed as polynomials in the basic variables $\hat{y}\equiv e^{2\pi i\nu/q_1 q_2}$ and $\hat{\pi}\equiv e^{\partial_\nu}$:
\begin{equation}
\hat{\pi}^{q_2} + \hat{y}^{q_2} - 1, \quad \hat{\pi}^{q_1} + \hat{y}^{q_1} - 1.
\end{equation}
With continuous squashing \cite{Dimofte:2014zga}, the difference operators can instead be written as
\begin{equation}
y + x^{-1} - 1, \quad \tilde{y} + \tilde{x}^{-1} - 1
\label{s3bdifference2}
\end{equation}
where $x = e^{2\pi b\nu}$, $\tilde{x} = e^{2\pi b^{-1}\nu}$, $y = e^{ib\partial_\nu}$, $\tilde{y} = e^{ib^{-1}\partial_\nu}$.  When $b^2 = q_1/q_2$, the two sets of operators in \eqref{s3bdifference1} and \eqref{s3bdifference2} are related by $\nu\to -i\sqrt{q_1 q_2}\nu$, giving the identifications
\begin{equation}
(\hat{y}_1, \hat{y}_2, \hat{\pi}_1, \hat{\pi}_2)\leftrightarrow (\tilde{x}, x, \tilde{y}, y).
\end{equation}
The latter set of operators satisfies the same mutually commuting algebras as the first:
\begin{equation}
xy = e^{2\pi ib^2}yx, \quad \tilde{x}\tilde{y} = e^{2\pi ib^{-2}}\tilde{y}\tilde{x} \quad (b\in \mathbb{C}).
\end{equation}
Note that $x$ and $\tilde{x}$ are related by complex conjugation when $\nu$ is real and $b$ is a phase, while $\hat{y}_1$ and $\hat{y}_2$ are not.  The existence of the difference operators \eqref{s3bdifference1} is related to a factorization property of the quantum dilogarithm at rational $b^2$ observed in \cite{Garoufalidis:2014ifa}, as discussed in \cite{Closset:2018ghr}.

Independently of representation, we can write
\begin{equation}
\mathbf{y}_1 = e^{2\pi i\boldsymbol{\nu}/q_1}, \quad \mathbf{y}_2 = e^{2\pi i\boldsymbol{\nu}/q_2}, \quad \boldsymbol{\pi}_1 = e^{2\pi iq_2\boldsymbol{\rho}}, \quad \boldsymbol{\pi}_2 = e^{2\pi iq_1\boldsymbol{\rho}}, \quad [\boldsymbol{\nu}, \boldsymbol{\rho}] = \frac{i}{2\pi}.
\end{equation}
The canonical commutation relation comes from the real symplectic form
\begin{align}
\omega_{1, \sigma} &= 2\pi\, d\boldsymbol{\nu}\wedge d\boldsymbol{\rho} \label{firstline} \\
&= \pi(1 - \sigma)\, d\boldsymbol{\nu}\wedge d\boldsymbol{\rho} + \pi(1 + \sigma)\, d\boldsymbol{\nu}\wedge d\boldsymbol{\rho} \\
&= \textstyle \frac{1}{4\pi}(4\pi^2 b^2(1 + \sigma)\, d\boldsymbol{\nu}\wedge d\boldsymbol{\rho} + 4\pi^2 b^{-2}(1 - \sigma)\, d\boldsymbol{\nu}\wedge d\boldsymbol{\rho}) \\
&= \textstyle \frac{1}{4\pi}((1 + \sigma)\, d\log\boldsymbol{\pi}_2\wedge d\log\mathbf{y}_2 + (1 - \sigma)\, d\log\boldsymbol{\pi}_1\wedge d\log\mathbf{y}_1), \label{lastline}
\end{align}
where we have used the standard parametrization for $k = 1$,
\begin{equation}
\sigma = \frac{1 - b^2}{1 + b^2},
\end{equation}
and the phase space is $\mathbb{R}^2$.  In the series of steps \eqref{firstline}--\eqref{lastline}, we have reversed the logic of \eqref{omegaksigma} so as to ``de-diagonalize'' the symplectic form and thereby expose the structure of the underlying $SL(2, \mathbb{C})$ Chern-Simons Lagrangian at level $(1, \sigma)$.  Note that the weights of the two terms in $\omega_{1, \sigma}$ are
\begin{equation}
1 - \sigma = \frac{2q_1}{q_1 + q_2}, \quad 1 + \sigma = \frac{2q_2}{q_1 + q_2}.
\label{thisgeneralizes}
\end{equation}
Despite their sum being real, the two terms in $\omega_{1, \sigma}$ are not related by complex conjugation because we have analytically continued away from $|b| = 1$ (imaginary $\sigma$).

\subsubsection*{Poincar\'e Homology Sphere}

The PHS is a Seifert manifold with exceptional fibers $(2, -1), (3, 1), (5, 1)$ and for which $\# = 0$.  We may set $\mathfrak{m}_q = 0$, so that we have the representations
\begin{equation}
\hat{y}_q = e^{2\pi i\nu/q}, \quad \hat{\pi}_q = e^{30\partial_\nu/q}
\end{equation}
for $q = 2, 3, 5$.  The tetrahedron partition function on this space is annihilated by the three difference operators $\hat{\pi}_q + \hat{y}_q - 1$, which can be written as polynomials
\begin{equation}
\hat{\pi}^{30/q} + \hat{y}^{30/q} - 1
\end{equation}
in the variables $\hat{y}\equiv e^{2\pi i\nu/30}$ and $\hat{\pi}\equiv e^{\partial_\nu}$.  Abstractly, we can write
\begin{equation}
\mathbf{y}_q = e^{2\pi i\boldsymbol{\nu}/q}, \quad \boldsymbol{\pi}_q = e^{2\pi i(30\boldsymbol{\rho}/q)}, \quad [\boldsymbol{\nu}, \boldsymbol{\rho}] = \frac{i}{2\pi}.
\end{equation}
The canonical commutation relation follows from the same symplectic form $\omega_{1, \sigma}$, which we can suggestively ``de-diagonalize'' as before by postulating that the multiplicative weight for a given fiber should be $\frac{2q}{\sum_i t_i}$, generalizing \eqref{thisgeneralizes}:
\begin{equation}
\frac{1}{4\pi}\sum_i \frac{2q_i}{\sum_j t_j}\, d\log\boldsymbol{\pi}_i\wedge d\log\mathbf{y}_i = 2\pi\sum_i \frac{t_i}{\sum_j t_j}\, d\boldsymbol{\nu}\wedge d\boldsymbol{\rho} = 2\pi\, d\boldsymbol{\nu}\wedge d\boldsymbol{\rho}.
\end{equation}
We expect a Hilbert space $L^2(\mathbb{R})$ whose elements are functions $f(\nu)$.

\subsubsection*{Integral Homology Sphere}

More generally, consider a Seifert integral homology sphere, for which the exceptional fibers are $(q_i, p_i)$ where $i = 1, \ldots, n$ and the $q_i$ are mutually coprime (among other conditions).  In all of these cases, $\# = 0$: hence the relevant phase space is $\mathbb{R}^2$ and the quantization of the dual TQFT resembles that of $SL(2, \mathbb{R})$ Chern-Simons theory.  The tetrahedron partition function is annihilated by the $n$ difference operators
\begin{equation}
\hat{\pi}_i + \hat{y}_i - 1 \quad (i = 1, \ldots, n),
\end{equation}
where $\hat{y}_i = e^{2\pi i\nu/q_i}$ and $\hat{\pi}_i = e^{t_i\partial_\nu}$.  The integers $t_i$ are fixed by the Chinese remainder theorem to satisfy $p_i t_i\equiv 1 \text{ (mod $q_i$)}$ and $t_i\equiv 0 \text{ (mod $q_j$)}$ for $j\neq i$, which ensures that the pairs $(\hat{y}_i, \hat{\pi}_i)$ form mutually commuting algebras (we may take all of the $t_i$ to lie between 0 and $q_1\cdots q_n - 1$).  All of these difference operators take essentially the same form as for lens spaces (keeping in mind that we consider only rational squashing).

\subsection{Line Operators} \label{lineops}

We now come to our final point.  While not directly related to quantization of the $\mathcal{M}_3$-TQFT, it lends a new perspective on the whole setup that may prove useful in turning some of our suggestive observations into sharp statements.  So far, we have motivated the difference equations from the bottom up.  From the top down, the algebra of line operators localized at special fibers makes clear physically that this kind of structure is inevitable.\footnote{I thank Tudor Dimofte for emphasizing this interpretation.}

The starting point is the observation that any 3D theory with abelian flavor symmetry provides a boundary condition for a 4D abelian gauge theory on the half-space $\mathcal{M}_3\times \mathbb{R}_+$, and the algebra of lines in the 4D theory acts on its $\mathcal{M}_3$ partition function.  In the particular setting of the 3D-3D correspondence, the setup is as follows.  Consider $M$ with boundary.  Near the boundary, $M$ looks like $\Sigma\times \mathbb{R}_+$.  The $(2, 0)$ theory on $\mathcal{M}_3\times \Sigma\times \mathbb{R}_+$ can be thought of as the class-$\mathcal{S}$ theory $S[\Sigma]$ on $\mathcal{M}_3\times \mathbb{R}_+$.  Then the relevant BPS line operators are the IR Wilson and 't Hooft lines of $S[\Sigma]$, supported on the singular fibers of $\mathcal{M}_3$ at distinct points of $\mathbb{R}_+$.  From this point of view, the number of elementary line operators is clearly the number of singular fibers times the dimension of the Coulomb branch of $S[\Sigma]$.  Commutation relations come from moving operators past each other in the $\mathbb{R}_+$ direction, so those associated to different singular fibers commute, whereas those on the same singular fiber do not.  The precise algebra of BPS line operators in 4D can be derived as in \cite{Gaiotto:2010be} (see \cite{Okuda:2014fja} for a clear review of these results).  The existence of a boundary in $\mathbb{R}_+$ further imposes Ward identities (linear relations) involving these line operators when brought to the boundary.

Henceforth, we restrict our discussion to the tetrahedron theory ($M = \Delta$) because, as mentioned earlier, the construction of an arbitrary theory from a collection of free chiral multiplets via standard operations on the UV Lagrangian leads to a parallel construction of the corresponding difference equations.  Hence we would like to interpret the difference equations as Ward identities for supersymmetric line operators in a pure 4D $\mathcal{N} = 2$ $U(1)$ gauge theory, generalizing similar analyses for $S^3_b$ \cite{Dimofte:2011ju}, $S^2\times S^1$ \cite{Dimofte:2011py}, and lens spaces \cite{Dimofte:2014zga} to arbitrary Seifert manifolds.

Since the magic happens in the neighborhood of a special fiber, our strategy is to combine the known action of line operators on holomorphic blocks with given complex structure \cite{Beem:2012mb} with the relation between blocks and Seifert fibering operators.  Letting $\tau$ denote the complex structure of the boundary torus, the Wilson line measures the holonomy of the background $U(1)$ flavor gauge field,
\begin{equation}
\nu = \oint_{S^1} A + \tau\oint_{\partial D^2} A,
\end{equation}
and the dual 't Hooft line shifts the magnetic flux:
\begin{equation}
dA = 2\pi\mathfrak{m}_0\delta^2(x - x_0) + 2\pi\sum_i \frac{\mathfrak{m}_i}{q_i}\delta^2(x - x_i), \quad c_1(L) = \frac{1}{2\pi}\int_{\hat{\Sigma}_g} dA.
\end{equation}
The Wilson line multiplies $B^\Phi(\nu, \tau)$ by $y^{-1}$ where $y = e^{2\pi i\nu}$, and the 't Hooft line multiplies $y$ by $\mathfrak{q}^{-1}$ where $\mathfrak{q} = e^{2\pi i\tau}$.\footnote{In the notation of \cite{Beem:2012mb}, $B_\Delta(x; \mathfrak{q}) = (\mathfrak{q}x^{-1}; \mathfrak{q})_\infty$; the Wilson line multiplies this by $x$ and the associated 't Hooft line multiplies $x$ by $\mathfrak{q}$.}  After an $SL(2, \mathbb{Z})$ transformation on the boundary, not to be confused with the action of electric-magnetic duality in 4D \cite{Witten:2003ya}, we obtain
\begin{equation}
B_{\tilde{g}}^\Phi(\nu, \tau) = (\tilde{\mathfrak{q}}\tilde{y}; \tilde{\mathfrak{q}})_\infty,
\end{equation}
and the actions of the line operators are then given in terms of the variables $\tilde{y}$ and $\tilde{\mathfrak{q}}$.  On the other hand, the chiral block with flux is defined by performing a large gauge transformation on the boundary torus \cite{Closset:2018ghr}:
\begin{equation}
B^\Phi(\nu, \tau)_{\mathfrak{m}}\equiv B^\Phi(\nu + \mathfrak{m}\tau, \tau)
\end{equation}
(with an arbitrary choice of zero for $\nu$).  The Wilson line multiplies $B^\Phi(\nu + \mathfrak{m}\tau, \tau)$ by $e^{-2\pi i(\nu + \mathfrak{m}\tau)}$, and the 't Hooft line multiplies $e^{2\pi i(\nu + \mathfrak{m}\tau)}$ by $e^{-2\pi i\tau}$.  Correspondingly, after both a large gauge transformation and an $SL(2, \mathbb{Z})$ transformation, we have
\begin{equation}
B_{\tilde{g}}^\Phi(\nu, \tau)_{\mathfrak{m}} = B^\Phi\left(\frac{\nu}{p\tau + q}, \frac{s\tau - t}{p\tau + q}\right)_{\mathfrak{m}} = (\tilde{\mathfrak{q}}(\tilde{\mathfrak{q}}^{\mathfrak{m}}\tilde{y}); \tilde{\mathfrak{q}})_\infty = \frac{(\tilde{\mathfrak{q}}\tilde{y}; \tilde{\mathfrak{q}})_\infty}{(\tilde{\mathfrak{q}}\tilde{y}; \tilde{\mathfrak{q}})_{\mathfrak{m}}},
\end{equation}
where the Wilson line multiplies this by $(\tilde{\mathfrak{q}}^{\mathfrak{m}}\tilde{y})^{-1}$ while the 't Hooft line multiplies $\tilde{\mathfrak{q}}^{\mathfrak{m}}\tilde{y}$ by $\tilde{\mathfrak{q}}^{-1}$.  The effect of taking $\tilde{\mathfrak{q}}^{\mathfrak{m}}\tilde{y}\mapsto \tilde{\mathfrak{q}}^{\mathfrak{m} - 1}\tilde{y}$ is to multiply $B_{\tilde{g}}^\Phi(\nu, \tau)_{\mathfrak{m}}$ by $1 - \tilde{\mathfrak{q}}^{\mathfrak{m}}\tilde{y}$ where
\begin{equation}
\tilde{\mathfrak{q}}^{\mathfrak{m}}\tilde{y} = \exp\left(\frac{2\pi i(\nu + \mathfrak{m}(s\tau - t))}{p\tau + q}\right)\xrightarrow{\tau\to 0} e^{2\pi i(\nu - t\mathfrak{m})/q}.
\end{equation}
Now, the effect of the 't Hooft line is equivalent to taking $\mathfrak{m}\mapsto \mathfrak{m} - 1$ or to taking $\nu\mapsto \nu + t$, in both cases up to an ambiguity by multiples of $q$.  After taking $\tau\to 0$, the resulting fibering operator is no longer a function of $\tilde{y}$ and $\tilde{\mathfrak{q}}$, so this ambiguity becomes important.  In either case (that is, $\mathfrak{m}\mapsto \mathfrak{m} - 1$ or $\nu\mapsto \nu + t$), we can fix the ambiguity by demanding that the Ward identity remain satisfied in the limit that the fiber becomes singular, but the former perspective ($\mathfrak{m}\mapsto \mathfrak{m} - 1$) has the advantage that the difference equations can be fixed already at the level of the individual fiber rather than at the level of the partition function.  To explain what we mean, note that in the second case ($\nu\mapsto \nu + t$), \eqref{Gqpshiftmt} and \eqref{Gqpshiftmq} give
\begin{equation}
\mathcal{G}_{q, p}^\Phi(\nu + t + mq)_{\mathfrak{m}} = (1 - e^{2\pi i\nu})^{mp - s}(1 - e^{2\pi i(\nu - t\mathfrak{m})/q})\mathcal{G}_{q, p}^\Phi(\nu)_{\mathfrak{m}}.
\end{equation}
Only after assembling the $\mathcal{G}_{q, p}^\Phi$ into $Z_{\mathcal{M}_3}$ do we expect the overall factor of $\Pi^\Phi(\nu) = (1 - e^{2\pi i\nu})^{-1}$ to cancel, as it must.  Furthermore, we prefer to shift $\mathfrak{m}$ because it makes the physical meaning of the 't Hooft line manifest.  We end up with
\begin{equation}
\hat{y}_q = e^{2\pi i(\nu - t\mathfrak{m})/q}, \quad \hat{\pi}_q = e^{-\partial_{\mathfrak{m}}},
\end{equation}
which satisfy $\hat{\pi}_q\hat{y}_q = e^{2\pi it/q}\hat{y}_q\hat{\pi}_q$.  This is exactly as in \eqref{onlyflux}.

\section{Outlook} \label{outlook}

In this paper, we have undertaken a preliminary investigation of the 3D-3D correspondence for (primarily) homology spheres, based on constraints from the $\sigma$-contour and the algebraic properties of Seifert fibering operators.  Our findings represent a small step toward identifying the $\mathcal{M}_3$-TQFT, for $\mathcal{M}_3$ a Seifert manifold.  Our analysis leaves many questions unanswered.

So far, our treatment of the state-integral model for the putative TQFT dual to $\mathcal{M}_3$ has been highly indirect, taking for granted the principle of gluing as symplectic reduction and the well-posedness of the affine symplectic action.  It would be nice to work out the proper definitions of the affine shifts and to identify the correct functional spaces in which the (truncated) TQFT wavefunctions are valued, as in \cite{Dimofte:2014zga}.  This would put the constructive definition of the TQFT on more solid footing.

It would also be nice to relax some of our seemingly inessential assumptions, such as the assumption that $\kappa < 0$, as well as to generalize all aspects of our analysis for integral homology spheres to rational homology spheres and beyond.  The generalization to $\kappa\geq 0$ would entail modifying the shape of the $\sigma$-contour, and it would remain to show whether such nonlinear deformations play nicely with the angle polytopes and positive angle structures of \cite{Dimofte:2014zga}.

It might also be interesting, though difficult, to generalize the $\sigma$-contour to nonabelian theories.  Such a generalization could have relevance for the program of finding nonabelian UV descriptions of theories in class $\mathcal{R}$, particularly the $T_n[M]$ theories.

Perhaps most of all, it would be desirable to obtain an explicit description of the TQFT dual to $\mathcal{M}_3$, and in particular to derive the difference equations and symplectic form from a Lagrangian description of this theory.  Once such details are ironed out, one could contemplate venturing beyond the prototypical setting of hyperbolic $M$ and the $T_2$ theories associated to two M5-branes.  We hope that such a detailed understanding can be achieved in future work.  An alternative route to the one that we have taken is to adapt the DGG construction of $M$ by gluing ideal tetrahedra, but at the level of individual fibering and handle-gluing operators rather than at the level of the Seifert partition function.  This is analogous to the holomorphic block decomposition, and may aid in building up the A-model data of $T[M]$ as well as possibly translating them into the language of complex Chern-Simons theory.

\section*{Acknowledgements}

This paper was heavily influenced by Brian Willett, whom I heartily thank for collaboration during the formative stages of this work, as well as for detailed answers to my many questions about the Seifert fibering formalism.  I also thank Tudor Dimofte for very helpful comments.  This work was partially supported by the NSF GRFP under Grant No.\ DGE-1656466 and by a Centennial Fellowship from the Graduate School at Princeton University.

\appendix

\section{Explicit Formulas} \label{explicit}

Here, we give explicit expressions for the handle-gluing and Seifert fibering operators of a 3D $\cN = 2$ gauge theory.  We assume that the theory has gauge group $G$ and flavor symmetry group $G_F$, that the matter transforms in a representation $R = \bigoplus_I (R_I\otimes S_I)$ of $G\times G_F$, and that the $U(1)_R$ charge of the $I^\textrm{th}$ term in the sum is $r_I$.  We take the chiral multiplets in the $U(1)_{-1/2}$ quantization, which entails UV Chern-Simons contact terms of level $-1/2$ for the symmetries under which they are charged.  In addition, we include bare levels $k^{ab}$, $k^{a\alpha}$, and $k^{\alpha\beta}$ for the $GG$, $GF$, and $FF$ Chern-Simons terms, respectively.  For simplicity, we suppress contact terms involving the $R$-symmetry and gravitational Chern-Simons terms.  Then we have for the handle-gluing operator that 
\begin{equation}
\cH(u, \nu) = e^{2\pi i\Omega(u, \nu)}H(u, \nu),
\label{handlehessrelation}
\end{equation}
where $H$ is given by \eqref{hessdef} in terms of the gauge flux operator
\begin{equation}
\Pi_a = e^{2\pi i(\sum_b k^{ab}u_b + \sum_\alpha k^{a\alpha}\nu_\alpha)}\prod_I\prod_{(\rho, \omega)\in R_I\times S_I} (1 - e^{2\pi i(\rho(u) + \omega(\nu) + \nu_R r_I)})^{-\rho_a}
\label{gaugeflux}
\end{equation}
and the ``effective dilaton'' $\Omega$ is given by\footnote{Notice that \vspace{\maxdimen} the $R$-charges are absent from the exponents of the flux operators appearing in the effective dilaton, in contrast with the conventions in \cite{Closset:2017zgf, Closset:2018ghr}.  This dependence instead arises in the operator $\cG_{q_0 = 1, p_0 = \mathbf{d}}$ below.  Namely, in the ``standard'' $R$-symmetry background, we have $\n_0^R = g - 1$, giving rise to the extra factor that usually appears in the effective dilaton.  However, it is important for us that one can also consider more general backgrounds with different choices of $\n_0^R$.} 
\begin{equation}
e^{2\pi i\Omega(u, \nu)} = \prod_I\prod_{(\rho, \omega)\in R_I\times S_I} (1 - e^{2\pi i(\rho(u) + \omega(\nu) + \nu_R r_I)})\prod_{\alpha\in \Delta} (1 - e^{2\pi i\alpha(u)})^{-1}.
\label{effectivedilaton}
\end{equation}
In addition, the fibering operators are given by
\begin{align}
\cG_{q_i, p_i}(u, \nu)_{\n_i, \m_i} &= \cG^{\cW}_{q_i, p_i}(u)_{\n_i}\cG^\text{CS}_{q_i, p_i}(u, \nu)_{\n_i, \m_i} \nonumber \\
&\phantom{==} \times \prod_I\prod_{(\rho, \omega)\in R_I\times S_I} \cG^\Phi_{q_i, p_i}(\rho(u) + \omega(\nu) + \nu_R r_I)_{\rho(\n_i) + \omega(\m_i) + \n_i^R r_I},
\end{align}
where we have defined the chiral multiplet contribution
\begin{equation}
\cG_{q_i, p_i}^\Phi(u)_{\n_i} = \Pi_{q_i, p_i}^\Phi(u)_{\n_i}\cG_{q_i, p_i}^\Phi(u)
\end{equation}
with\footnote{Recall that $(x; q)_n = (x; q)_\infty/(q^n x; q)_\infty$ (where $n\in \Z$) is the finite $q$-Pochhammer symbol.}
\begin{align}
\Pi_{q_i, p_i}^\Phi(u)_{\n_i} &\equiv (e^{2\pi iu/q_i}; e^{2\pi it_i/q_i})_{-\n_i}, \\
\cG_{q_i, p_i}^\Phi(u) &\equiv e^{\frac{p_i}{q_i}\left(\frac{1}{2\pi i}\operatorname{Li}_2(e^{2\pi iu}) + u\log(1 - e^{2\pi iu})\right)}\prod_{\ell=1}^{q_i-1} (1 - e^{2\pi i(u + t_i\ell)/q_i})^{\ell/q_i}, \label{Gqpchiral}
\end{align}
the vector multiplet contribution
\begin{equation}
\cG^{\cW}_{q_i, p_i}(u)_{\n_i} = \left(\frac{1}{\sqrt{q_i}}\cG^{(0)}_{q_i, p_i}\right)^{\operatorname{rank}(G)}\prod_{\alpha > 0} \cG^{\cW_0}_{q_i, p_i}(\alpha(u))_{\alpha(\n_i)} \label{vectormult}
\end{equation}
with
\begin{equation}
\cG^{\cW_0}_{q_i, p_i}(u)_{\n_i}\equiv (-1)^{\n_i(t_i + \ell_i^R t_i + 2\nu_R s_i)}\frac{\sin(\pi(u - t_i\n_i)/q_i)}{\sin(\pi u)}, \label{wboson}
\end{equation}
and the Chern-Simons contribution
\begin{align}
\cG^\text{CS}_{q_i, p_i}(u, \nu)_{\n_i, \m_i} &= \prod_a \cG_{q_i, p_i}^{G_a G_a}(u_a)_{\n_{ia}}^{k^{aa}}\prod_\alpha \cG_{q_i, p_i}^{F_\alpha F_\alpha}(\nu_\alpha)_{\m_{i\alpha}}^{k^{\alpha\alpha}} \\
&\times \prod_{a>b} \cG_{q_i, p_i}^{G_a G_b}(u_a, u_b)_{\n_{ia}, \n_{ib}}^{k^{ab}}\prod_{a, \alpha} \cG_{q_i, p_i}^{G_a F_\alpha}(u_a, \nu_\alpha)_{\n_{ia}, \m_{i\alpha}}^{k^{a\alpha}}\prod_{\alpha>\beta} \cG_{q_i, p_i}^{F_\alpha F_\beta}(\nu_\alpha, \nu_\beta)_{\m_{i\alpha}, \m_{i\beta}}^{k^{\alpha\beta}}. \nonumber
\end{align}
Here, we have introduced $t_i, s_i$ satisfying $p_i t_i + q_i s_i = 1$.  We have also introduced the convenient phase
\begin{equation}
\cG^{(0)}_{q_i, p_i} = \exp\left[\pi i\left(\frac{p}{12q} - \frac{1}{4q}\sum_{\ell=1}^{q-1} \cot\left(\frac{\pi\ell}{q}\right)\cot\left(\frac{\pi\ell p}{q}\right)\right)\right]
\end{equation}
and written the Chern-Simons contribution as a product of powers of the ``diagonal'' and ``mixed'' contributions
\begin{align}
\cG_{q_i, p_i}^{GG}(u)_{\n_i} &= (-1)^{\n_i(1 + t_i + \ell_i^R t_i + 2\nu_R s_i)}\exp\left(-\frac{\pi i}{q_i}(p_i u^2 - 2\n_i u + t_i\n_i^2)\right), \label{csdiagonal} \\
\cG_{q_i, p_i}^{G_a G_b}(u_a, u_b)_{\n_{ia}, \n_{ib}} &= \exp\left(-\frac{2\pi i}{q_i}(p_i u_a u_b - \n_{ia}u_b - \n_{ib}u_a + t_i\n_{ia}\n_{ib})\right). \label{csmixed}
\end{align}
In this paper, our interest is primarily in abelian theories, for which we may ignore the W-boson contributions \eqref{wboson} in \eqref{vectormult}, leaving only the prefactor due to the $\operatorname{rank}(G)$ abelian vector multiplets.

There may additionally exist $R$-symmetry and gravitational Chern-Simons terms, which contribute as overall factors.  The contributions of $RR$ and mixed gauge-$R$ Chern-Simons terms are as in \eqref{csdiagonal} and \eqref{csmixed}, with appropriate substitutions by $\nu_R$ and $\n_i^R$ from \eqref{rsymbackground}, raised to appropriate powers $k_{RR}$ and $k_{GR}$.  Up to a sign, the gravitational Chern-Simons term at unit level contributes the phase $\mathcal{G}_{q, p}^\text{grav} = (\mathcal{G}_{q, p}^{RR})^{-1/2}\mathcal{G}_{q, p}^{(0)}$.

\subsection*{Relation to Holomorphic Blocks}

In view of the explicit formulas above, we can derive the key relation \eqref{fiberingfromblock} for the free chiral multiplet. (Here, we essentially collect in a convenient place some arguments in Section 9 and Appendix F of \cite{Closset:2018ghr}.) From \cite{Beem:2012mb}, we have the asymptotic
\begin{equation}
(\mathfrak{q}y; \mathfrak{q})_\infty \xrightarrow{\tau\to 0} \exp\left(\frac{1}{2\pi i\tau}\operatorname{Li}_2(y) - \frac{1}{2}\log(1 - y) + O(\tau)\right).
\label{asymptoticbottom}
\end{equation}
To take the $\tau\to 0$ limit of $(\tilde{\mathfrak{q}}\tilde{y}; \tilde{\mathfrak{q}})_\infty$, we write
\begin{equation}
(\tilde{\mathfrak{q}}\tilde{y}; \tilde{\mathfrak{q}})_\infty = \prod_{\ell=0}^{q-1} (\tilde{\mathfrak{q}}^q(\tilde{\mathfrak{q}}^{-\ell}\tilde{y}); \tilde{\mathfrak{q}}^q)_\infty,
\label{qidentity}
\end{equation}
which allows us to leverage the fact that $\lim_{\tau\to 0} \tilde{\mathfrak{q}}^q = 1$.  For small $\tau$, we have
\begin{align}
\tilde{\mathfrak{q}}^q &= \exp\left[\frac{2\pi i\tau}{q}\left(1 - \frac{p\tau}{q}\right) + O(\tau^3)\right], \\
\tilde{\mathfrak{q}}^{-\ell}\tilde{y} &= e^{2\pi i(\nu + t\ell)/q}\left(1 - \frac{2\pi i(p\nu + \ell)\tau}{q^2}\right) + O(\tau^2), \\
\operatorname{Li}_2(\tilde{\mathfrak{q}}^{-\ell}\tilde{y}) &= \operatorname{Li}_2(e^{2\pi i(\nu + t\ell)/q}) + \frac{2\pi i(p\nu + \ell)\tau}{q^2}\log(1 - e^{2\pi i(\nu + t\ell)/q}) + O(\tau^2)
\end{align}
(the latter by virtue of $\operatorname{Li}_2'(z) = -\frac{1}{z}\log(1 - z)$), so that
\begin{gather}
(\tilde{\mathfrak{q}}\tilde{y}; \tilde{\mathfrak{q}})_\infty \xrightarrow{\tau\to 0} \label{asymptotictop} \\
\exp\left[\sum_{\ell=0}^{q-1} \left(\frac{p\tau + q}{2\pi i\tau}\operatorname{Li}_2(e^{2\pi i(\nu + t\ell)/q}) + \left(\frac{p\nu + \ell}{q} - \frac{1}{2}\right)\log(1 - e^{2\pi i(\nu + t\ell)/q})\right) + O(\tau)\right]. \nonumber
\end{gather}
Using the identities
\begin{equation}
\sum_{\ell=0}^{q-1} \operatorname{Li}_2(e^{2\pi i(\nu + t\ell)/q}) = \frac{1}{q}\operatorname{Li}_2(e^{2\pi i\nu}), \quad \sum_{\ell=0}^{q-1} \log(1 - e^{2\pi i(\nu + t\ell)/q}) = \log(1 - e^{2\pi i\nu}),
\end{equation}
both of which are consequences of \eqref{completeresidueproperty}, the ratio of \eqref{asymptotictop} and \eqref{asymptoticbottom} becomes
\begin{equation}
\frac{(\tilde{\mathfrak{q}}\tilde{y}; \tilde{\mathfrak{q}})_\infty}{(\mathfrak{q}y; \mathfrak{q})_\infty} \xrightarrow{\tau\to 0} \mathcal{G}_{q, p}^\Phi(\nu)
\end{equation}
with $\mathcal{G}_{q, p}^\Phi(\nu)$ as in \eqref{Gqpchiral}.

In fact, it is no more difficult to derive a more general relation that includes flux:
\begin{equation}
\mathcal{G}_{q, p}^\Phi(\nu)_{\mathfrak{m}} = \lim_{\tau\to 0}\frac{B_{\tilde{g}}^\Phi(\nu, \tau)_{\mathfrak{m}}}{B^\Phi(\nu, \tau)} = \lim_{\tau\to 0}\frac{(\tilde{\mathfrak{q}}\tilde{y}; \tilde{\mathfrak{q}})_\infty/(\tilde{\mathfrak{q}}\tilde{y}; \tilde{\mathfrak{q}})_{\mathfrak{m}}}{(\mathfrak{q}y; \mathfrak{q})_\infty}.
\end{equation}
This amounts to observing that
\begin{equation}
(\tilde{\mathfrak{q}}\tilde{y}; \tilde{\mathfrak{q}})_{\mathfrak{m}}^{-1} \xrightarrow{\tau\to 0} (e^{2\pi i(\nu - t)/q}; e^{-2\pi it/q})_{\mathfrak{m}} = (e^{2\pi i\nu/q}; e^{2\pi it/q})_{-\mathfrak{m}} = \Pi_{q, p}^\Phi(\nu)_\mathfrak{m}.
\end{equation}

\section{Comparison to DGG} \label{DGGcomparison}

Given the many notational differences, it is useful to emphasize the points of contact between our work and previous work, particularly that of DGG.

\subsection{Example: \texorpdfstring{$S_b^3$}{Sb3}}

\subsubsection{Useful Identities}

Let $Q\equiv b + b^{-1}$ and $c_b\equiv \frac{iQ}{2}$.  From \cite{Faddeev:2000if}, the noncompact quantum dilogarithm and double sine function satisfy the inversion formulas
\begin{align}
e_b(x)e_b(-x) &= e^{i\pi x^2 - i\pi(1 + 2c_b^2)/6}, \label{inv1} \\
e_b(x)^{-1} &= e^{-i\pi x^2/2}s_b(-x), \label{inv2} \\
s_b(x)s_b(-x) &= e^{i\pi(1 + 2c_b^2)/6}, \label{inv3}
\end{align}
where \eqref{inv3} follows from \eqref{inv1} and \eqref{inv2}.  We also have
\begin{equation}
e_b(x - ib^{\pm 1}/2) = (1 + e^{2\pi b^{\pm 1}x})e_b(x + ib^{\pm 1}/2).
\label{shift}
\end{equation}
The basic Fourier transform identity and its inverse are:
\begin{align}
\int dx\, e_b(x)^{\pm 1}e^{2\pi iwx} &= C^{\pm 1}e^{\mp i\pi w^2}e_b(\pm w\pm iQ/2)^{\pm 1}, \label{ft} \\
e_b(x)^{\pm 1} &= C^{\pm 1}\int dy\, e^{-2\pi ixy}e^{\mp i\pi y^2}e_b(\pm y\pm iQ/2)^{\pm 1}. \label{ftinv}
\end{align}
The constant $C$ will not be important for us.  Finally, we have the following functional form of the operatorial pentagon identity:
\begin{equation}
\int dz\, e_b(z + iQ/2)e_b(z - x - iQ/2)^{-1}e^{-2\pi iz(u + iQ/2)} = e_b(u - x)e_b(-x - iQ/2)^{-1}e_b(u)^{-1}.
\label{pentagon}
\end{equation}
The contour of integration is $\mathbb{R}$ in all cases.

\subsubsection{Conventions}

In the conventions of \cite{Dimofte:2011ju}, the tetrahedron theory is a free chiral with a particular choice of Chern-Simons contact term for the background $U(1)$ gauge field.  With real mass $m$ for the $U(1)$ flavor symmetry and $R$-charge $r$, we have up to constant factors that
\begin{equation}
Z_{S_b^3}(T_{\Delta, \Pi_Z}) = e^{\frac{i\pi}{2}X^2}s_b(X)\propto e_b(X), \quad X\equiv \frac{iQ}{2}(1 - r) - m.
\label{DGGresult}
\end{equation}
This is the result in the $U(1)_{1/2}$ quantization, to use the language of \cite{Closset:2018ghr}.

On the other hand, from \eqref{bethesum} and Appendix \ref{explicit}, we have for this theory that
\begin{equation}
Z_{S_b^3}^\Phi = \Pi^\Phi(\nu + \nu_R r)\mathcal{G}_{q_1, p_1}^\Phi(\nu + \nu_R r)\mathcal{G}_{q_2, p_2}^\Phi(\nu + \nu_R r), \quad \nu_R = \frac{q_1 + q_2}{2} = \frac{\sqrt{q_1 q_2}}{2}Q,
\label{Seifertresult}
\end{equation}
where $b^2 = q_1/q_2$ and $q_1 p_2 + q_2 p_1 = 1$.  The relation between \eqref{DGGresult} and \eqref{Seifertresult} can be deduced from a factorization formula of \cite{Garoufalidis:2014ifa} for the quantum dilogarithm at rational $b^2$, as formulated in Appendix D of \cite{Closset:2018ghr}:
\begin{equation}
e_b\left(\frac{i\nu}{\sqrt{q_1 q_2}} - \frac{iQ}{2}\right)^{-1} = \Pi^\Phi(\nu)\mathcal{G}_{q_1, p_1}^\Phi(\nu)\mathcal{G}_{q_2, p_2}^\Phi(\nu), \quad b^2 = q_1/q_2, \quad q_1 p_2 + q_2 p_1 = 1.
\end{equation}
Setting $\nu = -i\sqrt{q_1 q_2}m$, we get that
\begin{equation}
Z_{S_b^3}^\Phi(\nu + \nu_R r) = e_b(-X)^{-1}\propto e^{-i\pi X^2}e_b(X).
\end{equation}
Now we use the asymptotic\footnote{More precisely, we have $e_b(z)|_{|z|\to\infty}\sim e^{i\pi z^2 - i\pi(1 + 2c_b^2)/6}$ for real $b$ and $|{\arg z}| < \pi/2$ \cite{Faddeev:2000if}.  When $b = 1$, \eqref{ebasymptotic} follows from
\begin{equation}
e_{b=1}(z) = (1 - e^{2\pi z})^{iz}e^{-\frac{1}{2\pi i}\operatorname{Li}_2(e^{2\pi z})}
\label{b1dilogarithm}
\end{equation}
and properties of the ordinary dilogarithm, namely $\operatorname{Li}_2(0) = 0$ and the reflection identity
\begin{equation}
\operatorname{Li}_2(z) + \operatorname{Li}_2\left(\frac{1}{z}\right) = -\frac{\pi^2}{6} - \frac{1}{2}\log(-z)^2.
\end{equation}}
\begin{equation}
e_b(-X)^{-1}\to \begin{cases} 1 & \text{as $X\to\infty$}, \\ e^{-i\pi X^2} & \text{as $X\to -\infty$} \end{cases}
\label{ebasymptotic}
\end{equation}
to see that the result \eqref{Seifertresult} is indeed given in the $U(1)_{-1/2}$ quantization, as consistent with the fact that a Chern-Simons contribution of level $+1$ on $S_b^3$ is
\begin{equation}
e^{-i\pi\nu^2(p_1/q_1 + p_2/q_2)} = e^{-i\pi\nu^2/q_1 q_2} = e^{i\pi m^2}
\end{equation}
and the fact that the limits of large positive or negative real mass ($\operatorname{Im}(\nu)\to \pm\infty$) are equivalent to $X\to \pm\infty$, or $m\to \mp\infty$.  Therefore, adjusting \eqref{Seifertresult} to the $U(1)_{1/2}$ quantization as in \eqref{DGGresult} requires multiplying by an exponential in $\nu$.

\subsubsection{Mirror Check}

Invariance of the $\mathcal{M}_3$ partition function under cyclic permutations of the polarizations $\Pi_Z, \Pi_{Z'}, \Pi_{Z''}$ of the tetrahedron theory means invariance under the \emph{affine} $ST$-trans\-for\-ma\-tion $\rho = \sigma\circ S\circ T$ where $S = (\begin{smallmatrix} 0 & -1 \\ 1 & 0 \end{smallmatrix})$ and $T = (\begin{smallmatrix} 1 & 0 \\ 1 & 1 \end{smallmatrix})$, which acts as
\begin{equation}
\rho : \left(\begin{array}{c} Z \\ Z'' \end{array}\right)\mapsto \left(\begin{array}{c} Z' \\ Z \end{array}\right) = \left(\begin{array}{cc} -1 & -1 \\ 1 & 0 \end{array}\right)\left(\begin{array}{c} Z \\ Z'' \end{array}\right) + \left(\begin{array}{c} i\pi \\ 0 \end{array}\right)
\end{equation}
on the space of polarizations (note that the notations of \cite{Dimofte:2011ju} and \cite{Dimofte:2014zga} differ here).  Like $ST$, $\rho$ gives the identity when iterated thrice.  The actions of $S$, $T$, and affine shifts on $S_b^3$ partition functions are given in Section 6.1 of \cite{Dimofte:2011ju}.  In particular, starting with $Z_{S_b^3}(\tilde{m}) = e_b(iQ/2 - \tilde{m})$ where $\tilde{m}\equiv m + (iQ/2)r$ as in \eqref{DGGresult}, we have
\begin{align}
Z_{S_b^3}(\tilde{m}) &\xrightarrow{T} e^{-i\pi\tilde{m}^2}Z_{S_b^3}(\tilde{m}) \vphantom{\frac{}{1}} \\
&\xrightarrow{S} \int d\tilde{m}\, e^{-2\pi i\tilde{m}\tilde{m}'}e^{-i\pi\tilde{m}^2}Z_{S_b^3}(\tilde{m}) \\
&\xrightarrow{\sigma} \int d\tilde{m}\, e^{-i\pi\tilde{m}(\tilde{m} + 2\tilde{m}' - iQ)}Z_{S_b^3}(\tilde{m}), \label{rhoonZ}
\end{align}
where $\sigma : \tilde{m}'\mapsto \tilde{m}' - iQ/2$ and the contour of integration is $\mathbb{R}$.  The statement that $(\rho\circ Z_{S_b^3})(\tilde{m}') = Z_{S_b^3}(\tilde{m}')$ is then
\begin{equation}
\int d\tilde{m}\, e^{-i\pi\tilde{m}(\tilde{m} + 2\tilde{m}' - iQ)}e_b\left(\frac{iQ}{2} - \tilde{m}\right) = e_b\left(\frac{iQ}{2} - \tilde{m}'\right)
\label{mirrorcheck}
\end{equation}
(up to a constant), which is simply a rewriting of the standard identity \eqref{ftinv}.

Up to constants, and with no fluxes, we instead find that
\begin{equation}
\int du\, e^{-\frac{i\pi}{q_1 q_2}u^2}e^{-\frac{2\pi i}{q_1 q_2}u\zeta}Z_{S_b^3}^\Phi(u + \nu_R r) = e^{-\frac{i\pi}{q_1 q_2}\nu_R^2 r^2}e^{\frac{2\pi i}{q_1 q_2}\nu_R r\zeta}Z_{S_b^3}^\Phi(\zeta + \nu_R(1 - r))
\label{firstequality}
\end{equation}
where the contour of integration is given by $\operatorname{Re}(u) = \delta$ with
\begin{equation}
0 < \delta + \nu_R r < -\operatorname{Re}(\zeta) + \nu_R r < \nu_R.
\label{inequalities}
\end{equation}
We have used that
\begin{equation}
\kappa = -\frac{1}{2}\left(\frac{1}{q_1} + \frac{1}{q_2}\right) = -\frac{\nu_R}{q_1 q_2}, \quad c = \frac{p_1}{q_1} + \frac{p_2}{q_2} = \frac{1}{q_1 q_2}.
\end{equation}
Let us make the actions of $S$, $T$, and affine shifts of $\zeta$ manifest in the result of the Seifert fibering formalism.  The inequalities \eqref{inequalities} suggest that the natural variables to use are $\tilde{u} = u + \nu_R r$ and $\tilde{\zeta} = \zeta - \nu_R r$, in terms of which \eqref{firstequality} becomes
\begin{equation}
\int d\tilde{u}\, e^{-\frac{i\pi}{q_1 q_2}\tilde{u}^2}e^{-\frac{2\pi i}{q_1 q_2}\tilde{u}\tilde{\zeta}}Z_{S_b^3}^\Phi(\tilde{u}) = Z_{S_b^3}^\Phi(\tilde{\zeta} + \nu_R)
\label{secondequality}
\end{equation}
where the contour is along constant $\operatorname{Re}(\tilde{u})$ with
\begin{equation}
0 < \operatorname{Re}(\tilde{u}) < -\operatorname{Re}(\tilde{\zeta}) < \nu_R.
\end{equation}
The result \eqref{secondequality} can be written in a more symmetric way in terms of $\tilde{\zeta}' = \tilde{\zeta} + \nu_R$ as
\begin{equation}
\int d\tilde{u}\, e^{-\frac{i\pi}{q_1 q_2}\tilde{u}^2}e^{-\frac{2\pi i}{q_1 q_2}\tilde{u}(\tilde{\zeta}' - \nu_R)}Z_{S_b^3}^\Phi(\tilde{u}) = Z_{S_b^3}^\Phi(\tilde{\zeta}')
\label{thirdequality}
\end{equation}
where the contour is along constant $\operatorname{Re}(\tilde{u})$ with
\begin{equation}
\operatorname{Re}(\tilde{u}), \operatorname{Re}(\tilde{\zeta}')\in (0, \nu_R), \quad \operatorname{Re}(\tilde{u}) + \operatorname{Re}(\tilde{\zeta}') < \nu_R.
\end{equation}
In the form \eqref{thirdequality}, the $\rho$-action is completely transparent, and the interpretation is exactly as in \eqref{mirrorcheck}.

Notice that all of the work we did above was simply to make the action of affine shifts manifest: the effects of the other ``DGG operations'' on fibering operators are more or less immediate.  In particular, Section 5 of \cite{Closset:2018ghr} presents examples of non-affine symplectic transformations on fibering operators, and the superpotential has no effect on fibering operators aside from restricting global symmetry charges.  To take a step back from $S_b^3$ to general $\mathcal{M}_3$, the physical meaning of the affine shifts is as follows.  Real masses are complexified by $R$-charges of matter fields and the geometric parameter $\nu_R$ (in our contour conventions, the mass parameters $u + \nu_R r$, deformed by the $R$-charge, are actually real).  These mass parameters are identified with positions in $\mathcal{P}_{\partial M}$ or the corresponding Hilbert space $\mathcal{H}_{\partial M}$ in the $\mathcal{M}_3$-TQFT.  A classical shift of a position coordinate by $\pm i\pi$ shifts the $R$-charge of the corresponding operator by $\pm 1$, and hence the ``complexified'' mass parameter by $\nu_R$ \cite{Dimofte:2011ju}.  The affine shifts also entail shifts of the integration contour, to ensure convergence \cite{Dimofte:2014zga}.

Above, we have made explicit the meaning of the $\rho$-invariance of the tetrahedron theory in the Seifert formalism, starting with the formulas of Section \ref{duality1}.  One can do the same for the invariance of the bipyramid theory under 2-3 moves, starting with the formulas of Section \ref{duality2}.  The result is a variation on the pentagon identity \eqref{pentagon}.  We leave this as an exercise for the attentive reader.

\subsection{Example: PHS}

On the PHS, we have
\begin{align}
Z_\text{PHS}^\Phi(\nu) &= \Pi^\Phi(\nu)\mathcal{G}_{2, -1}^\Phi(\nu)\mathcal{G}_{3, 1}^\Phi(\nu)\mathcal{G}_{5, 1}^\Phi(\nu) \\
&= (1 - e^{2\pi i\nu})^{\nu/30 - 1}e^{\frac{1}{30}\frac{1}{2\pi i}\operatorname{Li}_2(e^{2\pi i\nu})}\prod_{q = 2, 3, 5}\prod_{\ell_q = 1}^{q-1} (1 - e^{2\pi i(\nu + \ell_q)/q})^{\ell_q/q}
\end{align}
in the $U(1)_{-1/2}$ quantization (see \eqref{PHSasymptotics}), where we have again used \eqref{Gqpchiral} (with $t\equiv 1\text{ (mod $q$)}$ in all cases relevant to us).

Let us again run through the statement of $\rho$-invariance of the tetrahedron theory.  From Sections \ref{duality1} and \ref{PHSexample}, the gauged/free chiral duality implies the relation
\begin{equation}
\int du\, e^{-\frac{i\pi}{30}u^2}e^{-\frac{2\pi i}{30}u\zeta}Z_\text{PHS}^\Phi\left(u + \frac{1}{2}r\right) = e^{-\frac{i\pi}{120}r^2}e^{\frac{i\pi}{30}r\zeta}Z_\text{PHS}^\Phi\left(\zeta + \frac{1}{2}(1 - r)\right),
\end{equation}
up to a constant factor.  According to \eqref{b1dilogarithm}, this can be seen as an identity involving the quantum dilogarithm at $b = 1$.  Note that since
\begin{equation}
\mathcal{G}_{q, p}^{U(1)_k}(u) = e^{-\frac{p}{q}k\pi iu^2},
\end{equation}
a $U(1)_k$ Chern-Simons contribution on the PHS takes the form
\begin{equation}
e^{\frac{1}{2}k\pi iu^2}e^{-\frac{1}{3}k\pi iu^2}e^{-\frac{1}{5}k\pi iu^2} = e^{-\frac{1}{30}k\pi iu^2}.
\end{equation}
Since $\nu_R = 1/2$ in this case, the contour of integration is given by $\operatorname{Re}(u) = \delta$ where
\begin{equation}
0 < \delta + \frac{1}{2}r < -\operatorname{Re}(\zeta) + \frac{1}{2}r < \frac{1}{2}.
\end{equation}
In terms of the variables $\tilde{u} = u + \frac{1}{2}r$ and $\tilde{\zeta} = \zeta + \frac{1}{2}(1 - r)$, we equivalently have
\begin{equation}
\int d\tilde{u}\, e^{-\frac{i\pi}{30}\tilde{u}^2}e^{-\frac{2\pi i}{30}\tilde{u}(\tilde{\zeta} - \frac{1}{2})}Z_\text{PHS}^\Phi(\tilde{u}) = Z_\text{PHS}^\Phi(\tilde{\zeta})
\end{equation}
where the contour is along constant $\operatorname{Re}(\tilde{u})$ with
\begin{equation}
\operatorname{Re}(\tilde{u}), \operatorname{Re}(\tilde{\zeta})\in \left(0, \frac{1}{2}\right), \quad \operatorname{Re}(\tilde{u}) + \operatorname{Re}(\tilde{\zeta}) < \frac{1}{2}.
\end{equation}
The effect of the affine $ST$-transformation is now manifest.

\bibliographystyle{JHEP}
\bibliography{seifert}

\end{document}